\documentclass[12pt,letterpaper]{article}%

\pdfoutput=1 

\usepackage{enumerate}
\usepackage{color,amssymb,amsmath}
\usepackage[pdftex]{graphicx}
\usepackage[normalem]{ulem} 
\usepackage{float}
\usepackage[square,numbers]{natbib}
\graphicspath{{./figures_linear/}{./figures_nonlinear/}{./figures_schematic/}}

\newcommand{\be}{\begin{equation} }
\newcommand{\ee}{\end{equation}}
\newcommand{\bse}{\begin{subequations} }
\newcommand{\ese}{\end{subequations}}
\newcommand{\bi}{\begin{itemize} }
\newcommand{\ei}{\end{itemize}}
\newcommand{\bmath}{\begin{displaymath} }
\newcommand{\emath}{\end{displaymath} }

\newcommand{\rd}{\mathrm{d}}

\newcommand{\pd}[2]{\frac{\partial #1}{\partial #2}}

\newcommand{\pdl}[2]{{\partial #1}/{\partial #2}}

\newcommand{\od}[2]{\frac{\dd{#1}}{\dd{#2}}}
\newcommand{\bm}[1]{\mbox{\boldmath$ #1 $}}

\newcommand{\dd}{\textrm{d}}

\newcommand{\Pe}{\mathit{Pe}\,}

\newcommand{\var}{\mathrm{var}}

\begin{document}

%%%% Article title to be placed here
\title{Ice stream formation}

\author{%%%% Author details
Christian Schoof$^{1}$, and Elisa Mantelli$^{2}$\\%}
%%%%%%%%% Insert author address here
%\address{
$^{1}$Department of Earth, Ocean and Atmospheric Sciences, \\
University of British Columbia, Vancouver, Canada\\
$^{2}$AOS Program, Princeton University, Princeton, USA}

\maketitle

%%%% Subject entries to be placed here %%%%
%\subject{glaciology, mathematical modelling, fluid dynamics}

%%%% Keyword entries to be placed here %%%%
%\keywords{ice streams, pattern formation, glacier sliding, free boundaries}

%%%% Insert corresponding author and its email address}
%\corres{Christian Schoof\\
%\email{cschoof@eoas.ubc.ca}}

\begin{abstract}
Ice streams are bands of fast-flowing ice in ice sheets. We investigate their formation as an example of spontaneous pattern formation, based on positive feedbacks between dissipation and basal sliding. Our focus is on temperature-dependent subtemperate sliding, where faster sliding leads to enhanced dissipation and hence warmer temperatures, weakening the bed further, although we also treat a hydromechanical feedback mechanism that operates on fully molten beds. We develop a novel thermomechanical model capturing ice-thickness scale physics in the lateral direction while assuming the the flow is shallow in the main downstream direction. Using that model, we show that formation of a steady-in-time  pattern can occur by the amplification in the downstream direction of noisy basal conditions, and often leads to the establishment of a clearly-defined ice stream separated from slowly-flowing, cold-based ice ridges by narrow shear margins, with the ice stream widening in the downstream direction. We are able to show that downward advection of cold ice is the primary stabilizing mechanism, and give an approximate, analytical criterion for pattern formation.
\end{abstract}

\section{Introduction} \label{sec:intro}

Ice streams are narrow bands of fast flow within otherwise more slowly-flowing ice sheets, often forming near the margin or grounding line of the ice sheet as outlets that can carry the majority of the ice discharged \cite{AlleyBindschadler2001}. Some ice streams are confined to topographic lows that channelize flow \cite{TrufferEchelmeyer2003}, but not all, and those that are not controlled by topography may occur in parallel arrays of roughly similar ice streams. Where bed topography is not the primary control, several positive feedback mechanisms have been suggested for the formation of an alternating pattern of fast and slow flow.

The viscosity of ice is temperature-dependent, and as a result there is a natural positive feedback between dissipation, reduced viscosity, and faster flow, studied in detail in \cite{Hindmarsh2004} and likely the cause of often pathological pattern formation in early thermomechanical ice sheet simulations (in the sense of grid-dependent results with no evidence of a short-wavelength cut-off, see \cite{Payneetal2000}). An alternative view holds that ice streams form as the result of high water pressure (or more precisely, lower effective pressure, the difference between normal stress and water pressure) at the ice sheet bed. This can cause pattern formation through a hydromechanical positive feedback in which a reduced effective pressure leads to faster sliding and more water production, which in turn requires effective pressure to be decreased further in order to open basal drainage conduits to evacuate that water \cite{FowlerJohnson1996,KyrkeSmithetal2014,KyrkeSmithetal2015}.

Situated somewhere between these two extremes is the possibility that sliding can occur at the ice sheet bed before the melting point has been reached, and there is no liquid water. Friction generated by this type of subtemperate sliding can be expected to be temperature-dependent \cite{Barnesetal1971,Fowler1986b}, leading to a potential positive feedback between raised temperature, faster flow, and enhanced dissipation of heat. This has been studied previously by \cite{Hindmarsh2009}, though using a simplified model and ostensibly as a way of emulating the dissipation-viscosity feedback described above.

Motivated by recent work \cite{Mantellietal2019} that demonstrates that subtemperate sliding must happen somewhere in the transition between a cold-based ice sheet that essentially does not slide, and a warm-based ice sheet sliding over a bed at the melting point, we revisit the pattern formation due to subtemperate sliding here. That said, our model turns out to be equally applicable to the hydromechanical patterning process described above.

The paper is organized as follows: first, we develop a novel model that combines thin-film balances in the main along-flow direction with the full mechanics and thermal physics relevant to lateral length scales comparable to ice thickness (\S \ref{sec:model}). This turns out to be key to capturing negative feedbacks that stabilize short wavelengths during pattern formation, and to allowing fully formed ice streams to emerge with a sharp lateral margin (henceforth, `margin') between the ice stream and the surrounding, slowly flowing ice ridge. In that sense, the model is the minimal model that allows these features to form self-consistently.

In order to isolate the basal dissipation feedback, we make ice viscosity independent of temperature. In fact, we use a constant viscosity model. Incorporating a temperature- and strain-rate-dependent viscosity like Glen's law \cite{CuffeyPaterson2010} into the model in future is straightforward in principle. Next, we look for steady state solutions of the model (\S \ref{sec:solution_all}\ref{sec:solution}), whose form allows unique solutions to be found by simple forward integration from a central ice divide \cite{Hutteretal1986,Yakowitzetal1986}. We show numerically that small perturbations of basal properties can become amplified in the downstream direction, but that this need not happen, and that it can occur in two flavours: the dissipation-temperature feedback in the region of subtemperate sliding, and, in the absence of the former, the hydromechanical feedback for temperate beds (\S\S \ref{sec:solution_all}\ref{sec:pattern}--\ref{sec:hydraulic_pattern}).

Next, we determine analytically a criterion for steady-state patterning to occur during subtemperate sliding (\S \ref{sec:stability}), with additional results regarding hydromechanical patterning confined to the supplementary material. We conclude by putting our results in the context of the existing literature (\S \ref{sec:discussion}) and identifying pressing areas for future research (\S \ref{sec:conclusion}). For a broader overview of the literature on ice stream modelling in the context of ice sheet simulations, we refer the interested reader to the introduction to \cite{Mantellietal2019}.

\section{Model Formulation} \label{sec:model}

%\begin{subequations}
We use a dimensionless curvilinear coordinate system $(x,y,z)$ in which the $z$-axis is oriented vertically, $x$ is oriented in the mean flow direction and $y$ is transverse to the mean flow direction. For any fixed $x$, $z = 0$ corresponds to the elevation of the bed, which we assume to be a given function of the downstream coordinate $x$ only (a model extension to beds with lateral variation is possible but not useful as a first step). The implied curvature of the coordinate system will turn out not to matter since the radius of curvature is comparable with the length of the ice sheet and curvature terms therefore scale as the ice sheet aspect ratio $\varepsilon$. The coordinates $(y,z)$ describing position the in plane transverse to the main flow direction are scaled with a typical ice thickness scale, and (despite the curvilinear nature of the coordinate system), the $(y,z)$-coordinate system is Cartesian for fixed $x$. The along-flow coordinate is scaled with the length of the ice sheet.

Let $\nabla_\perp = (\pdl{}{y},\pdl{}{z})$ be the gradient operator in the $(y,z)$-plane. Correspondingly, define $\bm{v}_\perp = (v,w)$ as a transverse velocity field, scaled by the usual shearing velocity scale \cite{FowlerLarson1978} multiplied by the ice sheet aspect ratio $\varepsilon$, while $u$ is the component of velocity in the main flow direction, scaled with the shearing velocity scale. In other words, the `axial' velocity $u$ is physically much larger than the secondary transverse flow velocity $\bm{v}_\perp$. Note also that we choose the horizontal component $v$ of transverse velocity to be parallel to the $y$-axis, but take $w$ to be the velocity component in the direction that is locally normal to the bed, rather than necessarily in the vertical direction. This slightly unorthodox choice simplifies the boundary conditions on the flow problem in transverse plane. A full derivation of the leading order model from first principles is given in the supplementary material (\S\S S1--S2); here we proceed simply to state the model.

At leading order, surface elevation $s$ above a fixed datum is a function of $(x,t)$ only, where $t$ is time scaled with the advective time scale for the ice sheet. The secondary flow velocity $\bm{v}_\perp$ acts to smooth out any leading-order lateral surface elevation variations extremely quickly, so we treat $s$ as independent of $y$. Let $b(x)$ be bed elevation, and $h(x,t) = s(x,t)-b(x)$ be ice thickness. Assuming the domain is periodic in $y$ with period $W$, $h$ satisfies
\begin{align}
 \pd{h}{t} + \pd{Q}{x} = &  \bar{a} \label{eq:mass_conservation} \\
 Q = & W^{-1} \int_0^W \int_0^h u \rd z \rd y, \label{eq:flux_def} \\
 \bar{a} = & W^{-1}\int_0^W a \rd y, \label{eq:average_balance}
\end{align}
where $a$ is specific surface mass balance.

The main along-flow velocity determining the ice flux $Q$ satisfies the antiplane version of Stokes' equations, with the flow driven by a gradient in cryostatic pressure. With a constant viscosity, $u$ satisfies
\begin{subequations}  \label{eq:antiplane}
\begin{equation} \nabla_\perp^2 u = \pd{(b+h)}{x} \label{eq:axial_flow} \end{equation}
The boundary conditions on $u$ ultimately couple the problem of finding $u$ and $h$ to the transverse flow. At the surface, vanishing stress dictates that
\begin{equation} \pd{u}{z} = 0 \label{eq:no_shear_surf} \end{equation}
at $z = h$. At the bed $z = 0$, we assume a general friction law that incorporates dependence on temperature and effective pressure,
\begin{equation} \pd{u}{z} = f(T,N,|u|)u/|u| \label{eq:sliding_law} \end{equation}
\end{subequations}
where $T$ is temperature scaled with the difference between melting point and a representative surface temperature, and $N$ is effective pressure (the difference between normal stress at the bed and water pressure), scaled to result in an $O(1)$ permeability in the drainage model that we will describe shortly.

The function $f(T,N,u)$ decreases with increasing $T$ and increases or remains constant with increasing  $N$ and $|u|$. $f$ must also be non-negative, and with $Q$ positive, it can be shown that $u$ will likewise be positive \cite{Schoof2006a}, so the modulus signs in \eqref{eq:sliding_law} are strictly speaking redundant. For temperatures below the melting point, $T < 0$, we assume a linear friction law in sliding speed $u$, with a temperature-dependent friction coefficient:
\begin{equation} \label{eq:subtemp_sliding_law} f(T,N,u) = \gamma(T)u, \end{equation}
where the function $\gamma$ takes an $O(1)$ value  $\gamma(0) = \gamma_0$ at $T = 0$, and increases for decreasing $T $. We define an associated temperature scale $\delta$ over which significant changes in friction coefficient occur as $\delta = -\gamma(0)/\gamma_T(0)$. The subscript $T$ here denotes differentiation with respect to temperature. We will later use
$$ \gamma(T) = \gamma_0 \exp(-T/\delta) $$
as a concrete example \cite{Fowler1986b}. Physically, a law of the form \eqref{eq:subtemp_sliding_law} can be justified for instance by shearing of a pre-melted water film at the interface between ice and bed, with the thickness of that film increasing as the melting point is approached \cite{Dashetal1995}. Form drag \cite{Fowler1981} eventually replaces shearing across the premelted film  as the main source of friction near the melting point, leading to a smooth friction law with friction increasing with sliding velocity and decreasing with temperature.

Note that previous work on subtemperate sliding in \cite{Mantellietal2019} explicitly considered the limit $\delta \ll 1$, which is however fraught with instabilities \cite{MantelliSchoof2019}. Here, we retain $\delta$ as a nominally $O(1)$ parameter, although we will be concerned primarily with the case of small $\delta$ eventually: as we will show, taking the limit $\delta \ll 1$ in the confines of our already reduced model will be valid provided $\varepsilon \ll \delta^2$, where $\varepsilon$ is the ice sheet aspect ratio.

Effective pressure does not enter into the friction law for $T < 0$, while at $T = 0$, we can identify $f(0,N,u)$ as a temperate, effective-pressure dependent friction law. In the limit of large effective pressure $N$, the temperate friction law should agree with the subtemperate version $\gamma(T)u$ as $T$ approaches the melting point,
\begin{equation} \lim_{N \rightarrow \infty} f(0,N,u) = \lim_{T\rightarrow T^-} \gamma(T)u. \label{eq:slide_continuous} \end{equation}
Some of the choices we consider later are the following: a simple linear law \cite{Weertman1957,NYe1969,Kamb1970} serves as a control case in which dissipation of energy couples back to the flow of ice only through basal temperature, but not through production of water (or latent heat) 
\begin{equation} f(0,N,u) = \gamma_0 u.\label{eq:Weertman} \end{equation}
Feedbacks between water production and ice flow require an $N$-dependent law. We use a modified version of the commonly used power law $f(0,N,u) = C u N$ \cite{Buddetal1979}, of the form
\begin{equation} f(0,N,u) = \frac{\gamma_0 u N}{N_s+N}. \label{eq:Budd}\end{equation}
Division by $(N_s+N)$ ensures that the condition \eqref{eq:slide_continuous} is met, and finite sliding velocities are possible even at infinite effective pressure. $N_s$ is a scale for the change from a friction law that is independent of $N$ when $N \gg N_s$ to one that is sensitive to $N$ for $N \lesssim N_s$.
In addition, we consider a regularized Coulomb friction law \cite{Schoof2005}
\begin{equation} f(0,N,u) =  \frac{\mu_0 \gamma_0 u N}{ \gamma_0 u + \mu_0 N}, \label{eq:Coulomb_regular} \end{equation}
where $\mu_0$ is a friction coefficient such that $f(0,N,u) \sim \mu_0 N$ when $\gamma_0u \gg \mu_0 N$. Another closely related choice would be the continuation of a linear, $N$-independent law up to a yield stress as considered in \cite{Schoof2010b,Tsaietal2015,ZoetIverson2020},
\begin{equation} f(0,N,u) =  \left\{ \begin{array}{l l} \gamma_0 u & \mbox{when } \gamma_0 u \leq \mu_0 N, \\
                                      \mu_0 N & \mbox{when } \gamma_0 u > \mu_0 N.
                                     \end{array} \right.
 \end{equation}
This is however qualitatively very close to \eqref{eq:Coulomb_regular} while having the disadvantage of not being smooth.
 
Temperature $T$ in turn depends on the secondary transverse flow: even though the transverse velocity is much smaller than the along-flow velocity, advection happens over much shorter distances, and the along-flow and transverse advection terms both appear at the same order. The heat equation becomes
\begin{equation} \label{eq:heat} \Pe \left( \pd{T}{t} + \bm{v}_\perp \cdot \nabla_\perp T + u \pd{T}{x} \right) - \nabla_\perp^2 T = \alpha |\nabla_\perp u|^2 \end{equation}
for $0 < z < h$, where the right-hand side is the appropriate leading-order shear heating term for a constant viscosity. $\Pe$ is the P\'eclet number appropriate for advection along the length of the ice sheet, and $\alpha$ is a dimensionless shear heating rate, or Brinkmann number. We treat $\Pe$ and $\alpha$ as $O(1)$ constants, as is appropriate for typical ice sheets. We do not incorporate a model for temperate ice formation here \cite{SchoofHewitt2016,HewittSchoof2017}; this will be the subject of a separate publication, but we note that none of the numerical solutions in \S \ref{sec:solution_all}\ref{sec:results} predict spurious positive temperatures that would indicate the production of temperate ice. The heat equation has a counterpart in the ice sheet bed,
\begin{equation} \label{eq:heat_bed} \Pe \pd{T}{t} - \nabla_\perp^2 T = 0 \end{equation}
for $z < 0$.

The transverse velocity satisfies a `two-and-a-half-dimensional' version of Stokes' equations.
\begin{equation} \label{eq:Stokes} \nabla_\perp^2 \bm{v}_\perp - \nabla_\perp p =  \bm{0}, \qquad \nabla_\perp \cdot  \bm{v}_\perp =  -\pd{u}{x},
\end{equation}
where the usual incompressibility condition is augmented by an apparent source (or more likely, loss) term due to accelerating flow in the $x$-direction.

Boundary conditions on $\bm{v}_\perp$ at the surface $z  = h$ are given by vanishing shear stress and the local kinematic boundary condition,
\begin{align} -\pd{h}{x}\pd{u}{y}+ \pd{v}{z} + \pd{w}{y} = & 0, \label{eq:shear_stress_surface} \\
 w = & u \pd{h}{x} -\pd{Q}{x} - (a-\bar{a}). \label{eq:normal_velocity_surface}
\end{align}
The second of these two arises from the width-integrated mass conservation equation \eqref{eq:mass_conservation} and the local kinematic condition, $ \pdl{h}{t} + u\pdl{h}{x} = w + a, $
eliminating $\pdl{h}{t} -a$ between the two. In general, we will assume that $a$ does not vary significantly in the transverse direction, so $(a-\bar{a}) = 0$. Its retention may however be relevant as a source of spatial `noise' in the forcing of the problem, and contribute to pattern formation.

The condition requiring normal stress to vanish at the surface turns into a diagnostic equation that determines the correction $\varepsilon^2 s_1(x,y,t)$ to the mean surface elevation $s(x,t)$: the corresponding lateral surface elevation gradient turns out to be necessary to drive the transverse flow $\bm{v}_\perp$, but need not be computed to determine that velocity field. We have
\begin{equation} s_1 = p - \pd{w}{z}, \label{eq:surface_correction} \end{equation}
where the right-hand side can be evaluated once  velocity and pressure have been determined.

At the base of the ice sheet $z = 0$, we have the same temperature-dependent friction law governing shear stress as for the antiplane flow problem \eqref{eq:antiplane}, and a condition of vanishing normal velocity
\begin{align} \pd{v}{z} + \pd{w}{y} = & f(T,N,|u|) \frac{v}{|u|}, \label{eq:lateral_friction} \\
 w = & 0.\label{eq:normal_velocity_base}
\end{align}
Note that $|u|$ appears as the argument in the friction law $f$ because sliding \emph{speed} is dominated by the axial flow in the $x$-direction.

The boundary conditions on the heat equation meanwhile take the form of a prescribed surface temperature
\begin{equation} T = T_s \label{eq:temp_surf} \end{equation}
at $z = h$, where a uniform surface temperature would permit us to impose a constant $T_s = -1$. Far below the bed, a prescribed heat flux is imposed,
\begin{equation} -\pd{T}{z} \rightarrow G \label{eq:geothermal} \end{equation} 
as $z \rightarrow -\infty$. At the bed $z = 0$, we have continuity of temperature and conservation of energy
\begin{align} [T]_-^+ = & 0 \label{eq:temp_bed} \\
 \pd{e}{t} + \pd{q_x}{x} + \pd{q_y}{y} + \left[-\pd{T}{z}\right]_-^+ = & f(T,N,|u|) |u| \label{eq:basal_energy}
\end{align}
where $[\cdot]_-^+$ denotes the difference between the limits of the bracketed quantities taken from above and below $z = 0$, $e$ is the latent heat content or enthalpy of the bed per unit area, and $q_x$ and $q_y$ are the components of latent heat flux along the bed. We also enforce that temperature cannot become positive,
\begin{equation} T \leq 0 \qquad \mbox{at } z =0.\label{eq:temp_bound} \end{equation}

Latent heat takes the form of liquid water, so $e$ is water content per unit area of the bed, while $q_x$ and $q_y$ are components of water flux. We choose a macroporous drainage parameterization \cite{FowlerJohnson1996,FlowersClarke2002} in which flux is linear in the hydraulic gradient, but with a permeability that depends on temperature and effective pressure:
\begin{equation} \label{eq:hydrology}  e =  \left\{ \begin{array}{l } 0,\\
                              \Phi(N),
                             \end{array} \right. \qquad
              q_x =  \left\{ \begin{array}{l}  0, \\
-\kappa(N)\pd{(h+ r^{-1}b)}{x} , 
                              \end{array} \right.  \qquad
              q_y =   \left\{ \begin{array}{l l}  0 & \mbox{if } T < 0, \\ 
-\kappa(N) \pd{(\sigma_{nn}-\beta N)}{y} & \mbox{if } T = 0, 
                              \end{array} \right.
                              \end{equation}
              $$                          \sigma_{nn} =  \left.\left( p - 2\pd{w}{z} \right)\right|_{z=0}
                              $$
where $\kappa$ and $\Phi$ are positive, decreasing functions of the effective pressure variable $N$. $\beta$ is the ratio the effective pressure scale to the deviatoric stress scale, and we have defined effective pressure as the difference between normal stress at the bed and water pressure in the bed. $r = \rho_i/\rho_w$ is the ice-to-water density ratio.

Note that the definition of $e$ together with the constraint \eqref{eq:temp_bound} ensures a Dirchlet condition on temperature
\begin{equation} T = 0 \qquad \mbox{where } e > 0.
\label{eq:temp_pre_Dirichlet} \end{equation}
It is important to stress that the hydrology model above is a free boundary problem, with a free boundary $\Gamma_m$ separating the temperate subdomains at the bed (sets of points $(x,y)$ at which $T(x,y,0) = 0$) from subtemperate subdomains (points at which $T(x,y,0) < 0$, see also \cite{Schoof2012b,Haseloffetal2015,Haseloffetal2018}). 
\eqref{eq:basal_energy} holds everywhere, but the effective pressure $N$ is only strictly speaking defined on the temperate subdomain, where it controls the flux $(q_x,q_y)$ and bed water content  $\Phi$. 

At the free boundary $\Gamma_m$, we distinguish between a growing temperate region and a shrinking one. Consider a part of the free boundary at $y_m(x,t)$, and suppose without loss of generality that the cold subdomain lies to the left of the boundary $y = y_m$. For a growing temperate region, we have  $V = \pdl{y_m}{t} < 0$, we assume that migration occurs only when heat flux is non-singular at $y = y_m$ (so there is no singular freezing rate $[-\pdl{T}{z}]_-^+$) as previously studied in \cite{Mantellietal2019,Schoof2012b,Haseloffetal2015,Haseloffetal2018}:
\begin{equation} \label{eq:no_singular_flux} \lim_{y \rightarrow y_m^+} \left(|y_m-y|^{1/2}\left[\pd{T}{z}\right]_-^+\right) \geq 0 \qquad \mbox{ if } V < 0. \end{equation}
Note that the local analysis around the transition point $y = y_m$ in \cite{Haseloffetal2018} is applicable here, which shows that we may not be able to prevent freezing near the margin, but we can insist that freezing rates not be singular (in which case the left-hand side of \eqref{eq:no_singular_flux} is zero). The condition \eqref{eq:no_singular_flux} is mathematically  analogous to prescribing a vanishing fracture toughness in crack propagation problems \cite{Zehnder2012}, although applied to the thermal rather than the mechanical problem. The condition must be stated explicitly as part of the model since one could otherwise construct a solution purely mathematically in which heat is siphoned out of the temperate region adjacent to the free boundary while the temperate region is widening: in other words, a locally infinite rate of basal freezing  occurs in these solutions adjacent to the edge of a widening region of temperate bed, with the frozen-on water supplied by the basal drainage system (see supplementary material \S S2). Effectively, in these solutions, water flows into areas that were frozen and forcibly warms them up in a similar way to magma being injected into a dyke \cite{ListerKerr1991}, instead of dissipation due to sliding or viscous deformation of ice causing the previously frozen bed to thaw. Such water flow however seems unphysicial to us unless driven by overpressurization and hydrofracturing.

As shown in \cite{Schoof2012b,Haseloffetal2015,Haseloffetal2018}, the thermal problem alone then furnishes the migration rate $V$, and the relevant boundary condition on the hydrology problem \eqref{eq:basal_energy} inside the temperate region arises simply from the weak form of \eqref{eq:basal_energy} as
\begin{equation} \lim_{y \rightarrow y_m^+} \left( V e - q_y + \pd{y_m}{x} q_x\right) = 0, \label{eq:mass_cons_margin} \end{equation}
ensuring conservation of energy at the boundary. For a shrinking temperate domain, a singular heat flux is possible and the local form of the temperature field is not constrained at the boundary (see appendix B of \cite{Schoof2012b}), and demand instead that bed water content reach zero 
\begin{equation} e = 0 \qquad \mbox{at }y =y_m \mbox{ if } V > 0\label{eq:freeze_margin} \end{equation}
in addition to \eqref{eq:mass_cons_margin}. It is worth stressing these constraints as some other hydrology models for partially temperate beds \cite{Wolovicketal2014} to the contrary assume implicitly that water flux can penetrate into the portions of the bed, effectively by imposing \eqref{eq:mass_cons_margin} combined with  \eqref{eq:freeze_margin} in cases where the temperate domain is expanding.

There is one remaining technical difficulty we need to address.
Note that the direction (that is, the sign) of the downstream flux $q_x$ is controlled purely by ice and bed geometry, and we assume that the flux $q_x$ and velocity $u$ are always oriented in the positive $x$-direction. From \eqref{eq:axial_flow}, the latter implies the surface slope $\pdl{(h+b)}{x}$ is negative, and provided $(r^{-1}-1)\pdl{b}{x} < - \pdl{(h+b)}{x}$, so is the gradient of the hydraulic potential in the $x$-direction in \eqref{eq:hydrology}$_2$. All this means is that we exclude retrograde slopes $\pdl{b}{x}$ that are steep enough to pond water permanently.

Where the bed first becomes temperate as we move along the $x$-axis, the boundary of the temperate domain is locally perpendicular to the $x$-axis, and \eqref{eq:mass_cons_margin} becomes instead
\begin{equation} V_x e + q_x = 0  \end{equation}
where $V_x$ is the rate at which the transition point moves along the $x$-axis. If $V_x \leq 0$ (a transition point that is static or moving upstream) then requires that $q_x = 0$ since $e \geq 0$ and $q_x$ cannot be negative. With a positive hydraulic gradient that is independent of $N$, a vanishing flux is however only possible if the permeability $\kappa(N) = 0$. For simplicity, we assume that $\kappa \rightarrow 0$ and $e \rightarrow 0$ as $N \rightarrow \infty$: vanishing downstream flux occurs at infinite effective pressure. This is of course a mathematical idealization: effective pressure does not really become infinite at the cold-temperate boundary, merely much larger than it is in the remainder of the temperate region (see the supplementary material \S S1.2 for further detail). As we assume that $e = \Phi(N)$ goes to zero as $N \rightarrow \infty$, note also that equation \eqref{eq:freeze_margin} corresponds to $N \rightarrow \infty$ at an inward-migrating margin.

The singular behaviour associated with letting $\kappa(N) \rightarrow 0$ however makes it challenging to use $N$ as the dependent variable computationally: consider for example a power law permeability $\kappa \sim N^{-k}$ for some $k > 1$. When dealing with the fluxes $q_x$ and $q_y$ near a cold-temperate boundary, we may have to deal with the product of a very large gradient $\pdl{N}{y}$ with a very small permeability $\kappa(N)$. To avoid these issues, we transform to an auxiliary hydrological variable as described in appendix \ref{app:hydrology}.

\section{Steady state solutions} \label{sec:solution_all}

\subsection{Method of solution} \label{sec:solution}

Section \ref{sec:model} is the minimal version of a systematically reduced model capable of capturing the thermally-controlled onset of sliding, if we treat all model parameters (other than the ice sheet aspect ratio $\varepsilon$, in which we have retained only leading order terms) formally as being $O(1)$. The model remains rather complicated and may appear to have few advantages over a standard ice sheet model using Stokes' equations for ice flow, for which there are established numerical methods. The primary reason for using our alternative model is the relatively easy computation of steady state solutions by a simple forward integration in $x$ from an ice divide.

This initial value problem in $x$ is structurally analogous to the solution of a two-dimensional ice sheet with subtemperate sliding in the limit of a small temperature activation scale $\delta$ in \cite{Mantellietal2019}, and to earlier work in \cite{FowlerJohnson1996,Yakowitzetal1986,Hutteretal1986}. By contrast, for a fully configured `standard' ice sheet model, we would have to rely on computationally costly forward integration in time until a steady state is reached, and the necessary spatial resolution could prove prohibitive for a Stokes flow solver.

Before we proceed, two technical points: First, the specification of margin migration physics in \eqref{eq:no_singular_flux}--\eqref{eq:freeze_margin} provides conditions that must hold at the free boundary $\Gamma_m$ depending on whether it is migrating into the cold or temperate bed portions, but does not uniquely specify how a static margin should behave. Here we make the following assumption, based on the time-like nature of $x$: where the margin moves into the cold bed moving downstream (rahter than forward in time), we assume that \eqref{eq:no_singular_flux}--\eqref{eq:mass_cons_margin} holds, while at locations where the margin moves into the temperate region on going downstream, \eqref{eq:mass_cons_margin}--\eqref{eq:freeze_margin} holds (with $V = 0$ in both cases).

Second, we have chosen to omit the normal stress term $\sigma_{nn}$ from \eqref{eq:hydrology}$_3$ purely for numerical reasons, since incorporation of this term requires higher smoothness of the Stokes flow solver used.  Omission of $\sigma_{nn}$ is equivalent to the widely-made, but in our case inaccurate, assumption of a cryostatic normal stress at the bed, and its implications are discussed further in supplementary material \S\S S3.3 and S5.7.

%Computation of gradients of normal stress along the bed requires higher order accuracy in the solution of the Stokes flow problem \eqref{eq:Stokes} than the computation of the transverse advection velocity in \eqref{eq:heat}, and that higher order accuracy has proved challenging in the solver we have developed. The normal stress term may be important in hydraulically-driven ice stream development, and future work will need to incorporate it: a linearized  version of the model that does incoroporate the normal stress term $\sigma_{nn}$ (see supplementary material \S S5.7)) suggests that we may expect pathological behaviour in the form of a resonance-type effect that occurs under certain conditions. That said, forward integration as an initial value problem should still be possible when these conditions are not met. Note also that the omission of the normal stress term

Superficially, it is useful to think of the steady state problem as akin to a reaction-diffusion problem, with $x$ acting as a time-like variable in the heat equation \eqref{eq:heat}, and basal dissipation acting as the reaction term. To understand in more detail how the problem can be solved as an initial value problem in $x$, observe the following: given current (at prescribed $x$) basal conditions specified by $T$ and $N$ and current thickness $h$, \eqref{eq:flux_def} and  \eqref{eq:antiplane} define $u$ and $\rd h/ \rd x$ given a known flux $Q = \int_0^z \bar{a}(x') \rd x'$, $\rd h / \rd x$, with $\rd h/\rd x$ independent of $y$ or $z$.
To evolve $T$, \eqref{eq:flux_def} and \eqref{eq:antiplane} have to be solved in concert with \eqref{eq:heat}--\eqref{eq:heat_bed} (omitting the time derivative), and boundary conditions \eqref{eq:temp_surf}, \eqref{eq:geothermal}, \eqref{eq:temp_bed} and either \eqref{eq:basal_energy} where $T < 0$ or \eqref{eq:temp_pre_Dirichlet} when $e >  0$. Where $e > 0$, we simultaneously need to solve  \eqref{eq:basal_energy} with the time derivative omitted as an evolution equation for $N$, with constitutive relations \eqref{eq:hydrology} and boundary condition \eqref{eq:mass_cons_margin}.

If $h$, $T$ and $N$ are given, then $u$ can be computed and all forcing terms in \eqref{eq:heat}, \eqref{eq:basal_energy} and their boundary conditions are known, bar the secondary flow velocity $\bm{v}_\perp$. The latter is determined by \eqref{eq:Stokes} for $\bm{v}_\perp$ and $p$ combined with boundary conditions \eqref{eq:shear_stress_surface}--\eqref{eq:normal_velocity_surface} (with $-\pdl{Q}{x} - (a-\bar{a})$ replaced by $-a$ in the latter for a steady-state solution), and \eqref{eq:lateral_friction}--\eqref{eq:normal_velocity_base}. These need to be solved simultaneously with the heat equation and the basal hydrology problem to evolve $T$ and $N$. Again, almost all forcing terms in the secondary flow problem for $\bm{v}_\perp$ and $p$ are known given the current $T$ and $N$, except $\pdl{u}{x}$: while we have a recipe for computing $u$, we do not yet have a means of computing its derivative in the time-like direction.

To see how this is no bar to forward integration, note that by differentiating both \eqref{eq:flux_def} and  \eqref{eq:antiplane} with respect to $x$, we obtain a problem relating $\pdl{u}{x}$ and $\rd^2 h /\rd x^2$, of linear elliptic type in $\pdl{u}{x}$,  to the unknown derivatives of basal conditions $\pdl{N}{x}$, $\pdl{T}{x}$ and the known derivative $\rd Q / \rd x = W^{-1} \int_0^W a \rd y$ (as well as the current state variables $T$ and $N$ and the current solution $u$ that can be computed from them):
 $$ \nabla_\perp^2 \pd{u}{x} = -\od{^2(h+b)}{x^2}, \quad \mbox{on }0 < z < h,  \qquad \pd{}{z}\left( \pd{u}{x} \right) = 0 \quad \mbox{at } z = h $$
$$ \pd{}{z}\left(\pd{u}{x}\right) = f_T(T,N,u)\pd{T}{x} + f_N(T,N,u)\pd{N}{x} + f_u(T,N,u)\pd{u}{x} \mbox{at } z = 0, $$
\begin{equation} \od{Q}{x} = W^{-1} \int_0^W \int_0^h u_x \rd z \rd y \label{eq:model_differentiate} \end{equation}
where we have made use of the assumption that $Q > 0$ and hence $u > 0$ to simplify matters, and we have used subscripts to indicate differentiation of $f$ with respect to the subscripted variable.. In other words, we can solve for $\pdl{u}{x}$ and $\rd^2 h / \rd x^2$ at the same time as finding $\pdl{T}{x}$, $\pdl{N}{x}$ and $\bm{v}_\perp$.

Key to the procedure is that $h$ and therefore its derivatives are functions of $x$ only and can therefore be solved for using the constraint of a known flux $Q$. This is ultimately what allows an initial value solver to be used: if we relax our geometry, then a two-dimensional boundary value problem of arises for $h(x,y)$ (or $s_1$, see \S \ref{sec:discussion} and the supplementary material \S S5.9--5.10), and the efficient numerical method proposed here no longer applies.

In practice, we semi-discretize the steady state model in $x$ using upwinded finite differences, equivalent to a backward Euler step in $x$, having made the change of variables for $N$ described in appendix \ref{app:hydrology}. We  use an operator splitting  to keep track of the parts of the bed $z = 0$ that are at or below the melting point (see also \cite{SchoofHewitt2016}). Each of the resulting partial differential equations is elliptic in one of the dependent variables. The supplementary material \S S3 details the full system of coupled partial differential equations, where we use $y$ and a stretched vertical coordinate $\zeta = z/h$ as independent variables.  We discretize fully using finite volumes in $(y,\zeta)$, and solve at each step using Newton's method to handle nonlinearities. The necessary divergence-form formulation for the heat equation in terms of $\zeta$ is given in appendix \ref{app:stretch} because it may be of independent interest in ice sheet modelling \cite{HindmarshHutter1988,Hindmarsh1999,GreveBlatter2009}. Similarly, we use a formulation of the compressible Stokes flow problem \eqref{eq:Stokes} in terms of a stream function as described in the supplementary material, \S S3.

Note that the forward integration in $x$ requires upstream boundary conditions, which we assume to be given by an ice divide. The construction of an ice divide solution without lateral structure, but taking account of subtemperate sliding, is described in appendix \ref{app:divide} (see also \cite{Hindmarsh1999}).
These initial conditions require ice thickness $h(0)$ to be prescribed at the ice divide. $h(0)$ is ultimately constrained by the need to satisfy boundary conditions at a downstream ice margin \cite{Hutteretal1986,Yakowitzetal1986,Mantellietal2019}. For a marine-terminating ice sheet margin location $x_g$ and divide thickness $h(0)$ are determined by two conditions at the downstream margin, which we can then equate with a grounding line (as was also done in \cite{Mantellietal2019}): flotation $h(x_g) =-r^{-1}b(x_g)$, and a flux condition $Q(x_g) = Q_g(h(x_g))$,
where the function $Q_g$ computed from a boundary layer theory for marine ice sheets \cite{ChugunovWilchinsky1996,NowickiWingham2008,Schoof2012b,Schoofetal2017}. Our primary interest here is not in conditions at the margin. In order to avoid an unnecessarily costly shooting solution for $h(0)$, we therefore use the following construction: given $h(0)$, the solution for $h$ is invariant under a shift of bed elevation $b \mapsto b + b_0$ for constant $b_0$. If we simply prescribe $h(0)$, then we can use the constraint $Q(x_g) = Q_g(h(x_g))$ to determine the location $x_g$, and subsequently compute the bed elevation shift required to satisfy the flotation condition at that location as $b_0 = rh(x_g) -b(x_g)$.

\subsection{Two dimensions} \label{sec:results}

\begin{figure}
 \centering
 \includegraphics[width=\textwidth]{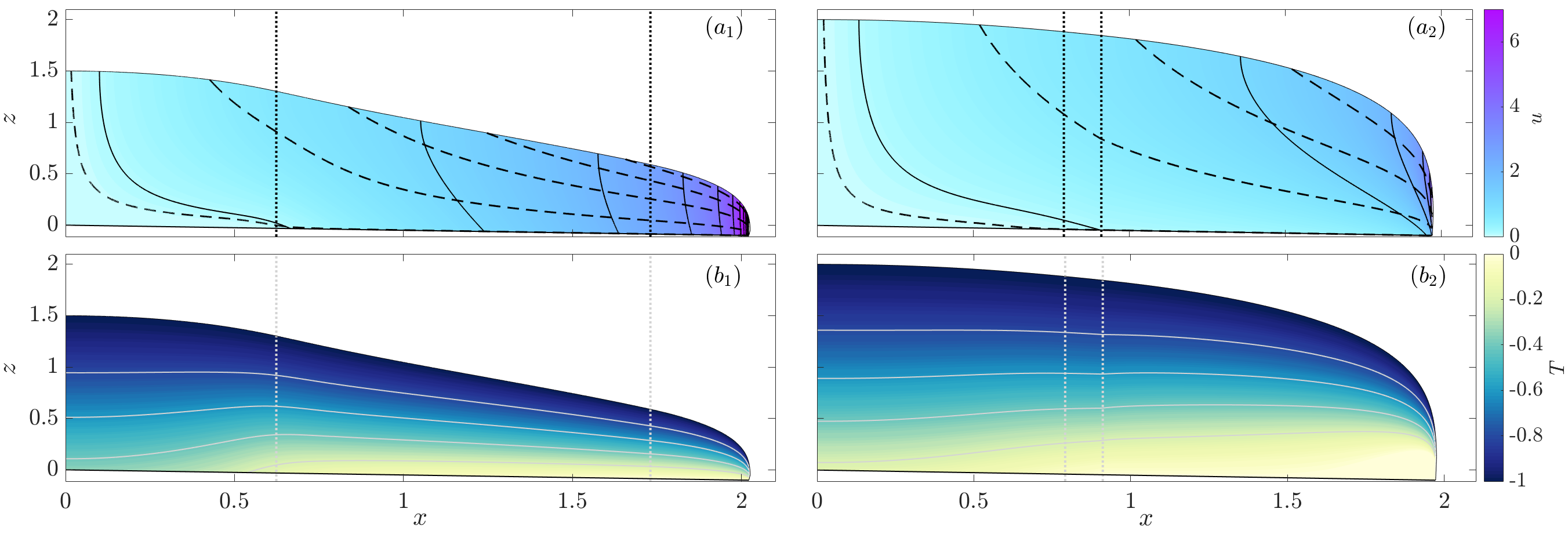}
\caption{Two-dimensional reference solutions. Panels a$_1$, b$_1$ show velocity profiles ($u$ given by solid contours, contour interval 1, background shading) and streamlines (dashed lines) for two different, two-dimensional steady state solutions. Vertical dotted lines demarcate the region in which there is significant sliding at subtemperate basal temperatures. Panels a$_2$, b$_2$ show the corresponding temperature profiles (white contours, background shading, contour interval of 0.2). Both examples use $Pe = a = 1$, $G = 0.5$, $T_s = -1$, $\rd b / \rd x = 0.05$, $\delta = 0.03$. $\gamma_0 = 0.1$, $h(0) = 1.5$ (column a, referred to in the text as reference case 1) and $\gamma_0 = 3$, $h(0) = 2$ (column b, reference case 2). Both examples use the temperate friction law \eqref{eq:Weertman}.} \label{fig:2D_results}
\end{figure}

Figure \ref{fig:2D_results} shows two examples of two-dimensional solutions with no lateral structure, obtained by requiring all dependent variables as well as the forcing term $a$ to be independent of $y$.  These solutions are a generalization of the results in figure 6 of \cite{Mantellietal2019}, where solutions are computed for the formal asymptotic limit $\delta \ll 1$ of highly temperature-sensitive sliding. In that case, a sharply defined subtemperate region separates a cold-bedded region upstream from a temperately-bedded region downstream. Within the subtemperate region, bed temperature is (to leading order in $\delta$) equal to the melting point, but sliding occurs at a rate that is slower than for fully temperate bed conditions: sliding velocity is instead controlled by the need to maintain energy balance. This is distinct from the temperate bed, which generally has a net positive energy balance, leading to melting.

Figure \ref{fig:2D_results} shows that, for small but finite $\delta$, we obtain a similar thermal and velocity structure to those computed in \cite{Mantellietal2019}. A more systematic  demonstration of convergence to the limiting form of \cite{Mantellietal2019} as $\delta \rightarrow 0$ is given in the supplementary material \S S3.  For small but finite $\delta$, we again find an extensive region of significant sliding upstream of the transition to a fully temperate bed, indicated by the dashed vertical lines, with bed temperatures close to but below the melting point. 

We use two different parameter combinations in figure \ref{fig:2D_results} as reference cases: column a shows an ice sheet in which sliding is relatively fast (with smaller $\gamma_0$), while column b shows an ice sheet in which sliding at the onset of the temperate bed is still slow (with a larger $\gamma_0$). The corresponding subtemperate regions are of very different extents, with a longer subtemperate region required to reach temperate conditions for the ice sheet that is able to slide faster. We refer to these two reference cases as `1' and `2', prespectively. In both, we use the simple linear friction law \eqref{eq:Weertman}.

\subsection{Three dimensions: pattern formation} \label{sec:pattern}

Next, we use the same parameter values but solve for the steady ice sheet in three dimensions, introducing a small amount of stochastic noise to the basal sliding coefficient in the subtemperate region, so that $ \gamma = \gamma_0(1+\epsilon(x,y))\exp(-T/\delta)$,
 where $\epsilon(x,t)$ is a small white noise term.

\begin{figure}
 \centering
 \includegraphics[width=\textwidth]{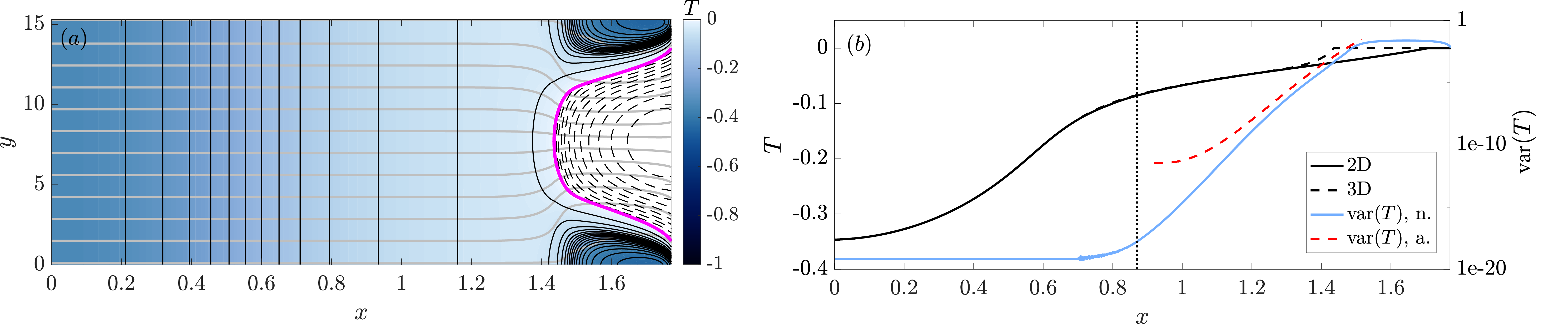}
\caption{A three-dimensional solution, same parameter choices as figure \ref{fig:2D_results}a. Panel a: contours of temperature $T$ (solid lines, also background shading) and $\Pi$ (dashed lines) defined by \eqref{eq:actual_transformation} with $k_0 = 4/3$ and $\Pi_0 = 0.1$, $\beta = 1$. Contour levels are 0.025 for $T$, 0.05 for $\Pi$, the subtemperate-temperate boundary is marked in pink. Solid grey lines are streamlines of velocity at the bed. Recall that the length-to-width ratio of the plot is not meaningful: $y$ is measured relative to an ice thickness scale length, $x$ relative to an ice sheet scale length: the evident ice stream pattern is highly elongated in reality. Panel b: Basal temperature against $x$ for the solution in figure \ref{fig:2D_results}a (solid black) and in panel a here (dashed black), and the variance in basal temperature with respect $y$ against $x$ (see equation \eqref{eq:variance}; note the logarithmic vertical scale on the right-hand axes). The red dashed curve shows an analytical prediction of the same quantity, see equation \eqref{eq:variance_analytical}, with the black dotted vertical line marking the location where $\Lambda$ defined in \eqref{eq:smalldelta_growth_rate_solution}.} \label{fig:bedmap_long}
\end{figure}

An along-flow profile as in figure \ref{fig:2D_results} no longer universally captures the structure that emerges. Figure \ref{fig:bedmap_long}a shows a map of basal temperature against $(x,y)$, obtained for case 1 with stochastic noise applied to $\gamma$. There remains an extended subtemperate region from $x \approx 0.6$ to $x \approx 1.35$, with basal temperatures slowly increasing as we move away from the ice divide. Instead of the transition to a simple temperate bed at $x \approx 1.75$ in figure \ref{fig:2D_results}a, a laterally-differentiated pattern rapidly emerges around $x \approx 1.4$ in the three-dimensional calculation, upstream of the subtemperate-temperate transition for the two-dimensional solution. Note that the emergence of this pattern may appear abrupt, but figure \ref{fig:bedmap_long}b reveals that lateral variations in bed temperature actually start to grow a significant distance upstream of the location where see them.

A finger-like region of temperate bed forms in  figure \ref{fig:bedmap_long}b, flanked by regions of basal temperatures that decrease with distance downstream, and are much colder than anywhere else under the ice sheet. The streamlines for flow at the ice sheet bed also indicate that the flow of ice rapidly converges into the temperate finger at the onset: this pattern is clearly an ice stream surrounded by ice ridges. Note that here, as in all other calculations in this paper showing formation of a pattern, we find a single temperate `finger' per domain width. That is no accident, as we will show in \S \ref{sec:stability} and discuss further in \S \ref{sec:discussion}.

The formation of a region of the finger of temperate bed is superficially simple to understand as being the result of a positive feedback between faster sliding at warmer temperatures and greater dissipation of heat at the bed (see \S \ref{sec:stability} below). The pattern formation process is broadly speaking the same as in \cite{Hindmarsh2009}, who however uses the limit of fast sliding throughout the ice sheet. This leads to less sharply defined margins, and also alters the effect of potential negative feedbacks suppressing pattern formation. We discuss these differences further in \S \ref{sec:discussion} and the supplementary material \S S5.6.

\begin{figure}
 \centering
 \includegraphics[width=\textwidth]{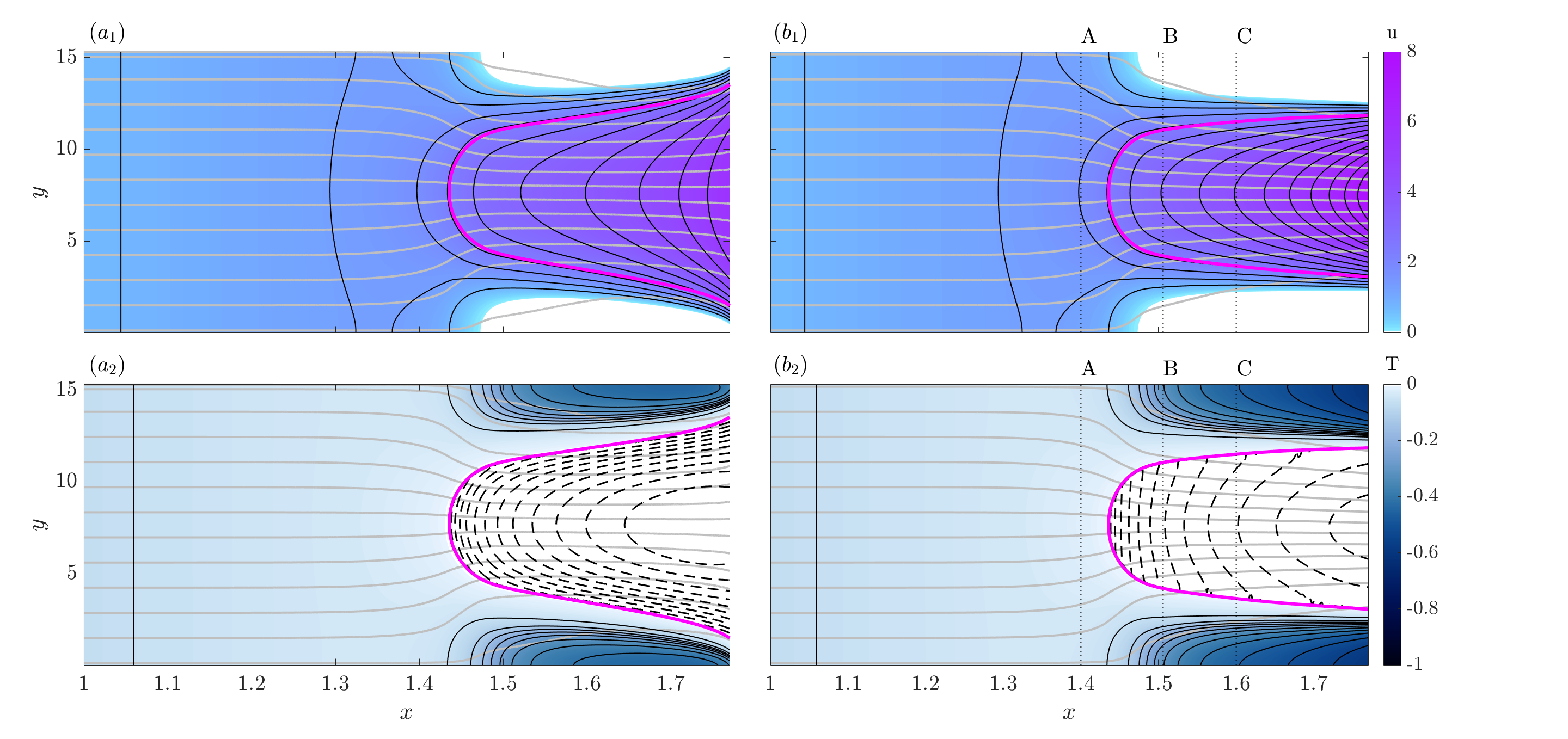}
\caption{Three-dimensional solutions with different sliding laws: column a) is the same as figure \ref{fig:bedmap_long}, column b) uses the regularized Coulomb friction law with $\mu_0 = 0.005$, $\beta = 10$, all other parameter values are identical. Row 2 uses the same plotting scheme as figure \ref{fig:bedmap_long}a with contour intervals of 0.05 for $T$ and $\Pi$, row 1 shows contours of axial velocity $u$ at the bed (as well as background shading), with contour interval 0.5. The pink curve is still the subtemperate-temperate boundary, grey streamlines show velocity at the bed.} \label{fig:bedmap_zoom}
\end{figure}

Figure \ref{fig:bedmap_zoom} shows a close-up of the ice-stream-ice-ridge structure for two cases, the case 1 calculation also shown in figure \ref{fig:bedmap_long}, and a second calculation with the same parameter choices, but the sliding law switched to a regularized Coulomb friction law in the temperate region. Since the solution is obtained by a forward integration in $x$, the plots are identical up to the point where the bed becomes temperate. In addition to basal temperature, each column now also shows basal velocity, confirming that sliding velocities are elevated in the ice stream and continue to increase in the downstream direction.  

Importantly, elevated velocities prevail in the region outside of the temperate bed demarcated by the pink lines: in the margins of the ice stream, there is a portion of bed with singificant subtemperate sliding (see also \cite{Haseloffetal2018}). Sliding is only significantly suppressed where a sharp lateral temperature gradient is evident at the lateral edge of the ice ridge (the edge of the dark blue areas in the second row of figure \ref{fig:bedmap_zoom}).

The two solutions, for the linear temperate friction law \eqref{eq:Weertman} on the left (column a) and for the regularized Coulomb friction law on the right (column b), differ in perhaps unexpected ways. The margins of the ice stream widen more significantly in the downstream direction for the linear friction law, in which the meltwater produced due to dissipation in the ice stream does \emph{not} feed back into the motion of the ice stream, while for the effective-pressure-sensitive Coulomb law,  ice velocity increases much more significantly along the ice stream trunk, but the margins widen only by a small amount. It is likely that the downstream acceleration of the ice stream is responsible, drawing in more ice from the surrounding ice ridges as it lowers the ice surface in the stream, and the attendant lateral heat transport prevents the margins from migrating outwards (see also \cite{Haseloffetal2015,Haseloffetal2018}).

\begin{figure}
 \centering
 \includegraphics[width=\textwidth]{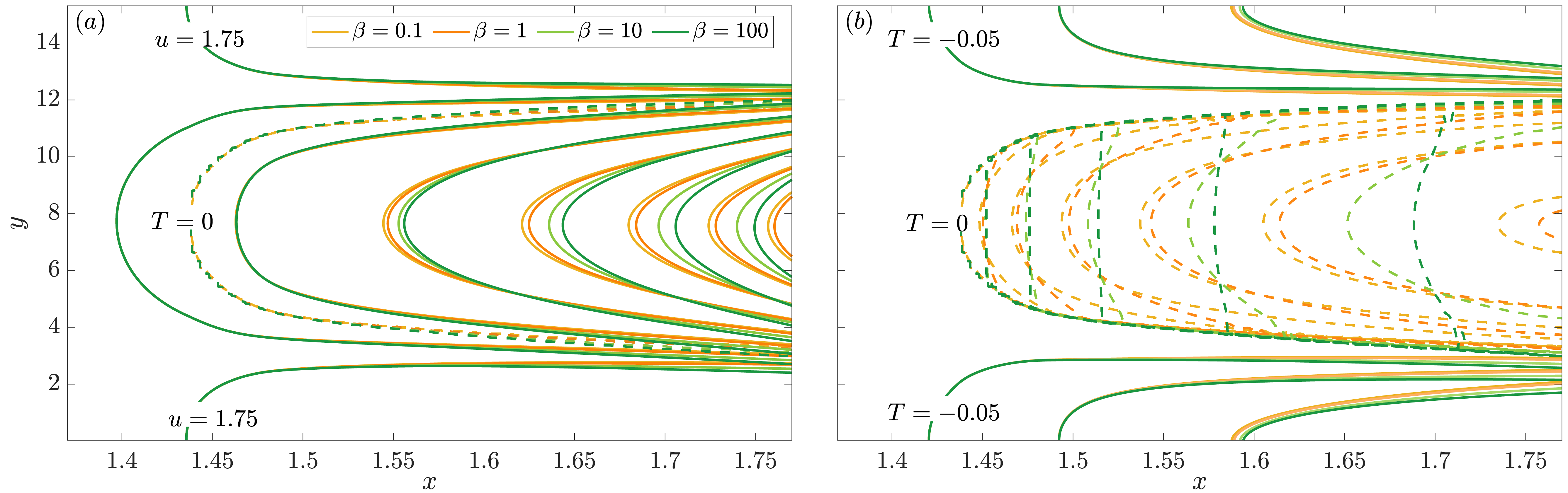}
\caption{The effect of basal hydrology: solutions of the same problem as column b) in figure \ref{fig:bedmap_zoom}, but with different values of lateral diffusivities $\beta$ as indicated in the figure legend. Panel a: contours of axial velocity as solid lines, contour inteval 1, dashed line is the subtemperate-temperate boundary. Panel b) contours of temperature as solid lines with contour interval 0,2, $\Pi$ as dashed lines with contour interval 0.1. Contour lines are colour-coded as indicated in the legend in panel a.} \label{fig:bedmap_Coulomb}
\end{figure}

In addition to the choice of sliding law, the hydrology model also affects the fully evolved ice stream, although to a much lesser extent. Figure \ref{fig:bedmap_Coulomb} shows solutions for the regularized Coulomb friction law 
in figure \ref{fig:bedmap_zoom}b), but with different choices of the diffusion parameter $\beta$ that controls how lateral effective pressure gradients drive the flow of water. As in figures \ref{fig:bedmap_long} and \ref{fig:bedmap_zoom}, dashed lines in panel b) show contours of the proxy variable $\Pi \sim N^{-(k_0-1)}$ defined in appendix \ref{app:hydrology} for a permeability function $\kappa(N) = N^{-k_0}$; plotting $\Pi$ rather than $N$ has the advantage that $\Pi$ remains bounded and contours do not become tightly bunched. Key here is to remember that large $\Pi$ corresponds to small $N$ and vice versa. The dashed line in panel a) is the $T = 0$ contour.

The effect of $\beta$ on hydrology is obvious in panel b): for a widening ice stream, the proxy $\Pi$ for effective pressure does not typically go to zero near the edge of the temperate region, though $\Pi$ is small there for small diffusivities $\beta$, implying that large effective pressures $N$ are reached. In that case, there are significant lateral gradients in $\Pi$ (or $N$) across the width of the ice stream. For large diffusivities, $\Pi$ (and therefore $N$) becomes nearly constant across the width of the ice stream: lateral drainage is highly effective then, and ensures insignificant variations in basal effective pressure.

As a corollary, we also see that $\Pi$ becomes noticeably larger in the centre of the ice stream for small $\beta$, corresponding to smaller effective pressure $N$, with correspondingly larger sliding velocities (panel a) there for beds with poor lateral drainage, although the velocity difference is quite muted due to the role of lateral shearing in the force balance of the ice stream. By contrast, ice streams with larger $\beta$ and effective lateral drainage widen more rapidly in the downstream direction, although again the effect is muted. This probably occurs because more efficient drainage reduces basal friction towards the edges of the ice stream and thereby concentrates dissipation in the margins themselves, facilitating outward migration \cite{Schoof2012b,Haseloffetal2015,Haseloffetal2018}

\subsection{Cross-sections} \label{sec:cross_section}

\begin{figure}
 \centering
 \includegraphics[width=\textwidth]{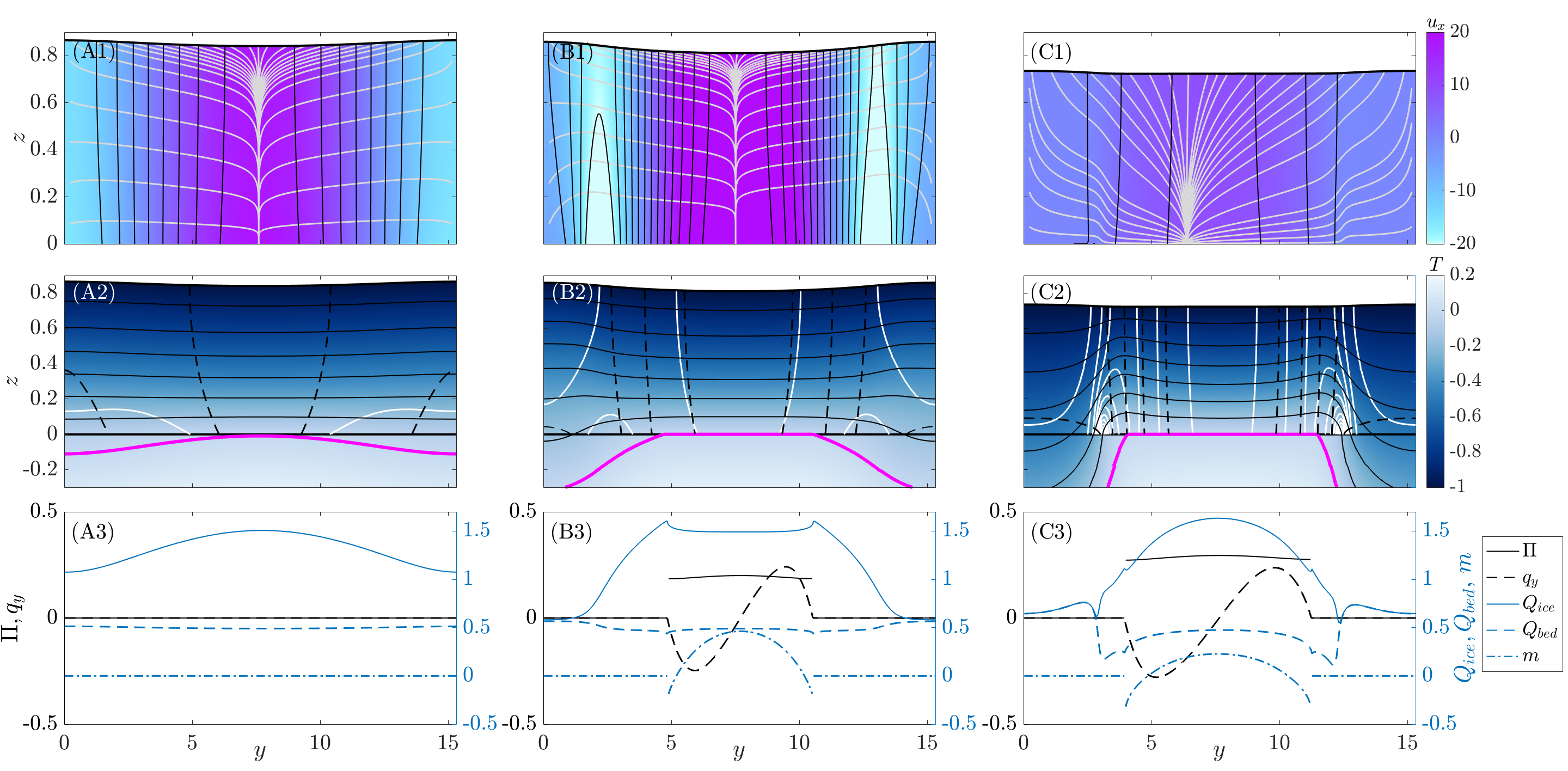}
\caption{Patterns of flow, dissipation and temperature in the ice, and fluxes at the bed. Each column corresponds to a cross-section indicated by the corresponding letter label in figure \ref{fig:bedmap_zoom}b. Row 1 (top): contours of $\pdl{u}{x}$ (black, contour interval 4,  and background shading), streamlines of $\bm{v}_\perp$, Row 2 (middle): contours of temperature $T$ (black, contour interval 0.15, and background shading), contours of $u$ (dashed black, contour interval 1) and $\alpha |\nabla_\perp u|^2$ (white, contour interval 0.5), pink is the $T = 0$ contour. Note that part of the bedrock domain is shown. Row 3 (bottom): basal heat flux $Q_{ice} - \lim_{z\rightarrow 0^+}\pdl{T}{z}+$ in the ice (solid blue) and $Q_{bed} = -\lim_{z\rightarrow 0^+} \pdl{T}{z}$ in the bed (dashed blue), melt rate $m = Q_{bed} - Q_{ice}  + \alpha f(T,N,u)|u|$ (dot-dashed blue), effective pressure proxy $\Pi$ (solid black), lateral $q_y$ (dashed black).} \label{fig:cross_section}
\end{figure}

We can gain greater insight into the nonlinear pattern forming process by looking  at  the patterns of axial flow and englacial dissipation as well as of the transverse secondary flow. Figure \ref{fig:cross_section} illustrates these for the ice stream in figure \ref{fig:bedmap_zoom}b, for which we plot cross-sections across the domain at the locations marked by letters A,B,C in column b) of figure \ref{fig:bedmap_zoom}. 

Each column of figure \ref{fig:cross_section} corresponds to one of these cross-sections, identified by the corresponding letter label. The top row displays isolines of the axial strain rate $u_x = \pdl{u}{x}$, also indicated by the background colour shading, while the solid white curves are streamlines of the transverse velocity field $\bm{v}_\perp$. The middle row shows isolines of temperature in black, also indicated by the background colour shading. The pink contour indicates the melting point $T = 0$. Dashed contours are velocity isolines, while white contours indicate strain heating rate $\alpha |\nabla_\perp u|^2$. In all cases, contour intervals are consistent between columns A, B, and C, and in order to illustrate the surface correction $s_1$ defined in equation \eqref{eq:surface_correction}, we use $(1+\varepsilon^2 s_1(x,y))z$ as the vertical coordinate for plotting purposes, arbitrarily using a fairly large value $\varepsilon = 0.015$ for clarity.

The bottom panel shows aspects of the energy balance of the bed: blue solid and dashed lines are heat fluxes $-\pdl{T}{z}$ at $z = 0$, computed in the ice and the bed, repsectively, while the dot-dashed blue line is the net melt rate $m = [\pdl{T}{z}]^+_- + \alpha f(T,N,u)|u|$. The solid black line is the effective pressure proxy $\Pi$, the dashed black line the later water flux $q_y$.

Near the onset of the ice stream (column A), the patterns of englacial flow and dissipation are simple, with a patch of slightly warmer bed in the middle of the domain leading to faster flow (panel A2) and enhanced dissipation (panel A3), where conductive flux $Q_{bed}$ (solid blue line) is elevated at the centre of the domain to balance the additional heat generated at the bed). The faster flow also drawing somewhat colder ice inwards and downwards through a convergent secondary flow (panel A1). Note that there is nothing inconsistent in the streamlines of $\bm{v}_\perp$ converging (panel A1) despite the flow being incompressible: these streamlines are not actual particle trajectories, as they ignore the simultaneous axial motion of the ice. Note that englacial dissipation also begins to be shifted towards the edges of the region of fast flow.

Somewhat further downstream, the bed has become fully temperate at the centre of the incipient ice stream (panel B2). This coincides with the formation of the very cold-based ice ridges, and the margin of the region of fast flow migrates inwards (even as the margin of the temperate bed migrates outwards, see figure \ref{fig:bedmap_zoom}b). As a result, the axial strain rate in the incipient ice ridges is locally substantially negative, especially around $y = 2.5$ and $y = 12.5$ in figure \ref{fig:cross_section}B1. The transverse streamlines are therefore locally warped upwards. This contrasts with the usual expectation of ice being drawn downwards as it traverses an ice stream margin \cite{Haseloffetal2015,Haseloffetal2018}, but is potentially consistent with specific field observations of englacial radar reflectors near the margins at the upstream end of the Northeast Greenland Ice Stream similarly being warped upwards \cite{Holschuhetal2019}. It is unclear however whether direct comparisons with field data are a good test of theory, since it is unclear whether the field data in \cite{Holschuhetal2019} or elsewhere reflect steady state conditions, or indeed whether spatial anomalies in $G$ that are not considered here play a significant role in individual locations (see also \cite{KyrkeSmithetal2015}).

In column B, significant  subtemperate sliding occurs over a region substantially larger than the temperate bed patch at the centre of the domain, as is reflected by the elevated basal heat flux $Q_{bed}$ in panel B3. Inside the temperate bed region, an active drainage system is established. The effective pressure proxy $\Pi$ has a slight gradient, and does not go to zero at the edge of the temperate region. Consequently, the bed retains a finite water content there, and basal shear stress is discontinuous across the subtemperate-temperate boundary. Water flows from the centre of the ice stream, where melt rates $m$ are positive, towards the margins, where there is a small but finite freezing rate (panel B3). This freezing rate is mathematically unavoidable in a margin  where basal shear stress is discontinuous and subtemperate sliding occurs (see \cite{Haseloffetal2018} and \S S2 of the supplementary material). 

By the time cross-section C is reached, there is a fully-established pattern of an ice ridge in which there is a simple transverse `shallow-ice' type shearing flow towards the margin, a clearly defined margin, and a lateral-shear-dominated ice stream as described by \cite{Haseloffetal2015}. In fact, for a wide domain, it can be shown that the model we use here is equivalent to the parameter regime described in appendix B of \cite{Haseloff2015}; the advantage of our model over that in \cite{Haseloffetal2015} is that we are able to capture the onset of the ice stream in addition to the fully evolved form.  

The axial strain rate $\pdl{u}{x}$ is now significantly smaller than in the region of rapid flow reorganization around cross-sections A and B, and shows a noticeable asymmetry (panel C1): the ice stream does not initiate perfectly symmetrically, and as a result its margins do not migrate outwards symmetrically, giving the now much weaker secondary flow a corresponding asymmetry, with convergence not centered on the middle of the stream. As a result of the weaker secondary flow, the surface correction $s_1$ is now also less pronounced than in the onset region, and has the familiar pattern of a convex surface over the ice ridges, and a flat surface over the ice stream.

The margins exhibit the strongly concentrated englacial dissipation (panel C2) familiar from previous studies \cite{Raymond1996,JacobsonRaymond1998,Schoof2004b,Schoof2012b,Suckaleetal2014,Haseloffetal2015,EllsworthSuckale2016,Haseloffetal2018,Meyeretal2018}, and advection due to the secondary flow is angled sharply downward through the margins (panel C1). The competition between these two and the effect of basal dissipation due to subtemperate sliding  ultimately control margin migration \cite{Haseloffetal2018}, although here in the sense of ice stream widening in the downstream direction, rather than in time. 

Two observations may be significant: first, the strong concentration of shear does not occur at the transition from a temperate to a subtemperate bed. Instead, it coincides with a sharp lateral temperature gradient between small negative temperatures near the edge of the ice stream, accompanied by fast subtemperate sliding, and the much lower basal temperatures of the ice ridge. Secondly, no temperate ice is formed at this location, with englacial temperatures remaining firmly below the melting point. Both observations are consistent with previous work on subtemperate sliding in shear margins \cite{Haseloffetal2018}, though it is conceivable that the absence of temperate ice is the result of the moderate width of the modelled ice stream \cite{Haseloffetal2019,Meyeretal2018,MeyerMinchew2018}

The temperate bed portion of the ice stream exhibits a similar though slightly more complicated pattern to cross-section B. Basal dissipation is concentrated at the centre of the ice stream, where net melt occurs, and a relatively weak later gradient in $\Pi$ suffices to drive water towards the margins, where there is a small but finite rate of melting, and $\Pi$ remains finite. at the cold-temperate transition.

\subsection{Other forms of patterning, or no pattern at all} \label{sec:hydraulic_pattern}

All `patterned' three-dimensional solutions that we have described so far are based on the two-dimensional reference case 1 of figure \ref{fig:2D_results}a. Even if we apply a stochastic perturbation to $\gamma$ in reference case 2 of figure \ref{fig:2D_results}b, no pattern emerges in three dimensions in the much shorter subtemperate area, in which sliding velocities are also much slower. If we continue the calculation as in figure \ref{fig:2D_results}b with a linear friction law \eqref{eq:Weertman}, no pattern emerges at all even downstream of the cold-temperate transition, and we simply recover the solution in figure  \ref{fig:2D_results}b with a small amount of noise.

\begin{figure}
 \centering
 \includegraphics[width=\textwidth]{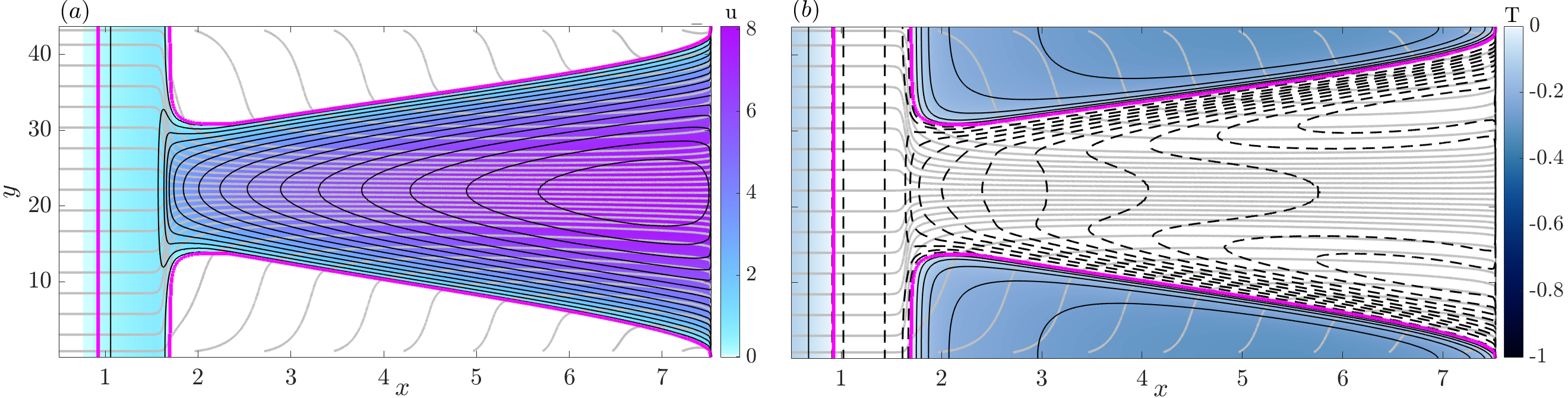}
\caption{Hydromechanical instability: same parameters as figure \ref{fig:2D_results}b but using the double power law \eqref{eq:Budd} and $N$ and $\tilde{\kappa}$ defined through \eqref{eq:actual_transformation} with $\Pi_0 =  0.1$, $N_s = 0.1$, $k_0 = 4/3$, $\beta = 0.1$. Same plotting scheme as figure \ref{fig:bedmap_zoom}, contour interval of 1 for $u$, 0.15 for $T$, 0.1 for $\Pi$} \label{fig:hydraulic}
\end{figure}

A pattern can still emerge from within the temperate area if we use an effective-pressure-dependent sliding law: figure \ref{fig:hydraulic} is analogous to figure \ref{fig:bedmap_zoom} but for reference case 2 with a power-law friction \eqref{eq:Budd}. The entire bed becomes temperate around $x \approx 0.9$, with no apparent patterning as described above. Around $x \approx 1.7$ a pattern then rapidly emerges, with a patch of low effective pressure $N$ (large $\Pi$) facilitating enhanced sliding and drawing in ice through the secondary transverse flow. The ice flow surrounding the patch of low effective pressure slows (as it must, with total ice flux being prescribed), dissipation there is reduced and the bed actually refreezes. 

Once again, an ice-stream-ice-ridge pattern forms, with very low bed temperatures in the ice ridges.  One notable feature of the solution is the formation of a double-peaked, off-centre maximum in the effective pressure proxy $\Pi$ in the fully evolved ice stream, corresponding to a double minimum in $N$. This is partly the result of the low diffusivity $\beta$ not smoothing basal effective pressure, but primarily results from englacial dissipation: strong shearing near the margins of the ice stream (which is substantially wider than in figure \ref{fig:bedmap_zoom}) warms the ice (though never to the melting point). Advection carries this warmer ice towards the centre of the ice stream, and the reduced conductive heat flux at the bed causes additional melting at an intermediate position between margin and ice stream centre, leading to the observed effective pressure distribution.
As in the case of the pattern emerging from subtemperate sliding, we can look at cross-sectional patterns of flow, dissipation and temperature in the ice to understand this pattern better, see supplementary material \S S3.
 
The formation of the ice-steam-ice-ridge pattern in this case is closely related to the mechanisms in \cite{FowlerJohnson1996,KyrkeSmithetal2014,KyrkeSmithetal2015}, as discussed further in \S \ref{sec:discussion}. Patterning in this instance is driven by a hydromechanical positive feedback between dissipation, increased discharge of additional meltwater requiring reduced effective pressure, and reduced effective pressure leading to faster sliding and increased dissipation. This feedback does not, however, unconditionally lead to the formation of a recognizable ice stream pattern: in the supplementary material \S S3, we showcase an example in which a slight lateral perturbation in effective pressure and velocity grows over a limited stretch of the ice sheet, only to disappear again further downstream, without ever forming a distinct ice stream margin. 

In fact, for many plausible parameter choices, we have found no pattern formation due to the the hydromechanical feedback at all. Overall, the question of negative feedbacks that act to suppress pattern formation remains open. We address this next.

\section{Stability analysis} \label{sec:stability}

The numerical solutions in the previous section raise the question of why (and where) patterning appears in some of the steady state solutions, but not others. Here, we show that patterning can be understood as a `spatial' instability. Rather than asking whether perturbations to a laterally uniform steady state grow in \emph{time}, we use the time-like nature of $x$ to identify conditions under which perturbations in upstream conditions (at the divide) or in forcing (such as the stochastic perturbations to $\gamma_0$ that we have used numerically) are amplified as we move \emph{downstream}. Similar notions of spatial stability have previously been used to study the growth of basal channel along the base of an ice shelf \cite{Dallastonetal2015} as well as ice stream formation using a simpler model than ours \cite{FowlerJohnson1996}.

We conduct our linear stability analysis in the limit of small $\delta$, and confine ourselves to the case of patterning in the subtemperate region. We will show that patterns can grow in the downstream direction if sliding velocities in the subtemperate region exceed a certain threshold. The hydraulically-controlled onset of patterning during temperate sliding is studied in more detail in the supplementary material (\S S5.7), as is the onset of patterning during subtemperate flow for the more general case $\delta = O(1)$ (\S\S S5.1--S5.4).

Recall that $\delta \ll 1$ is the bed temperature scale over which the ice sheet transitions from insignificant to fully temperate sliding. We can identify a similarly short thermal boundary layer length scale $z \sim \delta$ over which those temperature variations occur naturally near the bed, and a corresponding along-flow distance scale $x \sim \delta^2$ such that advection and diffusion balance in the basal thermal boundary layer for $Pe \sim O(1)$ (see also \cite{MantelliSchoof2019}). This turns out to be (formally) the length scale over which the onset of patterning takes place in figure \ref{fig:bedmap_long}. We can capture the dominant processes involved by rescaling
\begin{equation} Z = \delta^{-1}z,  \qquad X = \delta^{-2}x, \end{equation}
and putting $\Theta(X,y,Z) =\delta^{-1} T(x,z)$. $X$ is a local along-flow coordinate, while $Z$ is distance above the bed in the thermal boundary layer; $z$ will continue to describe distance above the bed in the `outer' region that occupies most of the ice thickness. Implicit here is that $\delta^2 \gg \varepsilon$ where $\varepsilon$ is the ice sheet aspect ratio, so that the local variable $X$ still describes displacements that are much larger than the ice thickness scale, and the shallow-in-$x$ model of \S \ref{sec:model} continues to apply. 

A rapid onset also alters the scale for the secondary ice flow, as faster transverse velocities are required to balance the potentially large velocity gradient $\pdl{u}{x} \sim O(\delta^{-2})$. We put $U_o(X,y,z) = u(x,y,z)$, $V_o(X,y,z) = \delta^2 v(x,y,z)$, $W_o(X,y,z) = \delta^2 w(x,y,z)$, $P_o = \delta^2 p(x,y,z)$ in the outer region, distant from the bed, and put $U_b(X,y,Z) = U_o(X,y,z)$, $V_b(X,y,Z) = V_o(X,y,z)$, $W_b(X,y,Z) = \delta^{-1} W_o(X,y,z)$ in the thermal boundary layer, the latter to account for the fact that vertical velocity vanishes at the bed. A rescaling of the mechanical problem to the thermal boundary layer variable $Z$ and matching with its outer version shows that $U_b = U_o(x,y,0)$, $V_b = V_o(x,y,0)$ and $W_b = W_{bZ}Z$, where $W_{bZ} = \pdl{W_b}{Z} = \lim_{z\rightarrow 0} \pdl{W_o}{z}$, and $U_b$, $V_b$, $W_{bZ}$ are all independent of $Z$ at leading order.

Ice thickness remains constant at leading order over the scale associated with the variable $X$, which really describes an internal layer with respect to the outer coordinate $x$ inside the ice sheet, but the surface slope may change by $O(1)$. Formally, this can be accounted for by putting $H(X) = h(x)$ and expanding $H(X) = \bar{H} + \delta^2 H_1(X)$, so that $\rd h / \rd x = \rd H_1 /\rd X$ while ice thickness remains constant at $\bar{H}$. In full, the mechanical problem consisting of \eqref{eq:antiplane}, \eqref{eq:subtemp_sliding_law},  \eqref{eq:Stokes}, \eqref{eq:shear_stress_surface}--\eqref{eq:normal_velocity_surface} and \eqref{eq:lateral_friction}--\eqref{eq:normal_velocity_base} becomes
\begin{subequations} \label{eq:transition_leading}
\begin{equation} \nabla_\perp^2 U_o = -\pd{b}{x} -\pd{H_1}{X} \quad \mbox{on } 0 < z < \bar{H}, \quad \pd{U_o}{z} = \Gamma(\Theta(0))U_o \quad \mbox{on } z = 0, \qquad \pd{U_o}{z} = 0 \quad \mbox{on } z = \bar{H}, \end{equation}
where $\pdl{b}{x}$ is constant at the inner horizontal scale associated with $X$ and $\Gamma(\Theta) = \gamma(T)$, and the rescaling ensures that $\rd \Gamma/\rd \Theta \sim O(1)$. 
In addition
\begin{equation} \nabla_\perp^2 (V_o,W_o) - \nabla_\perp P_o = 0, \quad \pd{V_o}{y} + \pd{W_o}{z} = - \pd{U_o}{X} \end{equation}
on $0 < z < \bar{H}$, subject to
\begin{equation} W_o = 0, \quad \pd{V_o}{z} + \pd{W_o}{y} = \Gamma(\Theta(0)) V_o/U_o \quad \mbox{at } z = 0, \quad W_o = \pd{V_o}{z} = 0 \quad \mbox{at } z = \bar{H} \end{equation}
while the surface perturbation $H_1(X)$ is independent of $(y,z)$ and determined by the constraint $Q = \int_0^W \int_0^{\bar{H}} U \rd z \rd y = $ constant.
\end{subequations}

At leading order, we obtain a thermal boundary layer problem of the form
\begin{subequations} \label{eq:thermal_boundary_layer}
\begin{align} Pe \left( U_b \pd{\Theta}{X} + V_b\pd{\Theta}{Y} + W_{bZ} Z \pd{\Theta}{Z}\right) - \pd{^2\Theta}{Z^2} = & 0 & \mbox{for } Z > 0 \label{eq:thermal_boundary_layer_heat} \\
  - \pd{^2\Theta}{Z^2} = & 0 & \mbox{for } Z < 0 \label{eq:thermal_boundary_layer_bed} \\
  \left[-\pd{\Theta}{Z}\right]_-^+ = & \Gamma(\Theta(0)) U_b^2, \qquad [\Theta]_-^+ = 0 & \mbox{at } Z = 0, \label{eq:thermal_boundary_layer_bc1}
\end{align}
\end{subequations}
The outer thermal problem is advection-dominated in the ice and diffusion-dominated in the bed. At leading order, matching of \eqref{eq:thermal_boundary_layer} with the far field therefore requires that
\begin{equation} U_b \pd{\Theta}{X} + V_b\pd{\Theta}{Y} + W_{bZ} Z \pd{\Theta}{Z} \sim 0 \qquad \mbox{and}  \qquad -\pd{\Theta}{Z} \sim -\lim_{z \rightarrow 0^-}\pd{\Theta}{z} = G \label{eq:thermal_boundary_layer_bc2} \end{equation}
as $Z \rightarrow +\infty$ and $Z \rightarrow -\infty$, respectively. 

The solutions without lateral structure computed in \S \ref{sec:solution_all}\ref{sec:results} are functions of $x$ and $z$,  and consequently are constant in the rescaled variable $X$ at leading order. If we take those two-dimensional solutions as a base state and consider their stability to pattern formation in three dimensions with $X$  as the time-like variable, we can therefore linearize and treat the base state as uniform. Denoting the base state by overbars and putting $\theta = -(h+b)_x$, evaluated locally, we have  $\bar{V}_b = \bar{W}_{bz} = 0$ and $\bar{\Gamma} \bar{U}_b = h \theta$, where $\bar{\Gamma} = \Gamma(\bar{\Theta}(0))$, as well as $\bar{\Theta} = \bar{\Theta}(0) - GZ$ for $Z < 0$ and $\bar{\Theta} = \bar{\Theta}(0) - (G + \bar{\Gamma}\bar{U}_b^2)Z$ for $Z > 0$. We perturb the base state as
$$ U_b = \bar{U}_b + U_b'\exp(\Lambda X + ik y), \qquad V_b = V_b'\exp(\Lambda X + iky), \qquad W_{z0} = \bar{W}_{z0} + W_{z0}'Z \exp(\Lambda X + iky), $$
\begin{equation} \Theta = \bar{\Theta}(Z) + \Theta'(Z) \exp(\Lambda X + iky), \end{equation}
and linearize in the perturbations, where we assume that $k \neq 0$; since the ice thicness variation term $H_1$ is independent of $y$, there is then no need at first order to incorporate $H_1$ in the linearization.

\begin{figure}
 \centering
 \includegraphics[width=0.66\textwidth]{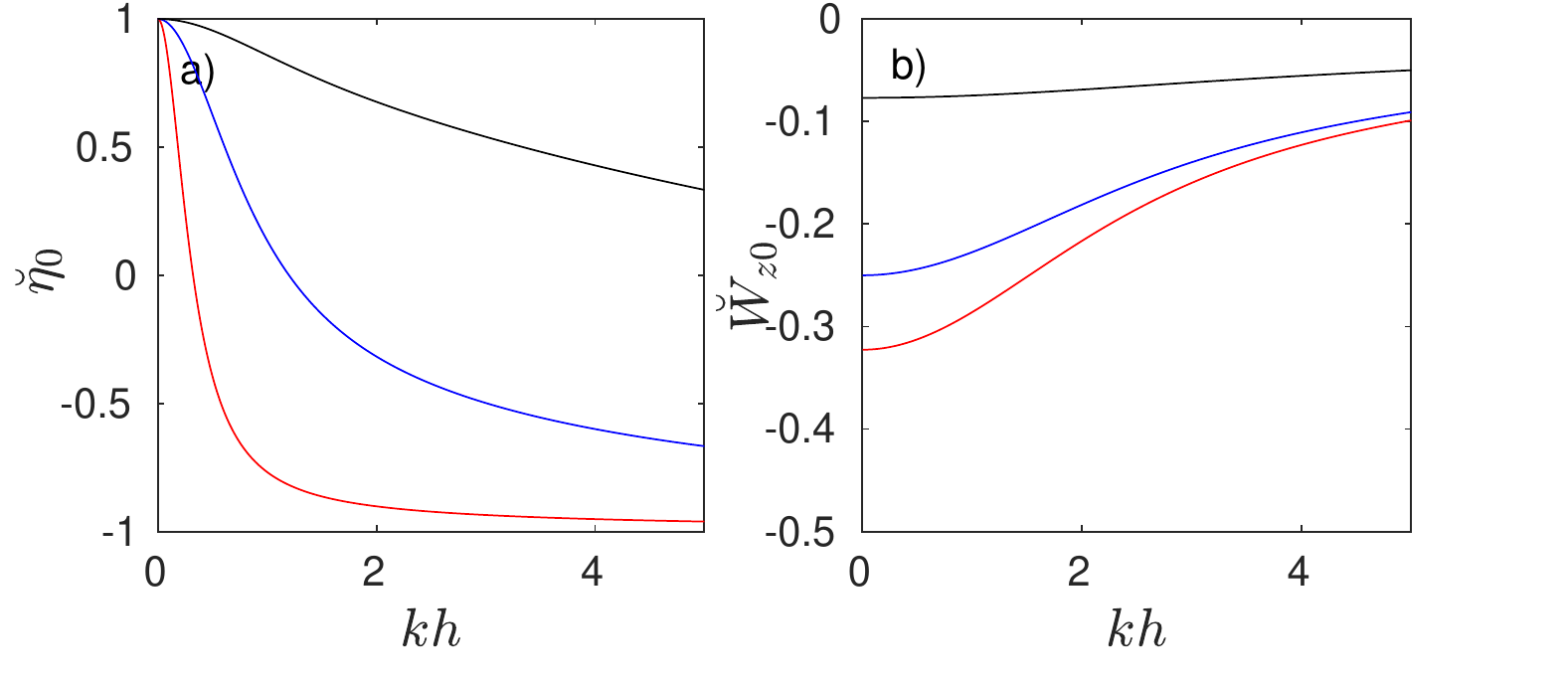}
 \caption{Plots of $\breve{\eta}_0 = \eta_0/(-\alpha \bar{\Gamma}_\Theta \bar{U}_b^2)$ (panel a) and $\breve{W}_{z0} = W_{z0}/(-\Bar{\Gamma}_\theta \bar{U}_b h)$ (panel b) as functions of $kh$ for $\bar{\Gamma}h$ = 0.1 (red), $\bar{\Gamma} h$ = 1 (blue) and $\bar{\Gamma} h = 10$ (black). Note that $\breve{\eta}_0$ and $\breve{W}_{z0}$ thus defined are functions of $\bar{\Gamma}h$ and $kh$ only. As $\bar{\Gamma}_\Theta < 0$, $\breve{\eta}_0$ and $\breve{W}_{z0}$ also have the same sign as $\eta_0$ and $W_{z0}$, respectively. Physically $\eta_0$ represents the feedback between bed temperature and basal dissipation for a given wavenumber, while $W_{z0}$ represents the effect of basal temperature on vertical advection near the bed.} \label{fig:transfer_functions}
\end{figure}

In the usual fashion, a positive growth rate  $\Re(\Lambda)$ indicates the growth of perturbations, albeit in the downstream direction rather than in time. Solving the appropriate linearized version of \eqref{eq:transition_leading}, we find that the perturbations in vertical velocity gradient and in basal dissipation can be expressed in terms of $\Theta'(0)$ as
\begin{equation}  W_{bZ} =  W_{z0} \Theta'(0), \qquad
 2\bar{\Gamma}\bar{U}_bU_b' + \Gamma' \bar{U}_b^2 =  \eta_0 \Theta'(0), \label{eq:stability_coefficient_definition}
\end{equation}
where $\Gamma' = \Gamma_\Theta \Theta'(0)$ is the perturbation in $\Gamma$, and
\begin{align}
  W_{z0} = &  \bar{\Gamma}_\Theta  \bar{U}_bh \frac{\sinh(kh)\cosh(k h)-kh}{2kh \sinh^2(kh) + \bar{\Gamma}h[ \sinh(kh) \cosh (kh) - kh]}, \label{eq:W_z0_def}\\
 \eta_0 = &  -  \alpha \bar{\Gamma}_\Theta  \bar{U}_b^2 \frac{ \bar{\Gamma}h \cosh(kh)-kh\sinh(kh)}{\bar{\Gamma} h\cosh(kh) + kh \sinh(kh)}. \label{eq:eta_0_def}
\end{align}
Here and below, we use the shorthand $\bar{\Gamma}_\Theta = \rd \Gamma / \rd \Theta |_{\Theta = \bar{\Theta}(0)}$. The functions $W_{z0}$ and $\eta_0$ are plotted in figure \ref{fig:transfer_functions}.

Since basal friction decreases with increasing temperature, we have $\bar{\Gamma}_\Theta< 0$. Defining the unperturbed basal heat flux above the bed as $ Q_0 = -\lim_{Z\rightarrow 0^+} \rd{\bar{\Theta}}/\rd{Z} = G + \bar{\Gamma}\bar{U}_b^2$,
 equal to the sum of geothermal heat flux and the unperturbed basal dissipation rate, the linearized heat equation in the basal boundary layer  \eqref{eq:thermal_boundary_layer_heat} becomes
\begin{subequations} \label{eq:smalldelta_stability}
\begin{equation} \Lambda \left( Pe \, \bar{U}_b \Theta'(Z) - Pe \, W_{z0} Q_0 Z \Theta'(0) \right) -  \od{^2\Theta'}{Z^2} = 0\end{equation}
for $Z > 0$,  while the heat equation  \eqref{eq:thermal_boundary_layer_bed} in the bed trivially yields $\rd \Theta' / \rd Z = 0$ for $Z < 0$. The boundary and  matching conditions \eqref{eq:thermal_boundary_layer_bc1}--\eqref{eq:thermal_boundary_layer_bc2} become
\begin{align}  \od{\Theta'}{Z} + \eta_0 \Theta' =  0 \qquad \mbox{at } Z = 0, \qquad \od{\Theta'}{ Z} \sim & W_{z0}Q_0/\bar{U}_b \qquad \mbox{as } Z \rightarrow \infty. \label{eq:small_delta_bc}
\end{align}
\end{subequations}

The eigenvalue problem \eqref{eq:smalldelta_stability} can also be derived as a parametric limit for $\delta \ll 1$ from a more general spatial stability problem for a parallel-sided slab of ice subject to subtemperate sliding at its base, as described in the supplementary material \S S5.5. \eqref{eq:smalldelta_stability} has solution
\begin{equation} \Theta(Z) = \Theta(0)\left[ \exp\left(-\sqrt{\Lambda Pe\, \bar{U}_b}Z\right) + \frac{W_{z0}Q_0 Z}{\bar{U}_b}\right], \label{eq:small_delta_eigfun} \end{equation}
where the eigenvalue $\Lambda$ satisfies $ \sqrt{\Lambda Pe\, \bar{U}_b} = \eta_0 + W_{z0}Q_0/\bar{U}_b$, and $\sqrt{\Lambda Pe  \, \bar{U}_b}$ must have positive real part (and therefore must be real and positive) in order to satisfy the matching condition \eqref{eq:small_delta_bc}$_2$. Hence we require
\begin{equation} \eta_0 + \frac{Q_0W_{z0}}{\bar{U}_b} > 0, \label{eq:instability_criterion_small_delta} \end{equation}
for a solution of this form to exist.

\begin{figure}
 \centering
 \includegraphics[width=\textwidth]{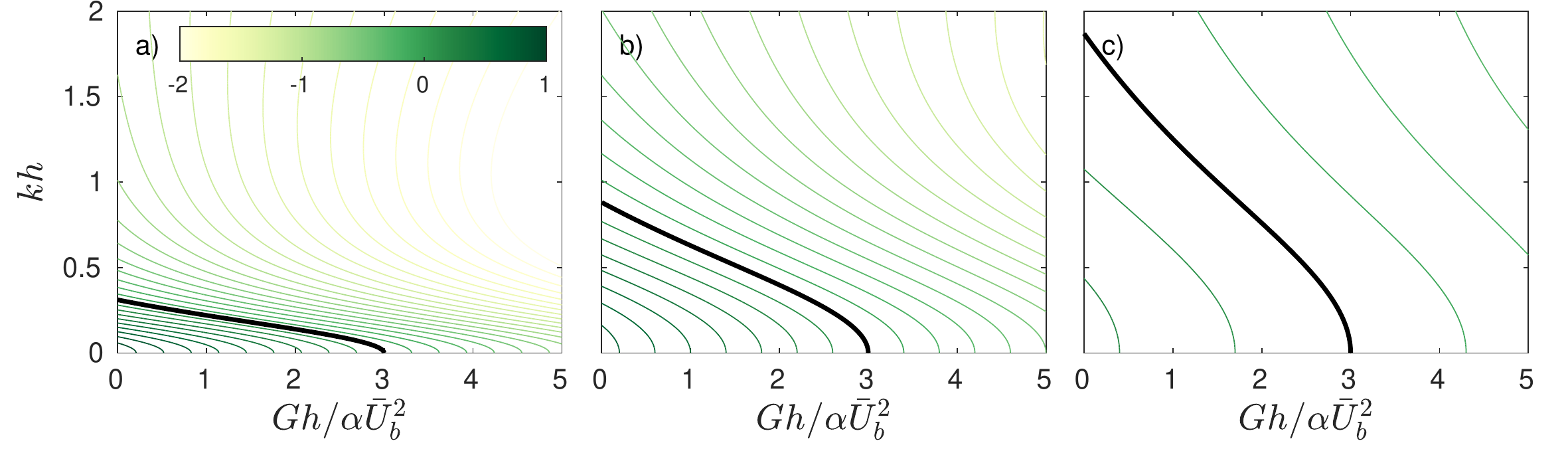}
\caption{Contours of $\breve{\eta}_0 + (\bar{\gamma}h + Gh/(\alpha \bar{U}_b)^2) \breve{W}_{z0} = (\alpha \bar{\Gamma}_\Theta \bar{U}_b^2)^{-1} (\eta_0 +Q_0 W_{z0}/\bar{U}_b)$  against $kh$ and $Gh/(\alpha \bar{U}_b^2)$ for $\bar{\gamma}h = 0.1$ (panel a), $\bar{\gamma}h = 1$ (panel b) and  $\bar{\gamma}h = 10$ (panel a). See figure \ref{fig:transfer_functions} for the definitions of $\breve{\eta}_0$ and $\breve{Q}_{z0}$. The solid black is the zero contour. A viable solution of the form \eqref{eq:small_delta_eigfun} exists if the contoured quantity is positive (below the solid black line), in which case the growth rate is proportional to the square of the contoured quantity, see equation \eqref{eq:smalldelta_growth_rate_solution}, and the solution is automatically unstable.} \label{fig:dispersion}
\end{figure}

When \eqref{eq:instability_criterion_small_delta} is satisfied, the eigenvalue is positive and instability results:
\begin{equation} \Lambda =  \frac{1}{Pe \, \bar{U}_b}\left(\eta_0 + \frac{W_{z0}Q_0}{\bar{U}_b}\right)^2.\label{eq:smalldelta_growth_rate_solution} \end{equation}
Note that $\eta_0$ is positive for small $k$ but changes sign at some finite $k$, while $W_{z0}$ is always negative (figure \ref{fig:transfer_functions}). Consequently, we have stabilization of small wavelengths, since \eqref{eq:instability_criterion_small_delta} is violated for large enough $k$. Recall that $\eta_0$ represents the feedback between raised basal temperature $\Theta'(0)$ and dissipation at the bed (see equation \eqref{eq:stability_coefficient_definition}). That feedback consists of two competing components: raising $\Theta'(0)$ has the direct effect of reducing the basal friction coefficient, and therefore of reducing basal dissipation (the second term on the left-hand side of \eqref{eq:stability_coefficient_definition}$_2$). It also has the indirect effect of allowing faster sliding, which increases dissipation (the first term on the right-hand side). At short transverse wavelengths, or large $k$, basal velocity variations are suppressed by lateral shear stress is in the ice \cite{Hindmarsh2006,SchoofHewitt2013}, and the negative feedback dominates. At larger transverse wavelengths, these basal velocity variations invariably become large enough to dominate the feedback mechanism.

The stability criterion \eqref{eq:instability_criterion_small_delta} shows that we must have a positive feedback with $\eta_0 > 0$ in order for instability to occur as described, but that is insufficient; $\eta_0$ must in fact exceed a positive threshold $-W_{z0}Q_{0}/\bar{U}_b$, which represents the stabilizing effect of downward advection of cold ice as the downstream velocity $\bar{U}_b$ increases.

Next, we extract from  \eqref{eq:instability_criterion_small_delta} an explicit criterion for instability in terms of model parameters. 
The dispersion relation  \eqref{eq:smalldelta_growth_rate_solution}  demonstrates that there is no wavelength selection at long wavelengths: if there is an unstable mode for a given set of parameters, the fastest growing wavenumber is always the limit $k \rightarrow 0$  (or $kh \rightarrow 0$, see figure \ref{fig:dispersion}). This is consistent with only a single ice stream forming in the domain in the numerical solutions (see also \S \ref{sec:discussion} and supplementary material \S S5.9--S5.10.
In the limit $k \rightarrow 0$, we have $\eta_0 \sim - \alpha\bar{\Gamma}_\Theta  \bar{U}_b^2 $, $W_{z0} \sim   \bar{\Gamma}_\Theta \bar{U}_b h/(\bar{\Gamma} h + 3 )$, as well as $Q_0 = \alpha \bar{\Gamma}\bar{U}_b^2 + G$, so that \eqref{eq:smalldelta_growth_rate_solution} gives $ \sqrt{Pe \, \bar{U}_b \Lambda} \sim  -\alpha \bar{\Gamma}_\Theta \bar{U}_b^2 \left[3/(\Gamma h) - G/(\alpha \bar{\Gamma}\bar{U}_b^2)\right]/\left[1+ 3/(\bar{\Gamma} h)\right]$
and we see that the criterion for instability (the right-hand side of the equation is positive) is that $Gh/(\alpha \bar{U}_b^2) < 3$, or
\begin{equation} \label{eq:instability_criterion}  \bar{U}_b^2 > \frac{Gh}{3\alpha}. \end{equation}
This is the criterion we seek: for a  given geothermal flux $G$, Brinkmann number $\alpha$ and ice thickness $h$, the subtemperate sliding velocity $\bar{U}_b$ needs to exceed a threshold value before the spatial pattering instability first occurs in the manner predicted above. 

Note that the instability criterion is based on the `viability' of the thermal boundary layer solution, through requiring that the eigenfunction solution \eqref{eq:small_delta_eigfun} can match with an advection-dominated outer region, with the exponential part of the solution decaying in $Z$. If there is no viable solution of the form \eqref{eq:small_delta_eigfun}, this does not mean that we cannot formulate a stability problem, or even that there is necessarily no instability. It does mean that any growing mode is no longer of a boundary-layer type, and does not grow at rapidly with $\Lambda \sim \delta^{-2}$: instead, we obtain slowly growing solutions that satisfy $T' = 0$ at $z = 0$ at leading order. These can in fact have eigenvalues with real parts. However, the corresponding e-folding lengths are then comparable with ice sheet length, contrasting with e-folding lengths of $\delta^2$ times ice sheet length for the boundary layer solution \eqref{eq:small_delta_eigfun}: in that case, an initial perturbation might grow by a multiple of itself over the full length of the subtemperate region, but is unlikely to lead to a fully formed ice stream.  The supplementary material provides further detail in \S S4.

The instability criterion \eqref{eq:instability_criterion_small_delta}  explains why there is an extended region of slower subtemperate sliding upstream of the rapid transition to patterned flow in figure \ref{fig:bedmap_long}: patterning does not occur as soon as appreciable sliding appears.  Recall however that figure \ref{fig:bedmap_long}a is misleading, since the point at which we perceive the pattern is a fairly large number of spatial $e$-folding lengths from the location at which pattern growth is initiated, depending on the level of noise in the system. This is illustrated in figure \ref{fig:bedmap_long}b, which shows the lateral variance  of basal temperature,
\begin{equation} \label{eq:variance} \var(T)(x) = W^{-1} \int_0^W \left[T(x,y,0) - W^{-1} \int T(x,y',0) \rd y' \right]^2 \rd y, \end{equation}
as a function of $x$. We see that $\var(T)$ begins to grow at $x \approx 0.7$, while the pattern does not reach an $O(1)$ amplitude, with the bed becoming partly temperate, until $x \approx 1.4$.

In fact, the dispersion relation \eqref{eq:smalldelta_growth_rate_solution} predicts that pattern growth should start at a location further downstream ($x \approx 0.9$) than the location where $\var(T)$ actually starts to exhibit approximately exponential growth at comparable rates to those predicted by  \eqref{eq:smalldelta_growth_rate_solution}. The red curve in \ref{fig:bedmap_long}a$_3$  shows a plot of 
\begin{equation} \label{eq:variance_analytical} \var_\infty(T)(x) = c \exp\left(2\delta^{-2} \int_{x_0}^x \Lambda(x') \rd x'\right) \end{equation}
against $x$, the constant $c$ chosen to make the plot visible and $x_0$ being the first location at which a viable solution \eqref{eq:smalldelta_growth_rate_solution} to the eigenvalue problem appears, having linearized  locally around the two-dimensional reference solution in figure \ref{fig:2D_results}a and chosen $k = 2\pi/W$ as the smallest non-zero to fit into the domain. The formula \eqref{eq:variance_analytical} is the form of the variance in $T$ one would expect from our linearized problem if we account for the fact that parameters in the linearization change slowly with $x$.

Clearly, the red curve in \ref{fig:bedmap_long}b  misses part of the original growth in the variance shown in blue.
The most likely explanation for why equation \eqref{eq:smalldelta_growth_rate_solution} underpredicts pattern growth   is that the asymptotic solution relies on separation into an inner region in which diffusion appears at leading order, and an outer region in which the temperature problem is advection-dominated: \eqref{eq:small_delta_bc} arises from a purely advective temperature field in the outer region, justified at leading order because  $\delta^{-2}Pe \gg 1$ is the effective P\'eclet number in the outer region. In advection-diffusion problems, the effects of diffusion can be significant even at relatively large P\'eclet numbers (see, e.g., \cite{Haseloffetal2018} for a glaciological example). We demonstrate in the supplementary material (\S S5.5) that this is also the case here, by comparing the asymptotic solution presented above with a stability analysis that resolves the full depth of the ice.
.

Nonetheless, the effectively conditional nature of the spatial instability  explains why, in some cases, there is no pattern formation at all in the subtemperate region: subtemperate sliding speeds may never get large enough to exceed the threshold value for initiating pattering, as is the case for reference case 2 in figure \ref{fig:2D_results}b.  In fact, even if that threshold is passed, unstable growth of the pattern may not need to lead to a fully formed ice stream if the region of subtemperate sliding is too short and therefore comprises too few e-folding length scales before the boundary with the temperate bed is reached.

\section{Discussion} \label{sec:discussion}

We have modelled the formation of ice streams through positive feedbacks between sliding and dissipation at the bed, using a novel thermomechanical model that combines `shallow ice' balances in the along-flow direction with the ability to resolve fully lateral shear stress and the secondary transverse flow required by mass conservation. The model is the natural generalization of early work on thermomechanical ice sheet flow by Hutter \emph{et al.} \cite{Hutteretal1986}  and Yakowitz \emph{et al.} \cite{Yakowitzetal1986}. Our focus has been on sliding that occurs at temperatures below the melting point, when basal friction is temperature-dependent. 

In that case, the feedback is simple: locally warmer temperatures reduce friction, and will lead to faster sliding. How much faster depends on the length scales involved: for warmer temperatures confined to a narrow region (or a periodic temperature perturbation with a short wavelength, comparable to ice thickness or smaller for a subtemperate flow with significant vertical shearing), the acceleration in flow will be suppressed by lateral shear stresses. Local warming of the bed results in a net increase in basal dissipation and therefore further heating if basal temperatures are elevated over a  sufficiently large span of the bed  (i.e., if the basal temperature perturbation has a sufficiently long lateral wavelengths): the reduction in friction in isolation has the effect of reducing dissipation. Faster sliding, not affected excessively by lateral shear stresses, is key to increased dissipation.

In a steady state setting, the positive feedback can lead to ice flow accelerating  with distance in the downstream direction to form a `finger' of faster-flowing ice surrounded by slower-moving ice: an incipient ice stream. In doing so, the accelerating flow will draw in colder ice from above and the sides, generating a negative cooling feedback that must be overcome. That is the basis of the pattern-forming instability criterion \eqref{eq:instability_criterion_small_delta}. Where this is satisfied, our model predicts a length for the size of the onset region as $\delta^2$ times ice sheet length, where in dimensional terms $\delta^{-1} = \gamma_T(T_s-T_m)/\gamma$, $\gamma_T$ being the (dimensional) temperature derivative of the basal friction coefficent $\gamma$ at the melting point $T_m$, and $T_s$ being surface temperature.
When pattering occurs over a sufficiently short distance downstream,  a fully evolved ice stream-ice ridge pattern forms downstream, with sharply defined margins that slowly move apart in the downstream direction (\S\S \ref{sec:solution_all}\ref{sec:pattern}--\ref{sec:cross_section}). 

Being able to resolve the margins helps reveal the significant (typically $\sim$ one ice thickness wide) region of subtemperate bed that lies inside the ice stream as defined by the velocity pattern (figures \ref{fig:bedmap_zoom} and \ref{fig:cross_section}). Such a zone of rapid subtemperate sliding at the edge of an ice stream was previously predicted by \cite{Haseloffetal2018}. Resolving the margins and the processes that cause their position to shift allows us to explore the effect of different parameterizations of basal friction and hydraulics on ice stream geometry: for instance, we find that drainage systems with more efficient lateral transport tend to generate ice streams that widen more significantly in the downstream direction, but have higher effective pressures and lower velocities in the ice stream center (figure \ref{fig:bedmap_Coulomb}).

The clearest precursor to our work is due to Hindmarsh \cite{Hindmarsh2009}. While he motivates his model by appealing to concentrated, temperature-dependent shear near the bed, 
Hindmarsh's mathematical formulation is a plug flow, implying fast sliding, with friction a relatively slowly varying function of temperature. These two constraints correspond to the limit of $\delta \sim O(1)$ and $\gamma_0 \ll 1$ in our model.

While our model can be solved and analyzed for Hindmarsh's parameter regime, our analysis has focused on the alternative regime of a shearing flow in which friction is highly sensitive to basal sliding ($\delta \ll 1$, $\gamma_0 \sim O(1)$). The latter choice of parameter regime conforms to the expectation that basal sliding is significant only at temperatures close to the melting point \cite{FowlerLarson1980,Fowler1986,Mantellietal2019,MantelliSchoof2019}.
There are some graphically obvious differences between the solutions, with our model allowing the formation of narrow margins and incorporating a hydrological component (which the model description in \cite{Hindmarsh2009} does not mention), the two limits used here and in Hindmarsh's prior work have less obvious but still important differences.

The most significant is the role of downward advection of cold ice in potentially preventing steady state patterns from forming in our analysis in \S \ref{sec:stability}. That downward advection is closely associated with a vertical ice column being able to accommodate at least some shearing. For a pure plug flow, an acceleration in the downstream direction is still balanced purely by a commensurate acceleration in the secondary, transverse flow, which is however then also a plug flow and has no significant vertical component (see supplementary material \S S5.6). As a result, the case of rapid subtemperate sliding is potentially more prone to pattern formation.

Where no pattern emerges in the region of subtemperate sliding, the hydrological component of our model also captures the alternative, hydromechanical feedback mechanism for ice stream formation due to \cite{FowlerJohnson1996} and \cite{KyrkeSmithetal2014,KyrkeSmithetal2015} within a unified framework. Here, bed temperature is fixed at the melting point. Instead, a region of depressed effective pressure will, for a typical basal friction law, lead to reduced friction, faster flow and, for a sufficiently wide such region, to greater basal dissipation in much the same way as the temperature-dissipation feedback. With a hydraulic model in which the additional melt that results must be evacuated through an enlarged set of basal conduits at a reduced effective pressure, a positive feedback results. This can likewise lead to the formation of an ice-stream-ice-ridge pattern.

One of the advantages of our work over previous models of hydromechanical pattern formation \cite{FowlerJohnson1996,KyrkeSmithetal2014,KyrkeSmithetal2015} is in fact our ability to capture the refreezing of the bed that occurs in the ice ridges, and the thermomechanics of the margins: one notable difference between our results and those in \cite{KyrkeSmithetal2014,KyrkeSmithetal2015} is that our ice stream widens in the downstream direction, while theirs narrows, which may be the result of our model incorporating englacial heating and a more careful treatment of heat transport: the thermal model in Fowler and Johnson \cite{FowlerJohnson1996} and Kyrke-Smith \emph{et al.} \cite{KyrkeSmithetal2014,KyrkeSmithetal2015} is based on a similarity solution for a two-dimensional thermal boundary layer due to \cite{Fowler1992}, without lateral heat transport. The validity of that thermal model, especially in the presence of a frozen bed under the ice ridges, and of narrow shear margins, is questionable, while their ice flow model is also not strictly suitable for narrow shear margins (see also \cite{Haseloffetal2015}). In comparing our results to those of Kyrke-Smith \emph{et al.}\cite{KyrkeSmithetal2014,KyrkeSmithetal2015}, note however also that our hydrological models differ somewhat.

A major disadvantage of our model is that it will spontaneously produce only a single ice stream per periodic domain width unless forced strongly with a shorter wavelength. This feature is shared with the model in \cite{FowlerJohnson1996} but not with  \cite{KyrkeSmithetal2014,KyrkeSmithetal2015,Hindmarsh2009}, and can most likely be traced to the ease with which the transverse, secondary flow is able to maintain a laterally flat upper surface of height $h(x)$ at leading order in our model: because we operate at lateral length scales comparable with ice thickness, the surface correction $s_1$ that drives the lateral pressure gradient driving the secondary flow (visualized in figure \ref{fig:cross_section}) never gets large enough to affect the downstream velocity $u$. The thickened ice in the ice ridges, now matter how wide these are in the model, therefore never has any propensity to accelerate that downstream flow and create a new region of faster flow through enhanced dissipation. This, and a possible fix (albeit one that is computationally costly), is discussed in greater detail in the supplementary material (\S\S S5.9--S5.10). There we argue that the appropriate lateral length scale at which pattern growth should be suppressed is the same as the along-flow length scale for ice stream onset, $\sim\delta^2$ times ice sheet length. If confirmed (which is beyond the scope of this paper), that would imply that the width of ice streams at their onset is determined by how temperature sensitive basal friction is.

The inclusion of lateral shear stresses and the Stokes flow model describing the transverse, secondary flow by considering lateral length scales comparable with ice thickness is key in two regards: first, it ensures that growth of short wavelengths is suppressed through lateral shear stresses (\S \ref{sec:stability}). Second, it ensures the ability to capture heat production in, and heat transport through, the narrow margins that form once the ice stream is fully established (figure \ref{fig:cross_section}). Note however that the model is does not include extensional stresses in the along-flow direction, by contrast with what are now standard formulations for ice stream flow \cite{Hindmarsh2009,SchoofHindmarsh2010,KyrkeSmithetal2013}.

Although we have no proof, the steady state version of our model appears well-posed as an initial value problem that can be integrated efficiently, and uniquely, from an ice divide (\S \ref{sec:solution_all}\ref{sec:solution} and supplementary material \S S3). Hanging over our results is however the spectre of temporal instabilities, and in fact, even the question of well-posedness as a time-dependent problem. A myriad of such instabilities was catalogued in \cite{MantelliSchoof2019} for ice flow with subtemperate sliding. Although they formally operate in a different parametric limit ($\delta^2 \sim \varepsilon$ instead of our $\delta^2 \gg \varepsilon$), the instability mechanism in \S 3 of \cite{MantelliSchoof2019} is likely to be relevant to our model in only slightly modified form. In abstract terms, the mechanism is one of amplification of travelling temperature waves in the ice through a phase lag between heat flux and basal temperature introduced by vertical diffusion of heat in the bed. That phase lag leads to a small horizontal and therefore vertical strain rate in ice near the bed that ends up amplifying the basal temperature gradient through advection. The basic physical ingredients for the same instability to occur are contained in our model, and the physical balances involved suggest that it should appear at wavelengths $\sim \delta^2$ times the ice sheet length, if the subtemperate bed is nearly spatially uniform at that scale. 

The fact that our patterned solutions involve largely featureless areas of significant subtemperate sliding upstream of ice steam onset therefore suggsts that we should expect to see a similar temporal instability play out there; 
this clearly calls for further study. In fact, careful analysis of the instability (\S S4 of the supplementary material to \cite{MantelliSchoof2019}) suggests that the instability is suppressed at short wavelengths because the horizontal strain rates involved require extensional stresses that cannot be sustained at short along-flow wavelengths. The model developed in the present paper, being of `shallow ice' type in $x$, omits these extensional stresses, and may therefore not include a key stabilizing mechanism required to control temporal instabilities at short wavelengths. This suggests that the full set of Stokes equations may be necessary to model time-dependent ice stream onset due to subtemperate sliding feedbacks.

As already discussed in \cite{Mantellietal2019}, the plug flow model of \cite{Hindmarsh2009} does not appear to suffer from the same instability, since Hindmarsh's patterns are computed by forward integration in time and yet are steady, without any wave-like or oscillatory features. It is likely that this is again the result of not including vertical shearing in his model: it is relatively easy to show that the extensional strain rates required in \S3 of \cite{MantelliSchoof2019} cannot be generated in a pure plug flow since the depth-integrated flux must remain divergence-free. This does however also serve as a note of caution when treating concentrated basal shearing as a form of sliding with the ice acting as a pure plug flow, since the shear layer could be susceptible to temporal instability that a pure plug flow model may not capture.

There are two other ice sheet models that have reported ice stream flow that appears to converge robustly under grid refinement \cite{BuelerBrown2009,BrinkerhoffJohnson2016}. These are more difficult to compare with our work, primarily because they retain a temperature-dependent viscosity combined with vertical shearing in the ice. As a result, pattern formation in their results may be driven by the dissipation-viscosity feedback. Their sliding formulations also differ from ours: Brinkerhoff and Johnson\cite{BrinkerhoffJohnson2016} uses a piecewise constant friction law of the form \eqref{eq:Weertman}, with different values of $\gamma_0$ for $T < 0$ and $T = 0$. In the confines of our model, it is difficult to see how such a friction law alone would lead to a feedback that generates patterns.

The model due to Bueler and Brown \cite{BuelerBrown2009} poses some thornier problems.
Instead of regularizing the transition from no slip to temperate sliding using a subtemperate sliding law (see also \cite{FowlerLarson1980,Fowler2001}), Bueler and Brown `blend' solutions of two thin-film models across an abrupt change in basal boundary conditions. As with a subtemperate law, the purpose is to avoid the sharp vertical advection that causes the immediate refereezing for a hard switch between no slip to fully temperate sliding.. In our view, it is the (somewhat \emph{ad hoc}) choice of the blending function $f$ (their equation (22)) that solves the refreezing problem, rather than the retention of extensional stresses in one of their thin-film flow models: \cite{Mantellietal2019} show that the retention of such stresses does not generally solve the problem of refreezing in flow across an hard switch from no slip to fully temperate sliding. Further work is required to determine how the ice streams in \cite{BuelerBrown2009} relate to those in other models relying on more directly physics-based descriptions such as \cite{Hindmarsh2009}.

\section{Conclusions} \label{sec:conclusion}

We have shown that temperature-dependent, subtemperate sliding can lead to pattern formation in ice sheets, with clearly defined ice streams emerging. This behaviour is not universal, and if sliding velocities in the region of subtemperate sliding remain too slow, the onset of fully temperate sliding can remain laterally uniform. The model we have developed is also able to capture the formation of ice streams through hydromechanical feedbacks in temperate sliding within the same framework.

Numerous questions remain to be resolved. Chief among these is the question of temporal stability: our model can be solved straightforwardly in steady-state form, and we have not addressed temporal instabilities here, though prior work in \cite{MantelliSchoof2019} suggest that they should be an issue. In fact, it is possible that a more sophisticated model may be necessary for time-dependent calculations, solving a fully three-dimensional Stokes flow model at least near the onset of ice stream flow to capture the full zoology of instabilities in \cite{MantelliSchoof2019}. 

In turn, that poses the question of the form of a  tractable model that can capture ice stream onset in the context of continental-scale ice sheet simulations: solving the Stokes equations at sub-ice-thickness resolution as required here is not feasible for such simultations at present. Adaptive meshing around ice stream onset regions may be one approach, or alternatively it may be possible to treat the onset region as an internal layer within a model that relies on thin-film formulations for ice flow, in a form similar to internal layer models for ice stream shear margins that have been developed recently \cite{Haseloffetal2015,Haseloffetal2018}.

\appendix

\section{Stretched vertical coordinate system} \label{app:stretch}

In the numerical solution of the model of \S \ref{sec:model} we apply a vertical coordinate stretching in order to be able to use finite volumes with regularly-shaped volumes. We define the standard vertical coordinate transformation $\zeta = z/h$, $X = x$, $Y = y$, $\tau = t$ \cite{HindmarshHutter1988}, leading to the differentiation rules
$$ \pd{}{x} = \pd{}{X} - \frac{\zeta}{h}\pd{h}{X}\pd{}{\zeta}, \qquad \pd{}{y} = \pd{}{Y}, \qquad \pd{}{z} = \frac{1}{h}\pd{}{\zeta}, \qquad \pd{}{t} = \pd{}{\tau} - \frac{\zeta}{h}\pd{h}{\tau}\pd{}{\zeta}. $$
For the majority of the equations in the model, the transformation leads only to trivial changes, as the equations do not contain derivatives with respect to $x$, or contain $x$-derivatives of quantities that do not depend on $z$ or $\zeta$, namely $h$ and $\Pi$. The only exceptions are \eqref{eq:mass_conservation} and \eqref{eq:heat}. The use of finite volumes requires that we keep the transformed versions of these equations in divergence form, which can be shown to be
\begin{align} Pe\left[\pd{(hT)}{\tau} + \pd{(uhT))}{X} + \pd{(hvT)}{Y} + \pd{(w_{eff}T)}{\zeta}\right] & \nonumber  \\ +\left[\pd{}{y}\left(h\pd{T}{y}\right) + \pd{}{\zeta}\left(\frac{1}{h}\pd{T}{\zeta}\right)\right] = &  \alpha h \left[\left(\pd{u}{y}\right)^2 + \frac{1}{h^2}\left(\pd{u}{\zeta}\right)^2\right], \\ \pd{(uh)}{X} + \pd{(hv)}{y} + \pd{w_{eff}}{\zeta} = & 0, 
\end{align}
where
\begin{equation} w_{eff} = w - \zeta \left(u\pd{h}{X}+ \pd{h}{\tau}\right). \end{equation}
Further detail about the numerical implementation can be found in the supplementary material.

\section{Transformation of the basal hydrology model} \label{app:hydrology}

In order to handle the difficulties of a vanishing bed permeability that requires infinite effective pressures, we use a transformation to a new pressure-like variable $\Pi$, $N = \mathcal{N}(\Pi)$
with a suitably chosen function $\mathcal{N}$. The basal hydrology problem \eqref{eq:hydrology}$_1$
then becomes
\begin{subequations} \label{eq:hydrology_alt}
 \begin{equation}
 \label{eq:water_storage_alt} e =  \tilde{\Phi}(\Pi), \qquad   q_x =  -\tilde{\kappa}(\Pi)\pd{(h+r^{-1}b)}{x}  \end{equation}
 \begin{equation}  q_y =    -\tilde{\kappa}(\Pi)\pd{\sigma_{nn}}{y} + \beta  \tilde{\kappa}(\Pi)\pd{\mathcal{N}}{\Pi} \pd{\Pi}{y}, \qquad \sigma_{nn} =  \left.\left( p - 2\pd{w}{z} \right)\right|_{z=0} \end{equation}
 where $\tilde{\Phi}(\Pi) = \Phi(\mathcal{N}(\Pi))$, $\tilde{\kappa}(\Pi) = -\kappa(\mathcal{N}(\Pi))$.  
 Assuming $\kappa$ is monotonically decreasing, $\mathcal{N}$ can be chosen to satisfy the differential equation
 \begin{equation} -\kappa(\mathcal{N}(\Pi))\od{\mathcal{N}}{\Pi} = \kappa_2 \end{equation}
for a constant, positive $\kappa_2$.  Defining $\tilde{\kappa}(\Pi)= \kappa(\mathcal{N}(\Pi))$, this allows us to  re-write $q_y$ in the general form
\begin{equation}   q_y =   -\tilde{\kappa}(\Pi)\pd{\sigma_{nn}}{y} - \beta\kappa_2 \pd{\Pi}{y} \label{eq:lateral_flux_alt2} \end{equation}
\end{subequations}
This transformation makes sense if $\mathcal{N}$ is strictly monotone (and therefore invertible), maps $N = \infty$ to $\Pi = 0$ (which implies that $\pdl{\mathcal{N}}{\Pi} < 0$), and if $\tilde{\kappa}$ is an increasing function of $\Pi$ that has a finite limit as $\Pi \rightarrow 0$. We assume that all of these are the case. 
As a concrete example,  consider a power law $\kappa(N) = N^{-k_0}$ with $k_0 > 1$, in which case 
\begin{equation} \label{eq:sample_transformation} \mathcal{N}(\Pi) = \Pi^{-1/(k_0-1)}, \qquad \tilde{\kappa} = \Pi^{k_0/(k_0-1)},\qquad \kappa_2 = \frac{1}{k_0-1}. \end{equation} 
Computationally, we regularize \eqref{eq:sample_transformation} to remain differentiable at $\Pi$ as
\begin{equation}  \mathcal{N}(\Pi) = (\Pi^2+\Pi_0^2)^{-k_0/[2(k_0-1)]}\Pi, \qquad \tilde{\kappa}(\Pi) =  (\Pi^2+\Pi_0^2)^{1/[2(k_0-1)]}\Pi, \label{eq:actual_transformation} \end{equation}
With these assumptions, the hydrology model in isolation turns into a nonlinear diffusion problem for $\Pi$ with $x$ as the time-like direction and $y$ remaining space-like, provided we maintain a positive geometric potential gradient ($- \pdl{(h+r^{-1}b)}{x} > 0$) and permeability decreases with increasing effective pressure  ($\pdl{\tilde{\kappa}}{\Pi} > 0$). Note also that the assumption that water storage $\Phi$ in \eqref{eq:hydrology}$_1$ goes to zero as $N \rightarrow \infty$ now also translates into $e = \Phi \rightarrow 0$ as $\Pi \rightarrow 0^+$. Consequently, the boundary condition \eqref{eq:freeze_margin} at an inward-migrating margin translates into prescribing $\Pi = 0$ there. Likewise, the Dirichlet condition \eqref{eq:temp_pre_Dirichlet} becomes
$ T = 0$ at $z = 0$ where $\Pi > 0$.

One of the corollaries is that we cannot handle permeabilities $\kappa(N)$ that increase with effective pressure, as is characteristic of channel-like drainage conduits \cite{Rothlisberger1972,Schoof2010,Peroletal2015,Plattetal2016,Meyeretal2018}. That should however not be surprising, as the spontaneous localization of channelized drainage onto single conduits is at odds with the distributed drainage assumed by a macroporous drainage model of the form \eqref{eq:hydrology}.

\section{An ice divide solution} \label{app:divide}

At an ice divide $x = 0$, we assume symmetry in $x$ and hence $u = 0$, $\pdl{h}{x} = \pdl{b}{x} = 0$. Assume also there is no lateral structure, so $u = u(z)$, $w = w(z)$ while $v = 0$, $a = \bar{a}$ and $T_s$ is also independent of $y$. Below, we show how a steady-state solution for velocity and temperature can be found for a given ice divide thickness $h(0)$. Instead of solving \eqref{eq:antiplane} for $u$, we differentiate with respect to $x$, so
$$ \pd{^2}{z^2} \pd{u}{x} = \pd{^2(h+b)}{x^2} $$
on $0 < z < h$ subject to $\pdl{^2u}{z\partial x} = 0$ at $z = h$, $\pdl{^2u}{z\partial x} = \gamma(T_b)\pdl{u}{x}$ at $z = 0$, where $T_b = T|_{x=0,z=0}$ assuming subtemperate basal conditions. This effectively \eqref{eq:model_differentiate}$_{1-3}$, and has solution
$$ \pd{u}{x} = -\left[\frac{z(2h-z)}{2}+ \gamma(T_b)^{-1}\right]\pd{^2(h+b)}{x^2} $$
Equation \eqref{eq:mass_conservation} in steady state therefore becomes, on differentating $Q$ explicitly
\begin{equation} \bar{a}  = \pd{Q}{x} = \int_0^h \pd{u}{x} \rd  z=  -\left[\frac{h^3}{3} + \gamma(T_b)^{-1}h^2\right]\pd{^2(h+b)}{x^2},  \end{equation}
which therefore determines $\pdl{^2(h+b)}{x^2}$ and hence $\pdl{u}{x}$ for a given basal $T_b = T(0,y,0)$ and $h$; the latter is given, but we still need to solve for the former. The steady state heat equation \eqref{eq:heat} becomes
\begin{equation} Pe w \pd{T}{z} - \pd{^2 T}{z^2} = 0 \label{eq:divide_heat} \end{equation}
subject to $T = T_s|_{x = 0}$ at $z = h$ and $-\pdl{T}{z} = G$ at $z = 0$. The vertical velocity can be computed from $\pdl{u}{x}$ as
$$ w(z) = \int_0^z \pd{u}{x} \rd z' = -\left[\frac{3hz^2-z^3}{6} + \gamma(T_b)^{-1} z \right]\pd{^2(h+b)}{x^2} $$
and  \eqref{eq:divide_heat} can be reduced to (see also \cite{Hindmarsh1999})
$$ T(z)  = T_b  -Gz  +  \int_0^z \exp\left(-Pe \int_0^{z'} w(z'') \rd z''\right) \rd z' $$
by separation of variables. To complete the solution of the divide problem, we need to solve the rootfinding problem for $T_b$ that results from putting $T(h) = T_s$. This can in general only be done numerically, but once $T_b$ is known, the complete solution can be reconstructed.

%\ethics{No experimental data, human or animal subjects were used in the article.}

%\dataccess{This article has no additional data.}

%\aucontribute{Both authors contributed equally to this work.}

%\competing{We declare no competing interests.}

%\funding{

\vspace{12pt}

\noindent
{CS acknowledges NSERC Discovery Grant RGPIN-2018-04665. Computing resources were provided by Compute Canada. EM is supported by award NA18OAR4320123 from the National Oceanic and Atmospheric Administration, U.S. Department of Commerce. The statements, findings, conclusions, and recommendations are those of the author(s) and do not necessarily reflect the views of the National Oceanic and Atmospheric Administration, or the U.S. Department of Commerce.
%}
%\ack{
CS would like to thank the Department of Geology at the University of Otago for their hospitality during part of this work. EM acknowledges a visiting scholar position at the Department of Geophysics at Stanford University.}
%}

%\disclaimer{Insert disclaimer text here.}

%\bibliography{references}

\end{document}

\maketitle

\noindent

\section{Construction of the model from first principles} \label{sec:model_construction}

\subsection{The basic model} \label{sec:model_basic}

The underlying, dimensional model is a three-dimensional Stokes flow  occupying a bounded domain $\Omega$ with a fixed lower surface $z = b(x,y)$ and a free surface upper surface $z = s(x,y,t)$, and an associated heat equation that has a counterpart in the unbounded `bed' domain below $\Omega$:
\begin{subequations} \label{eq:basic}
\begin{align} \label{eq:Stokes_basic1} \nabla \cdot \mu \left[ \nabla \bm{u}  + \left(\nabla \bm{u}\right)^\mathrm{T}\right] - \nabla p + \rho \bm{g} = & \bm{0} & \mbox{for } b < z < s \\
\label{eq:Stokes_basic2} \nabla \cdot \bm{u} = & 0, & \mbox{for } b < z < s, \\
 \rho c \pd{T}{t} + \rho c \bm{u} \cdot \nabla T - k \nabla^2 T = & \frac{\mu}{2} \left[ \nabla \bm{u} + \left(\nabla \bm{u}\right)^\mathrm{T}\right] : \left[ \nabla \bm{u} + \left(\nabla \bm{u}\right)^\mathrm{T}\right]  & \mbox{for } b < z < s,\label{eq:heat_basic}  \\
 \rho c \pd{T}{t} - k \nabla^2 T = & 0 & \mbox{for } z < b. \label{eq:heat_basic_bed}
\end{align}
Here $\bm{u}$ is velocity, $p$ pressure and $T$ is temperature. $\rho$ and $\bm{g} = (0,0,-g)$ denote density and acceleration due to gravity, $\mu$ fluid viscosity, which we will shortly assume to be constant for simplicity, $c$ is heat capacity and $k$ thermal conductivity, which we again assume to be the same for the ice and bed. `$:$' stands for the usual double contraction over both indices in the product of two tensors. Note that do not include a model for temperate ice formation in \eqref{eq:heat_basic} here; this is clearly a desirable addition for future work \citep{Schoof2012,Haseloffetal2015, SchoofHewitt2016,HewittSchoof2017}.

We assume that sliding occurs through a temperature- and water-pressure-dependent sliding law at the lower boundary $\partial \Omega_b$. Assuming that the $xy$-plane is horizontal, we define bed and surface normals as
\begin{equation} \bm{n}_b = \frac{1}{\sqrt{1 + \left(\pd{b}{x}\right)^2 + \left(\pd{b}{y}\right)^2}}\left( - \pd{b}{x} , - \pd{b}{y}, 1\right), \qquad \bm{n}_s = \frac{1}{\sqrt{1 + \left(\pd{s}{x}\right)^2 + \left(\pd{s}{y}\right)^2}}\left(- \pd{s}{x} , - \pd{s}{y} ,1\right). \end{equation}
Shear stress at the bed is given by
\begin{equation} \bm{\tau}_b = \left(\bm{I} - \bm{n}_b \bm{n}_b\right) \cdot \mu\left.\left[  \nabla \bm{u} + \left(\nabla \bm{u}\right)^\mathrm{T}\right]\right|_{z=b} \cdot \bm{n}_b, \end{equation}
where $\bm{I}$ is the identity tensor and $\bm{I}-\bm{n}_b\bm{n}_b$ is the projection onto the tangent plane to $\partial \Omega_b$. Sliding velocity is
\begin{equation} \bm{u}_b =  \left(\bm{I} - \bm{n}_b \bm{n}_b\right) \cdot \left. \bm{u}\right|_{z =b}, \end{equation}
and we assume a friction law of the form
\begin{equation} \bm{\tau}_b = f(T,N,|\bm{u}_b|)\frac{\bm{u}_b}{|\bm{u}_b|}, \end{equation}
where $T$ is temperature and $N$ is effective pressure, the difference between between normal stress at the bed and basal water pressure. For temperatures below the melting point $T_m$, there is no dependence on $N$ as there is no `free' water at the ice-bed interface (there will typically be a pre-melted water film, whose thickness is however fully determined by the temperature deficit $T_m-T$). In that case, we assume we assume a simple multiplicative friction law of the form:
\begin{equation} \label{eq:subtemperate_basic} f(T,N,|\bm{u}_b|)= \gamma(T)|\bm{u}_b| \qquad \mbox{for } T < T_m\end{equation}
where $\gamma$ is a decreasing function of $T$. A concrete example that we will use in practice  is 
\begin{equation} f(T,N,u) = C \exp(-T/T_0)u, \label{eq:slide_exponential} \end{equation}
where $T_0$ is a temperature scale over which significant sliding appears as the melting point is approached. In practice, $T_0$ may be quite small, a fact we explore in the main paper through the parameter $\delta$ defined below. At the melting point, we consider more general sliding laws
\begin{equation} f(T,N,|\bm{u}_b|) = f_m(N,|\bm{u}_b|) \qquad \mbox{for } T = T_m, \end{equation}
where $f_m$ increases with its first argument and is non-decreasing in the second, and approaches zero as sliding speed $|\bm{u}_b|$ does. We also expect that, for large effective pressures $N$, the friction law behaves as  
$$f_m(N,|\bm{u}_b|) \sim \gamma(T_m)|\bm{u}_b|, $$
so there is no jump in basal friction when the bed reaches the melting point but there are limited amounts of liquid water. We will make this statement more specific later.

In addition to a sliding law, we assume that melt rates at the bed are small, so
\begin{equation} \bm{u}_b \cdot \bm{n} = 0 \qquad \mbox{at } z = b. \end{equation}
Conservation of energy at the bed equates the imbalance of heat fluxes into and out of the bed with a rate of change of energy storage at the bed, and a latent heat flux along the bed:
\begin{equation}
\label{eq:bed_enthalpy} \rho_w L \left(\pd{e}{t} + \nabla_b \cdot \bm{q}_b\right) = \bm{\tau}_b\cdot \bm{u}_b - \left[ - k\nabla T \cdot \bm{n}_b \right]^+_-
\end{equation}
where $e$ is water content per unit area of the bed, $\nabla_b \cdot \bm{q}_b$ denotes the divergence of the bed-parallel water flux $\bm{q}_b$, the divergence being taken on the curved surface $\partial \Omega_b$ defined by $z = b(x,y)$. $[\cdot]_-^+$ denotes the difference between the bracketed quantity evaluated as limits taken from above and below $\partial \Omega_b$. $\rho_w$ is the density of water and $L$ latent heat of fusion per unit mass of water. $\rho_w L \bm{q}_b$ can be identified with a latent heat flux along the bed, and $\rho_w L e$ with latent content per unit area of bed. We assume latent heat content and flux vanish below the melting point,
\begin{equation} e = 0, \qquad \bm{q}_b = \bm{0} \qquad \mbox{when } T < T_m \mbox{ at } z = b. \end{equation}
At the melting point, we assume that both are related to the effective pressure, and can be written as
\begin{equation} e = e_m(N), \qquad \bm{q}_b = -\kappa(N) \nabla_b \phi_m \label{eq:macroporous} \end{equation}
where $e_m$ is a decreasing function of $N$, denoted by $\Phi$ in the main text; we resort to the alternative notation $e_m$ here in order not to cause confusion with the hydraulic potential $\phi_m$.  $\nabla_b$ is the gradient of a hydraulic potential $\phi = p_w + \rho_w g b$ defined on the manifold $\partial \Omega_b$. Here, $p_w$ is basal water pressure. As a hydrology model, \eqref{eq:macroporous} takes the simplest conceivable `macroporous' form \citep[e.g.][]{FlowersClarke2002,Haseloffetal2019}, with a permeability $\kappa$ that depends on effective pressure, and we ignore some of the complications that can arise from the freezing of porous beds \citep{Haseloff2015}. As we will see below, it is essential that $\kappa$ should decrease with increasing $N$ throughout, which rules out channel-like drainage conduits \citep[e.g.][]{Schoof2010b,Peroletal2015,Plattetal2016,Meyeretal2018}.
$N$ is linked to $p_w$ through the definition
$$ N = \left.\left\{ p - \bm{n}_b \cdot \mu \left[\nabla \bm{u} + \left(\nabla \bm{u}\right)^\mathrm{T}\right] \cdot \bm{n}_b \right\} \right|_{z=b} - p_w $$
as the difference between normal stress at the bed and basal water pressure. Hence
\begin{equation} \phi_m =  \left.\left\{ p - \bm{n}_b \cdot \mu \left[\nabla \bm{u} + \left(\nabla \bm{u}\right)^\mathrm{T}\right]  \cdot \bm{n}_b   \right\} \right|_{z=b} - N + \rho_w g b. \end{equation}
We assume that $\kappa$ is a decreasing function of $[N]$. As we will discuss later, the model we derive below requires that $\kappa$ should approach zero for some $N$, and for the sake of simplicity we will assume that $\kappa \rightarrow 0$ as $N \rightarrow \infty$, $\kappa \rightarrow \infty$ as $N \rightarrow 0$.

We augment the enthalpy equation \eqref{eq:bed_enthalpy} with the constraint that $T$ cannot become positive
\begin{equation} T \leq 0 \qquad \mbox{on } \partial \Omega_b. \end{equation}
Equation \eqref{eq:bed_enthalpy} is a partial differential equation posed on the two-dimensional manifold $\partial \Omega_b$, but one which itself involves a one-dimensinoal free boundary $\Gamma_m$ embedded in $\partial \Omega_b$, namely that between the subset $\partial \Omega_{b,m}$ of $\partial \Omega_b$ on which $T = T_m$, and its relative complement $\partial \Omega_{b,-}$ on which $T < T_m$. The appropriate boundary condition at $\Gamma_m$ is
\begin{equation} \label{eq:no_flow_basic}
 \left[ \bm{q}_b - e \bm{v}_m\right]_-^+ \cdot \bm{n}_m = 0,
\end{equation}
where $\bm{v}_m \cdot \bm{n}_m$ is the rate at which the free boundary is migration, and $[\cdot]_-^+$ is the jump in the bracketed quantity between the two sides of $\Gamma_m$. Naturally, $\bm{n}_m$ itself is confined to the tangent plane to $\partial \Omega_b$. In other words, we consider \eqref{eq:bed_enthalpy} to hold in weak form everywhere on $\partial \Omega_b$, but it acts as a differential equation for $N$ only on $\partial \Omega_{b,m}$, while $N$ is not defined elsewhere. This mirrors the formulation for temperate ice models \citep{SchoofHewitt2016}. 

We require additional information to fix the actual rate of margin migration. As investigated in further detail in \citet{Schoof2012}, \citet{Haseloff2015} and \citet{Haseloffetal2015,Haseloffetal2018}, we require that an outward-migrating margin (with $\bm{v}_m\cdot \bm{n}_m > 0$ if $\bm{n}_m$ points into $\partial \Omega_{b,-}$) does not correspond to a singular heat flux out of the bed on the temperate side:
\begin{equation} \lim_{ n_m \rightarrow 0} n_m^{1/2} \left[ -\nabla T \cdot \bm{n}_b\right]_-^+ \geq 0, \end{equation}
where $n_m$ is distance from the free boundary $\Gamma_m$ along the bed, approached from within $\partial \Omega_{b,0}$. For an inward migrating margin ($\bm{v}_m\cdot \bm{n}_m > 0$), we demand instead that the bed water content goes to zero at the boundary
\begin{equation} \lim_{n_m\rightarrow 0} e = 0. \end{equation}

At the upper surface $\partial \Omega_s$ defined by $z = s$, we have vanishing traction
\begin{equation}  \mu \left[\nabla \bm{u} + \left(\nabla \bm{u}\right)^\mathrm{T}\right] \cdot \bm{n}_s - p \bm{n}_s = 0, \label{eq:surface_traction_basic}\end{equation}
and a kinematic boundary condition
\begin{equation} \pd{s}{t} + u \pd{s}{x} + v \pd{s}{y} = w + a, \end{equation}
where $\bm{u} = (u,v,w)$, and $a$ is a prescribed surface mass balance. In addition, the heat equation is subject to a boundary condition that we prescribe here in the form of specified temperature
\begin{equation} T = T_s \end{equation}
To close the problem, we also require boundary conditions below the bed, where we specify a far-field geothermal heat flux,
\begin{equation}  - k \pd{T}{z} = q_{geo}. \end{equation} 
\end{subequations}

\subsection{Non-dimensionalization and simplification} \label{sec:leading_order}

For simplicity, assume that the bed elevation function $b$ depends only on $x$, which allows us to treat the $x$-axis as being oriented in the `main' flow direction.  The scales chosen below reflect this: we scale position $x$ with the length of the ice sheet, and the corresponding velocity component $u$ with the same balance velocity as in a shallow ice model, which also dictates a scale for ice thickness and therefore vertical position $z$. Naturally we also assume a scale separation between ice sheet thickness and ice sheet length, relying on the small aspect ratio of ice sheets.

In the transverse direction, we however allow for the possiblity that structure emerges at length scales comparable to ice thickness: as we show in the main body of the paper, this is indeed the shortest transverse length scale at which instabilities occur due to thermomechanical feedbacks during the temperature-controlled onset of sliding. While the flow of ice is predominantly aligned in the $x$-direction, short-wavelength lateral patterning causes a secondary transverse flow, induced by along-flow changes in the main ice flow velocity $u$: the incompressibility of ice dictates that the secondary flow compensate for such velocity gradients in $x$. (Note that, as an alternative to making $b$ a function of $x$ only, we could assume that $b$ is periodic in $y$, with lateral structure at length scales comparable to ice thickness, but this would not add significant insight.) 

We assume scales $[x]$ for along-flow position and $[a]$ for accumulation rate are given, and define scales $[z]$, $[t]$, $[u]$, $[v]$, $[p]$, $[T]$ and $[N]$ for ice thickness, time, downstream velocity, transverse velocity, pressure, temperature and effective pressure through
$$ [u][z] = [a][x], \qquad [u][t] = [x], \qquad \mu[u]/[z]^2 = \rho g [z]/[x], \qquad [v]/[z] = [u]/[x], \qquad [p] = \rho g [z], $$
\begin{equation} [T] = T_m - T_{s0}, \qquad \rho_w L \kappa([N])\rho g [z]/[x]^2 = k [T]/[z], \end{equation}
where $T_{s0}$ is a typical surface temperature. Hence
$$ [z] = \left(\frac{\mu[a][x]^2}{\rho g}\right)^{1/4}, \qquad [u] = \left(\frac{\rho g [a]^{3/4}[x]^2}{\mu}\right)^{1/4}, \qquad [t] = \left(\frac{\mu[x]^2}{\rho g [a]^{3/4}}\right)^{1/4},  \qquad [v] = [a], $$
while no closed-form expression for $[N]$ is available without specifying the form of the permeability function $\kappa$. These scales $[z]$, $[t]$, $[u]$ and $[v]$ are those for a shallow ice model with a constant viscosity \citep[e.g.][]{FowlerLarson1978,MorlandJohnson1980,Hutter1983} if we regard $[v]$ as a scale for vertical velocity. As in shallow ice theory, we can define an aspect ratio $\varepsilon$ for the ice sheet, as well as effective P\'eclet and Brinkmann numbers $Pe$ and $\alpha$,
\begin{equation}
\varepsilon = \frac{[z]}{[x]}, \qquad Pe = \frac{\rho c u [z]^2}{k [x]}, \qquad \alpha  = \frac{\mu[u]^2}{k[T]}.
\end{equation}
In common with typical thermomechanical ice sheet models, we assume that $\varepsilon \ll 1$ but $Pe = O(1)$ and $\alpha = O(1)$.  Three additional dimensionless parameter emerges in the model as a dimensionless geothermal heat flux, the ratios of ice to water density, and of normal stress to effective pressure. We write these here in the form
\begin{equation} G = \frac{q_{geo}[z]}{k[T]}, \qquad r = \frac{\rho}{\rho_w}, \qquad \beta = \frac{[N]}{\rho g [z] \varepsilon^2} \end{equation}
in the anticipation that $\beta = O(1)$, but discuss different regimes for $\beta$ below.

In addition, the friction law and basal water storage capacity will, in general, engender additional dimensionless parameters, whose form depends on the specifics of the sliding law. The relevant water storage parameter is
\begin{equation} \nu = \frac{\rho_w L e_m'([N])[z]}{k[T]} \end{equation}
where the prime denotes differentiation. Two obvious slding parameters that are central to our model are the following: first, a dimensionless friction parameter, defined in terms of the multiplicative friction law \eqref{eq:subtemperate_basic} as
\begin{equation} \gamma_0 = \frac{\gamma(T_m)[u][x]}{\rho g [z]^2}, \end{equation}
which measures the amount of friction generated by a sliding velocity comparable to shearing velocities relative to the driving stress. Formally, we assume that $\gamma_0^{-1} \sim O(1)$: in other words, significant sliding is possible before the melting point is reached, but not sliding velocity that are large compared with shearing across the thickness of the ice.  Second, we find a dimensionless thermal activation parameter,
\begin{equation} \delta = \frac{\gamma(T_m)}{\gamma_T(T_m)[T]},
\end{equation}
where the subscript $T$ on $\gamma_T$ denotes differentiation. The smaller $\delta$, the more sensitive friction is to changes in temperature (since $\gamma_T(T_m) = \delta \gamma(T_m)/[T]$) compared with the natural temperature scale $[T]$. The case we are primarily interested in here is that of small $\delta$, though the analysis in the main paper reveals that the model we are about to construct as a leading order model requires that we restrict ourselves to
$$ \varepsilon \ll \delta^2. $$
Additional parameters will arise in the friction law depending on the form of $f_m(N,u_b)$, which we discuss when we specify different friction laws.

With these scales and parameters in hand, we define dimensionless counterparts to the variables involved in the model (as well as the data functions $b$, $a$ and $T_s$):
$$  u = [u]u^*, \qquad v =[v]v^*, \qquad w = [v]\left(w^* + u^*\pd{b^*}{x^*}\right), \qquad x = [x]x^*. \qquad y = [z]y^*, \qquad z = [z](z^*+b^*) $$
$$ p = [p]p^*, \qquad T = T_m + [T]T^*, \qquad N = [N]N^*, \qquad s = [z](b^* + h^*), $$ 
\begin{equation} b = [z]b^*, \qquad a = [a]a^*, \qquad T_s = T_m + [T]T_s^*. \end{equation}
Note that, alongside a scaling, we have introduced a curvilinear transformation in which $z^*$ depends on $x$ and $z$, and similarly the dimensionless `vertical' velocity $w^*$ corresponds to the actual vertical velocity minus a correction  term $u\pdl{b}{x}$. The latter corresponds to the projection of the horizontal velocity onto the plane perpendicular to the bed at leading order in $\varepsilon$: effectively, $w^*$ is the upward component of velocity in a plane normal to the bed, rather than in a vertical plane, and therefore accounts for bed slope. This device makes the leading order dimensionless model significantly simpler to write down. In addition, we switch from ice surface elevation $s$ to a scaled ice thickness $h^*$ as the primary dependent variable describing the evolving geometry of the domain.

As described, we assume that $b^* = b^*(x^*)$ is independent of $y^*$, which effectively turns out to mean that the bed does not have significant transverse features with wavelengths comparable to ice thickness. To the order indicated, the model \eqref{eq:basic} becomes the following: The Stokes equations \eqref{eq:Stokes_basic1}--\eqref{eq:Stokes_basic2} are, omitting the asterisk decorations on dimensionless variables,
\begin{subequations}
 \begin{align}
  \label{eq:Stokes_scaled2} \pd{u}{x} + \pd{v}{y} + \pd{w}{z} = & O(\varepsilon^2), \\
  \pd{^2u}{y^2} + \pd{^2u}{z^2} - \pd{p}{x} + \pd{p}{z}\pd{b}{x} = & O(\varepsilon^2), \\
  \varepsilon^2\left(\pd{^2v}{y^2} + \pd{^2v}{z^2}\right) - \pd{p}{y} = & O(\varepsilon^4), \label{eq:Stokes_scaled1b} \\
  \varepsilon^2\left[\pd{^2w}{y^2} + \pd{^2w}{z^2} + \left(\pd{^2 u}{y^2} + \pd{^2 u}{z^2}\right) \pd{b}{x} \right] - \pd{p}{z} - 1 = & O(\varepsilon^4), \label{eq:Stokes_scaled1c}
 \end{align}
for $0 < z < h$, where we have used \eqref{eq:Stokes_scaled2} to simplify the remaining equations. These are subject to the boundary conditions \eqref{eq:surface_traction_basic}, now rendered at the surface $z = h$ in the form
\begin{align} - \pd{u}{y} \pd{h}{y} + \pd{u}{z} = & O(\varepsilon^2), \\
 - \pd{u}{y} \pd{h}{x} - 2\pd{h}{y}\pd{v}{y} + \pd{v}{z} + \pd{w}{y} = & O(\varepsilon^2), \\
 p - \varepsilon^2\left[2\pd{w}{z} + \pd{u}{z}\left(\pd{b}{x} - \pd{h}{x}\right) - \left(\pd{v}{z}+\pd{w}{y} + \pd{u}{y}\pd{b}{x}\right)\pd{h}{y}\right] = & O(\varepsilon^4), \label{eq:normalstress_scaled} \\
 \pd{h}{t} + u \pd{h}{x} + v \pd{h}{y} = w + a, \label{eq:kineamtic_scaled}
\end{align} 
At the bed $z = 0$, the boundary conditions on \eqref{eq:Stokes_scaled2}--\eqref{eq:Stokes_scaled1c} become
\begin{align}
 \pd{u}{z} = & f(T,N,|u|)\frac{u}{|u|} + O(\varepsilon^2), \label{eq:sliding_scaled} \\
 \pd{v}{z} + \pd{w}{y} = & f(T,N,|u|)\frac{v}{|u|} + O(\varepsilon^2), \\
 w = & 0. \label{eq:nopenetration_scaled}
\end{align}
Here $f$ is a dimensionless counterpart to its original namesake, bearing in mind that the dimensionless melting point is $T = 0$. For instance \eqref{eq:subtemperate_basic} would become (with the definitions of the parameters $\gamma_0$ and $\delta$ previously given
$$ f(T,N,u) = \gamma_0 \gamma(T/\delta)u \qquad \mbox{for } T < 0 $$
for some dimensionless function $\gamma$ that has $\gamma_T(0) = 0$. In particular,  the exponential sliding law \eqref{eq:slide_exponential} specifically becomes
$$ f(T,N,u) = \gamma_0 \exp(-T/\delta)u \qquad \mbox{for } T < 0. $$
The parameters $\gamma_0$ and $\delta$ are therefore absorbed into the sliding law and somewhat `invisible' in the model in its most general formulation.

The heat equation \eqref{eq:heat_basic} and \eqref{eq:heat_basic_bed} meanwhile becomes
\begin{align} 
 Pe \left( \pd{T}{t} + u \pd{T}{x} + v\pd{T}{y} + w \pd{T}{z}\right) - \left(\pd{^2 T}{y^2} + \pd{^2T}{z^2} \right) = & \alpha \left[\left(\pd{u}{z}\right)^2 + \left(\pd{u}{z}\right)^2\right] + O(\varepsilon^2) & \mbox{for } 0 <z <h, \label{eq:heat_scaled1}  \\
 Pe \pd{T}{t}  - \left(\pd{^2 T}{y^2} + \pd{^2T}{z^2} \right) = & O(\varepsilon^2) \label{eq:heat_scaled2} & \mbox{for } z < 0
\end{align}
with boundary conditions
\begin{align} \nu \pd{e}{t} + \pd{q_x}{x} + \pd{q_y}{y} = & \alpha f(T,N,|u|)|u| + \left[\pd{T}{z}\right]_-^+  + O(\varepsilon^2) & \mbox{at } z = 0, \label{eq:basal_energy_scaled}\\
 T = & T_s & \mbox{at } z = h, \\
- \pd{T}{z} \rightarrow & G & \mbox{as } z \rightarrow -\infty \label{eq:geothermal_scaled}
\end{align}
where $e$ is a suitably-scaled version of the original dimensional water storage capacity: reintroducing asterisks to distinguish the two, this would be $e^* =  [N] e/e_m'([N])$. The omitted $O(\varepsilon^2)$ correction term absorbs the curvature terms that appear in the scaled version of the operator $\nabla_b$. Note that the main paper focuses on steady state solutions, and the parameter $\nu$ consequently disappears. In reality, if $\nu$ is large, it indicates that water storage is more important than drainage during transients (in the sense that the effective pressure has been chosen incorrectly to ensure sufficient permeability of the bed, rather than to account for the large storage capacity of the bed, and we may have something closer to the situation modelled by the `undrained bed' models of \citet{Tulaczyketal2000b} and \citet{Robeletal2014}).

The fluxes $q_x$ and $q_y$ take the form
\begin{align}
 q_x = & -\kappa(N)\pd{}{x}\left(p + r^{-1} b - \varepsilon^2 \beta N \right), \label{eq:flux_downstream_scaled} \\
 q_y = & -\kappa(N)\pd{}{y}\left( \varepsilon^{-2}p - 2\pd{w}{z} - \beta N\right), \label{eq:flux_transverse_scaled}
\end{align}
where the correction term $\varepsilon^2 \beta N$ has been retained in the definition of $q_x$ to illustrate different parameteric limits in $\beta$.
\end{subequations}

We will discuss those limits shortly. First, we simplify \eqref{eq:Stokes_scaled2}--\eqref{eq:nopenetration_scaled} by expanding
$$ p = p_0 + \varepsilon^2 p_1 + O(\varepsilon^4), \qquad h = h_0 + \varepsilon^2 h_1 + O(\varepsilon^4) $$
From \eqref{eq:Stokes_scaled1b}, \eqref{eq:Stokes_scaled1c} and \eqref{eq:normalstress_scaled}, we find that
\begin{equation} p_0 = h_0-z, \qquad \pd{h_0}{y} = 0, \end{equation}
so $h_0 = h_0(x,t)$ and
\begin{equation} \pd{p_0}{y} = 0, \end{equation}
The lateral secondary flow serves to keep the upper surface flat at leading order. Expanding to higher order and using these results, the entire flow model \eqref{eq:Stokes_scaled2}--\eqref{eq:nopenetration_scaled} can be simplified significantly to yield, omitting the $O(\varepsilon^2)$ error terms
 \begin{align}
  \label{eq:Stokes_final2} \pd{u}{x} + \pd{v}{y} + \pd{w}{z} = & 0, \\
\label{eq:Stokes_final1a}  \pd{^2u}{y^2} + \pd{^2u}{z^2} - \pd{(h_0+b)}{x} = & 0, \\
\left(\pd{^2v}{y^2} + \pd{^2v}{z^2}\right) - \pd{p_1}{y} = & 0, \label{eq:Stokes_final1b} \\
  \pd{^2w}{y^2} + \pd{^2w}{z^2} + \pd{(h_0+b)}{x} \pd{b}{x} - \pd{p_1}{z}  = & 0, \label{eq:Stokes_final1c}
 \end{align}
for $0 < z < h_0$, and
\begin{align}  \pd{u}{z} = & 0, \label{eq:bed_friction_final} \\
 - \pd{u}{y} \pd{h_0}{x} + \pd{v}{z} + \pd{w}{y} = & 0, \\
 -h_1 + p_1 - 2\pd{w}{z} = & 0, \label{eq:surface_correction_supp}  \\
 \pd{h_0}{t} + u \pd{h_0}{x} + v \pd{h_0}{y} = w + a, \label{eq:kinematic_final}
\end{align} 
at $z = h_0$, with the boundary conditions at the  bed $z = 0$ unchanged from \eqref{eq:sliding_scaled}--\eqref{eq:nopenetration_scaled}, except that we drop the $O(\varepsilon^2)$ correction terms. The thermal part of the model, \eqref{eq:heat_scaled1}--\eqref{eq:geothermal_scaled} is likewise unchanged, and only the flux definitions \eqref{eq:flux_downstream_scaled} and \eqref{eq:flux_transverse_scaled} are altered to become
\begin{align}
 q_x = & -\kappa(N)\pd{}{x}\left(h_0 + r^{-1} b - \varepsilon^2 \beta N \right), \nonumber \\
 q_y = & -\kappa(N)\pd{}{y}\left( p_1 - 2\pd{w}{z} - \beta N\right), \label{eq:flux_transverse_final}
\end{align}
at $z = 0$, where we have used the fact that $\pdl{p_0}{y} = 0$ to simplify the expression for $q_y$.

Our final scaling assumption is that effective pressure $[N]$ is much less than overburden $\rho  g [z]$, or in terms of the definition of $\beta$, 
$$\varepsilon^2 \beta \ll 1,$$
permitting us to simplify the hydraulic potential gradient in the definition of the flux $q_x$ to a purely geometrical term,
\begin{equation} q_x =  -\kappa(N)\pd{(h_0 + r^{-1} b)}{x} \label{eq:x_flux_final}
\end{equation}
This still leaves us with mutliple possibilities: $\beta$ could still be large, small or $O(1)$. No further simplifications are however required since there is a single occurrence of $\beta$ in the model, and we can simply assume the distinguished limit $\beta = O(1)$ to cover all three possibilities.

At this point, we are also in a position to be more definite in our assumptions about $\kappa$ and the temperate friction law $f_m(N,u)$ (both being in dimensionless form). By assuming that $[N] \ll \rho g [z]$, we are effectively able to allow infinite effective pressures $N \rightarrow \infty $: in reality, there is an upper bound of $N_{max} \sim  \rho g h$ that corresponds to vanishing water pressure. In scaled terms, that upper bound corresponds to $N \sim \beta^{-1}\varepsilon^{-2} \gg 1$. 

We suppose that the permeability function $\kappa$ and water storage function $e$ becomes very small for such large $N$; in fact, as described in the main paper, satisfying the boundary equation \eqref{eq:no_flow_basic} with the reduced flux \eqref{eq:x_flux_final} requires that $\kappa$ can become zero. Again as described in the main paper, we will assume that there is an invertible transformation from $N$ to an auxiliary variable $\Pi$,
\begin{subequations} \label{eq:hydrology_transform}
\begin{equation} N = \mathcal{N}(\Pi) \end{equation}
such that 
\begin{equation} -\kappa(\mathcal{N}(\Pi)) \od{\mathcal{N}}{\Pi} = \kappa_2 = \mbox{constant} > 0, \end{equation}
while 
\begin{equation} \tilde{\kappa}(\Pi) = \kappa(\mathcal{N}(\Pi))\end{equation}
is such that
\begin{equation} \od{\tilde{\kappa}}{\Pi} > 0, \end{equation}
and
\begin{equation} \mathcal{N}(\Pi) \rightarrow \infty, \qquad \tilde{\kappa}(\Pi) \rightarrow 0 \end{equation}
\end{subequations}
as $\Pi \rightarrow 0$. Note that the constraint $N \rightarrow \infty$ as $\Pi \rightarrow 0$ implies that $\tilde{\kappa}(\Pi)$ must go to zero sufficiently rapidly as $\Pi \rightarrow 0$. Linear convergence to zero suffices, and is generally computationally advisable since it ensures a non-singular Jacobian when $\Pi = 0$ somewhere in the domain.

 Two simple examples that allow for such a transformation are a power law $\kappa(N) = N^{-k_0}$ with  $k_0 > 1$ and an exponential law $\kappa(N) = \exp(-k_0N)$. For the power law we get
\begin{equation} \label{eq:sample_transformation_final} \mathcal{N}(\Pi) = \Pi^{1/(k_0-1)}, \qquad \kappa_2 = \frac{1}{k_0-1}, \qquad \tilde{\kappa}(\Pi) = \Pi^{k_0/(k_0-1)}, \end{equation}
which leaves $\tilde{\kappa}$ with faster-than-linear convergence to zero as $\Pi \rightarrow 0$. Alternatively, $\kappa(N) = \exp(-kN)$ leads to $\Pi = k_0^{-1}\exp(-kN)$ and therefore
$$ \mathcal{N}(\Pi) = -k_0^{-1}\log(k\Pi), \qquad \kappa_2 = 1, \qquad \tilde{\kappa}(\Pi) = k_0\Pi, $$
where $\tilde{\kappa}$ is linear.

In practice, we simply replace the constitutive relations in our hydrology model generically by
$$ q_x = -\tilde{\kappa}(\Pi)\pd{(h_0+r^{-1}b)}{x}, \qquad q_y = -\tilde{\kappa}(\Pi)\pd{}{y}\left.\left(p_1-2\pd{w}{z}\right)\right|_{z = 0} - \beta \kappa_2 \pd{\Pi}{y}, $$
$$ e = \tilde{\Phi}(\Pi) \qquad \mbox{for } T = 0, $$
with $\tilde{\kappa}$ and $\kappa_2$ and the mapping $\mathcal{N}(\Pi)$ (which is necessary to express the basal friction law in terms of $\Pi$) satisfying the constraints in \eqref{eq:hydrology_transform}.

As described in the main paper and in section \ref{sec:model_basic}above, we limit ourselves to cases where the geometrical hydraulic gradient $-\pdl{(h_0+r^{-1}b)}{x}$ remains positive, in which case \eqref{eq:bed_enthalpy} is a diffusion problem in $\Pi$ with $x$ as the time-like variable. In order to prevent the problem becoming one of backward diffusion, we require not only  $-\pdl{(h_0+r^{-1}b)}{x} > 0$ but also that $\pdl{\tilde{\kappa}}{\Pi} > 0$, which implies that $\kappa(N)$ decreases with $N$. A notable exception to such a constitutive relation is provided by channel-like drainage conduits \citep[e.g.][]{Schoof2010,Werderetal2013}, which (somewhat unsurprisingly, since they lead to collapse of the drainage system onto single drainage conduits) correspond to backward diffusion. We exclude such drainage systems from the model for that reason.

In-built here is the expectation that $N \rightarrow \infty$, $\Pi \rightarrow 0$ in locations where the bed first becomes temperate, and the friction law can be expressed in terms of $\Pi$ through the substitution $N = \mathcal{N}(\Pi)$. We assume that the friction law is continuous across such cold-temperate boundaries, and consequently require that  $\lim_{T \rightarrow 0^-} f(T,N,u) = \lim_{N \rightarrow \infty} f_m(N,u)$. The main paper elaborates on several examples of friction laws modified from the literature that satisfy this requirement.

We have almost completed the derivation of the model used in the main paper. All that is required to make the two agree is to recognize that the pressure variable $p$ used in the main paper is equal to the reduced pressure $p_1'$ defined through
$$ p_1' = p_1 - \pd{(h_0+b)}{x}\pd{b}{x}(h_0-z), $$
to drop the subscript $_0$ on $h_0$, to replace the variable name $h_1$ by $s_1$, and to recognize that, if we assume the domain to be periodic in $y$ with period $W$, integration of \eqref{eq:Stokes_final2} in conjunction with \eqref{eq:kinematic_final} and \eqref{eq:nopenetration_scaled} gives
\begin{equation} \pd{h_0}{t} + \pd{Q}{x} = \bar{a} \label{eq:masscons_diff}\end{equation}
where
\begin{equation} Q = W^{-1}\int_0^W \int_0^h u \rd z \rd x, \qquad \bar{a} = W^{-1}\int_0^W a \rd y. \label{eq:masscons_int} \end{equation}
Note that the correction term $ -\pdl{(h_0+b)}{x}\pdl{b}{x}(h_0-z)$ is independent of $y$ and vanishes at $z = h_0$, and therefore leaves $h_1$ and $q_y$ unaltered on replacing $p_1$ by $p_1'$.

\section{A note on the cold-temperate boundary}

The local behaviour of the temperature field can be captured by a minor modification of the local solution section S2 of the supplementary material to \citet{Haseloffetal2018}, accounting for the fact that there is non-zero friction and dissipation on the temperate side of the transition. If, as in the main paper here and in \citet{Haseloffetal2018}, we consider the cold region lying to the left of the transition point and we position the origin of a polar coordinate system $(r,\vartheta)$ in the $(y,z)$-plane at the transition $y = y_m(x,t)$ (so $y = y_m(x,t) + r\cos(\vartheta)$, $z = r \sin(\vartheta)$), then the temperature field to an error of $O(r^2\log(r)^2)$ is the same as in \citet{Haseloffetal2018},
\begin{equation} T(r,\vartheta) \sim a_0 r^{1/2} \sin\left(\frac{\vartheta}{2}\right) + a_1 r^{3/2} \sin\left(\frac{3\vartheta}{2} \right) + b_1 r \sin(\vartheta) - \left\{ \begin{array}{l l} \alpha \gamma_0 \bar{u}^2 r \sin(\vartheta)  & 0 < \vartheta < \pi \\ 0, & \pi < \vartheta < 2 \pi. \end{array} \right.  \label{eq:temp_margin} \end{equation}
Here $\bar{u}$ is sliding velocity at the transition, and we have made use of the fact that the temperature at the transition is the melting point $T = 0$, and hence the dissipation rate on the cold side of the transition is $\gamma_0 \bar{u}^2$.

\begin{figure}
 \centering
 \includegraphics[width=0.7\textwidth]{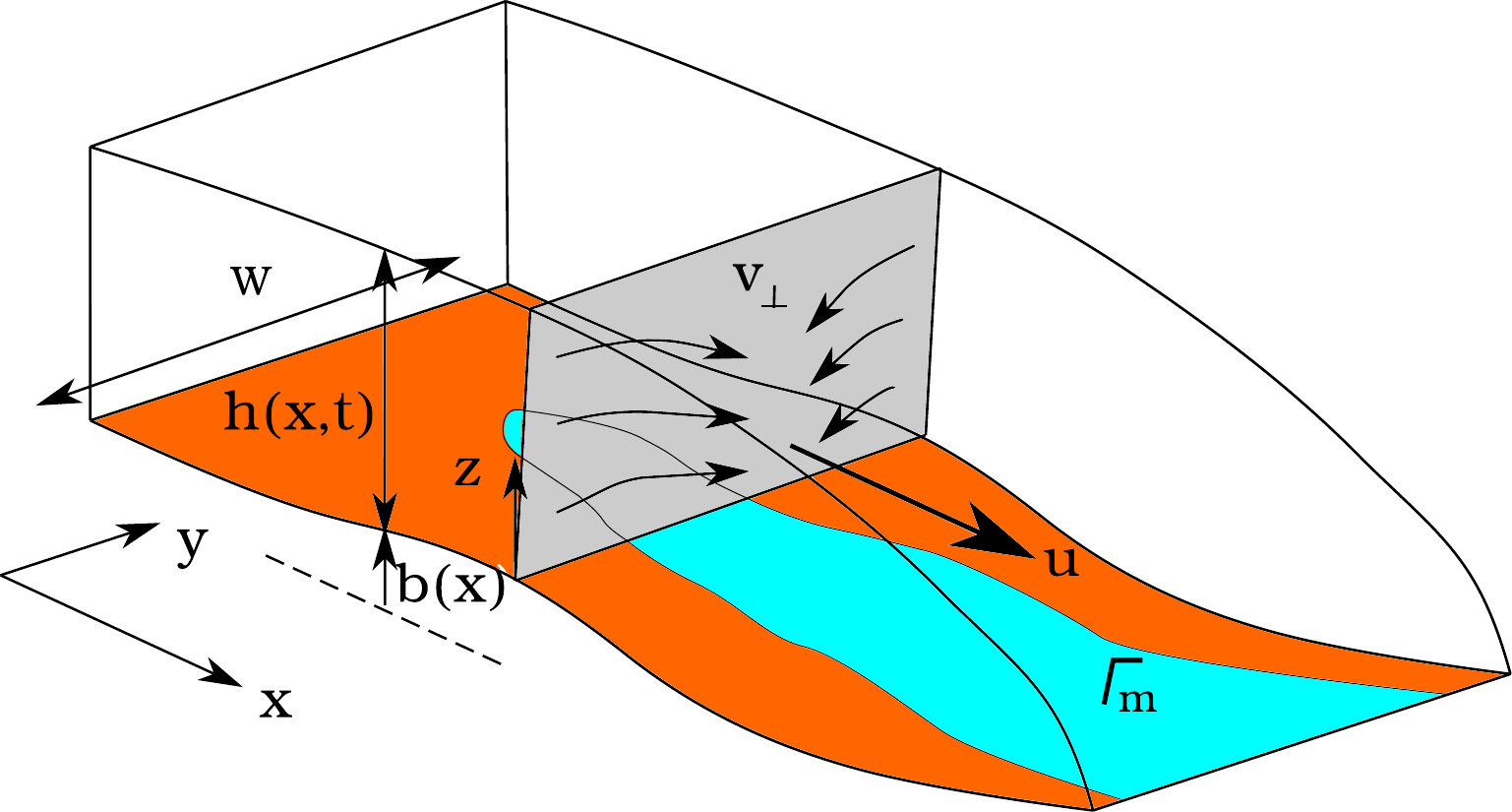}
 \caption{Geometry of the leading order problem, with the notation used in the main paper . Note that the grey plane (tangent to the secondary velocity field $\bm{v}_\perp$) is locally perpendicular to the bed, and not exactly parallel to the $z$-axis. This ensures that $w = 0$ at the bed. The temperate portion of the bed, where $T = 0$ is shown in light blue, the cold portion in brown.}
\end{figure}

With this in mind, we have basal temperature on the cold side $\vartheta = \pi$ as
$$ T(r,\pi) \sim a_0 r^{1/2} - a_1 r^{3/2}, $$
while net heat flux out of the bed on the warm side $\vartheta = 0$ is
$$ \left[-\pd{T}{z}\right]_-^+ = \left. - \frac{1}{r}\od{T}{\theta}\right|_{2 \pi}^0 \sim - a_0 r^{-1/2} - 3 a_1 r^{1/2} + \gamma_0 \bar{u}^2 $$
As in \citet{Haseloffetal2018} and \citet{Schoof2012}, we see immediately that we must have $a_0 \leq 0$ to ensure a non-positive temperature $T(r,\pi)$ on the cold side of the transition, $\vartheta = \pi$, and if $a_0 = 0$, then $a_1 \geq 0$. 

The net heat flux on the warm side now appears in a conservation equation for bed enthalpy. If the transition is migrating at a speed $V = \rd y_m/\rd t < 0$, then locally, using $y' = y- y_m(x,t) = r$,
$$ -V \pd{e}{y'} + \pd{q_y}{y'} -  \pd{y_m}{x}\pd{q_x}{y'} - a_0 y'^{-1/2} - 3 a_1 y'^{1/2} + \alpha \gamma_0 \bar{u}^2  \sim \alpha f(0,N,\bar{u}) \bar{u}. $$
The assumption that $\gamma_0 \bar{u} = \lim_{N\rightarrow \infty} f(0,N\bar{u})$, combined with friction being a non-decreasing function of $N$, implies that $ f(0,N,\bar{u}) \bar{u} < \gamma_0\bar{u}^2$. The weak form of the conservation equation
$$ \pd{e}{t} + \pd{q_y}{y} + \pd{q_x}{x} + \left[ -\pd{T}{z}\right]_-^+ = \alpha f(0,N,u)u $$
meanwhile requires that, at the transition itself (since bed enthalpy $e$ and bed enthalpy fluxes $q_x$ and $q_y$ vanish for $y' < 0$)
\begin{equation} \lim_{y' \rightarrow 0^+} \left( V e- q_y + \pd{y_m}{x} q_x \right) = 0, \label{eq:Rankine_Hugoniot_supp}  \end{equation}
as stated in the main paper. Note that  $q_y - (\pdl{y_m}{x}) q_x$  is the enthalpy flux \emph{perpendicular} to the transition \emph{curve} $y = y_m(t)$.

It is then formally possible to have a singular heat flux near the transition, with $a_0 < 0$, provided we put
$$  -V e + q_y -  \pd{y_m}{x} q_x \sim 2 a_0 y'^{1/2} + 2 a_1 y'^{3/2} + f(0,N\bar{u})\bar{u} - \alpha \gamma_0 \bar{u}^2 , $$
regardless of the sign of $V$. 

Suppose however that $V < 0$, with the transition advancing into the cold portion of the bed. Since $e > 0$ and $a_0 \leq 0$, a singular heat flux in the margins with $a_0 \neq 0$ then implies that the transition-normal enthalpy flux $q_y - (\pdl{y_m}{x})q_x$ must account for the singularity,
$$ q_y -\pd{y_m}{x} q_x \sim 2 a_0 y'^{1/2}, $$
requiring an infinite lateral hydraulic gradient. Moreover, this situation requires \emph{removing} enthalpy at a locally infinite rate from the temperate side of the transition to warm the cold side to the melting point and allow the transition to migrate. In turn that rate of enthalpy removal implies a locally infinite freezing rate $-2a_0 y'^{1/2}$ on the warm side.

Although that singular freezing rate can mathematically be balanced by a singular divergence of the lateral enthalpy transport rate $q_y - (\pdl{x_m}{y'})q_x$, the model in the main paper assumes that no such singular freezing rate can occur while the cold-temperate transition is migrating into the cold region, encapsulated in the statement that
$$ \lim_{y' \rightarrow 0^+} \left(y'^{1/2}\left[\pd{T}{z}\right]_-^+\right) \geq 0 \qquad \mbox{ if } V < 0; $$
in the notation used here, this statement simply amounts to putting $a_0 = 0$. Note that, as in \citet{Haseloffetal2018}, we can make the freezing rate non-singular, but even with $a_0 = 0$, there remains a non-singular freezing rate $\alpha (\gamma_0 \bar{u}^2 - f(0,N,\bar{u})\bar{u}) > 0$.

We hold the view that a singular freezing rate is what would occur if water were forced into the cold portion of the bed without that bed having first been warmed by dissipation; it is effectively what we would expect to see in situations like the injection of magma into a dyke \citep{ListerKerr1991}, which however requires hydrofracture and overpressurization of the intruding fluid, so we reject that possibility here.

The picture for a cold-temperate transition migrating into the temperate bed region is different; here $V > 0$ and the lateral enthalpy gradient $e$ can absorb the singular freezing rate, and the balance $e \sim -2 V^{-1} a_0 y'^{1/2} > 0$ is possible with non-zero $a_0 < 0$, see also the appendix to \citet{Schoof2012}. As discussed in the main paper, this must be supplemented with the requirement that bed water content $e = \Phi(N)$ go to zero at an inward-migrating margin, and with the form of constitutive relations we have assumed here (where $\Phi \rightarrow 0$ as $N \rightarrow \infty$), this translates into the condition
\begin{equation} \Pi = 0 \label{eq:margin_freeze_supp} \end{equation}
at $y = y_m$.

\section{Method of solution}

Here we set out the method by which we compute  steady state solution of the leading order model. Our first step is to make sense of the compressible Stokes flow problem \eqref{eq:Stokes_final2} and \eqref{eq:Stokes_final1b}--\eqref{eq:Stokes_final1c}.

\subsection{The compressible Stokes flow problem}

As in an incompressible Stokes flow problem, we can reduce the Stokes flow problem to the biharmonic equation. Reverting to the notation in the main paper, we have
\begin{equation} \label{eq:Stokes_transverse} \nabla_\perp^2 \bm{v}_\perp - \nabla_\perp p = \bm{0}, \qquad \nabla_\perp \cdot \bm{v}_\perp = - \pd{u}{x}, \end{equation}
We write $\bm{v}_\perp$ in terms of a Helmholtz decomposition as the sum of a solenoidal and an irrotational part,
\begin{equation} \label{eq:Helmholtz} \bm{v}_\perp = \nabla_\perp \times \Psi + \nabla \phi \end{equation} 
where, in the usual way, the two-dimesional curl is to be understood in the sense of $\Psi$ being a pseudovector perpendicular to the $yz$-plane,
$$ \nabla_\perp \times \Psi = \left(\pd{\Psi}{z},-\pd{\Psi}{y}\right) $$

Substitution into \eqref{eq:Stokes_transverse}$_2$ gives
\begin{equation} \nabla_\perp^2 \phi = -\pd{u}{x} \label{eq:Laplace_phi} \end{equation}
while substitution into \eqref{eq:Stokes_transverse}$_1$ and taking the curl of the result gives
\begin{equation} \nabla_\perp^4 \Psi = 0, \end{equation}
which we will write in mixed form
\begin{subequations} \label{eq:biharmonic}
\begin{align}
 \nabla_\perp^2 \omega = &  0, \\
 \nabla_\perp^2 \Psi = & \omega,
\end{align}
\end{subequations}
where $\omega = \nabla_\perp \times \bm{v}_\perp = \pdl{v}{z} - \pdl{w}{y}$ is vorticity. This leaves the boundary conditions on $\Psi$ and $\phi$ to be determined.

We can in fact choose the boundary conditions on $\phi$ at will provided the resulting problem has a solution, and pick the following:
\begin{subequations} \label{eq:bc_phi}
\begin{align} 
\pd{\phi}{z} = & u \pd{h}{x} - \pd{Q}{x} - (a  - \bar{a}) & \mbox{at } z = h \label{eq:surface_gauge} \\
\pd{\phi}{z} = & 0 & \mbox{at } z = 0
\end{align}
\end{subequations}
while also requiring $\phi$ and $\pdl{\phi}{y}$ to be continuous at the lateral boundaries $y = 0,\, W$. This choice imposes Neumann conditions at the bottom and top of the rectangular domain and periodic conditions on flux at the edges of the domain. Before proceeding, we therefore have to confirm that a solvability condition on the Poisson equation \eqref{eq:Laplace_phi} is satisfied by our choice of boundary conditions. This is the objective of the next paragraph.

\begin{subequations}
The relevant solvability condition for \eqref{eq:Laplace_phi}, obtained by integrating the equation over the rectangle $0<y<W$, $0<z<h$ and applying the divergence theorem, is that
\begin{equation} \int_0^W \left.\pd{\phi}{z}\right|_{z=h} \rd y  = \int_0^W \int_b^s -\pd{u}{x} \rd z \rd y \label{eq:solvability} \end{equation}
Note that, using the definition of $Q$,
\begin{align*} \pd{Q}{x} =&  W^{-1} \pd{}{x} \int_0^W \int_0^h u \rd z \rd y \\
 = & W^{-1} \int_0^W  u|_{z=h} \pd{h}{x}  \rd y +   W^{-1}  \int_0^W \int_0^h \pd{u}{x} \rd z \rd y.
\end{align*}
Note also that $\pdl{Q}{x}$ is independent of $y$, so $\int_0^W \pdl{Q}{x} \rd x = W \pdl{Q}{x}$. Hence, substituting the boundary condition \eqref{eq:surface_gauge} in the first term on the left-hand side of \eqref{eq:solvability} gives
\begin{equation}  \int_0^W \left.\pd{\phi}{z}\right|_{z=h} \rd y =  \int_0^W u|_{z=h} \pd{h}{x} \rd y -  \int_0^W  u|_{z=h} \pd{h}{x}  \rd y -  \int_0^W \int_0^h \pd{u}{x} \rd z \rd y,
\end{equation}
bearing in mind that $\int_0^W a - \bar{a} \rd y = 0$. As a result, \eqref{eq:solvability} indeed holds.
\end{subequations}

With this choice of gauge for $\phi$, the boundary conditions for $\Psi$ at the surface $z = h$ become
\begin{subequations} \label{eq:bc_Psi}
\begin{align} \pd{\Psi}{y} = & 0 \label{eq:Psi_top} \\
\omega = &  \pd{u}{y}\pd{h}{x}  \label{eq:omega_top}
\end{align}
while at the base of the ice $z  = 0$, the equivalent conditions are 
\begin{align} \pd{\Psi}{y} = & 0 \label{eq:Psi_bottom} \\
\omega =& \frac{f\left(T,N,|u|\right)}{|u|}\left(\pd{\Psi}{z} + \pd{\phi}{y}\right) \label{eq:omega_bottom}
\end{align}
\end{subequations}
The above uses the fact that $\pdl{^2\Psi}{y^2} = 0$ on both surface as a result of the boundary condition $\pdl{\Psi}{y} = 0$, and hence 
\begin{align} \pd{v}{z} + \pd{w}{y} = & \pd{^2\Psi}{z^2} - \pd{^2\Psi}{y^2} + 2\pd{^2\phi}{y\partial z} \\
= &  \pd{^2\Psi}{z^2} + \pd{^2\Psi}{y^2} + 2\pd{u}{y}\pd{s}{x} \\
= & \omega +  2\pd{u}{y}\pd{h}{x}.
\end{align}

Here we run into the need for an additional constraint: the conditions \eqref{eq:Psi_top} and \eqref{eq:Psi_bottom} in fact assert that $\Psi$ is constant along the top and bottom boundaries, but the value of those constants is not defined. Like $\phi$, $\Psi$ is a gauge variable and defined only up to the addition of an arbitrary constant. Hence we can require
\begin{equation} \Psi = 0 \end{equation}
on $z = h$ in lieu of \eqref{eq:Psi_top}. We are then not at liberty to demand the same at $z = 0$. Instead, we can write
\begin{equation} \Psi = \Psi_b \end{equation}
on $z = 0$ where $\Psi_b$ is a constant that must be constrained implicitly by a further equation, derived as follows. Take the $y$-component of \eqref{eq:Stokes_transverse}$_1$ with \eqref{eq:Helmholtz},
\begin{equation} \label{eq:Stokes_Helmholtz_x} \nabla_\perp^2 \left(\pd{\Psi}{z} + \pd{\phi}{y}\right) - \pd{p}{y} = 0 \end{equation}
Define a lateral average $\bar{\Psi} = W^{-1}\int_0^W \Psi \rd y$. Averaging both sides of \eqref{eq:Stokes_Helmholtz_x} and using the periodicity of the problem with no applied transverse pressure gradient gives simply
\begin{equation} \pd{^3 \bar{\Psi}}{z^3} = 0. \end{equation}
so $\bar{\Psi} = a + bz + cz^2$. Similarly, averaging \eqref{eq:omega_top} yields
$$ \pd{^2\bar{\Psi}}{z^2} = 0 $$
at $z = h$, or $c = 0$ in the formula for $\bar{\Psi}$. In turn, averaging \eqref{eq:omega_bottom} therefore yields the constraint
$$ W^{-1}\int_0^W f\left(T,N,\pd{\Psi}{z} + \pd{\phi}{y}\right) \rd y =  \pd{^2\bar{\Psi}}{z^2} = W^{-1} \int_0^W \omega \rd y = 0, $$
which is sufficient in principle to determine $\Psi_b$. In computational practice, we impose that $\pdl{\Psi}{y} = 0$ along $z = 0$ and require that the integral over $\omega$ along the bed vanishes. Physically, this constraint is nothing more complicated than that the average of the transverse shear stress at the bed is zero, because there is no applied lateral pressure gradient.

In order to solve the hydrology portion of the model, we also need to solve for the pressure variable $p$ from \eqref{eq:Stokes_transverse}. Numerically, it is advantageous to re-write this in the following form: take the divergence of \eqref{eq:Stokes_transverse}$_1$ and use \eqref{eq:Stokes_transverse}$_2$,
$$ \nabla_\perp^2 p = - \nabla_\perp^2 \pd{u}{x} $$ 
or, on defining
\begin{subequations} \label{eq:Laplace_p}
\begin{equation} p = -\pd{u}{x} + p', \end{equation}
we get
\begin{equation} \nabla_\perp^2 p' = 0. \end{equation}
Neumann boundary conditions at the upper and lower surfaces $z = b,\,s$ can be derived once the vorticity $\omega$ is known, using the $z$-component of \eqref{eq:Stokes_transverse}$_1$ and the definition of $p'$,
$$ \pd{p'}{z}  =  \nabla_\perp^2 w + \pd{^2 u}{x\partial z} = -\pd{\omega}{y} + \nabla_\perp^2 \pd{\phi}{z} + \pd{^2 u}{x\partial z} $$
\begin{equation} 
\pd{p'}{z}  =  -\pd{\omega}{y}
\end{equation}
\end{subequations}
on account of $\nabla_\perp^2 \phi = -\pdl{u}{x}$, with periodic boundary conditions on $p'$ at the lateral boundaries of the domain.

Again, we need to confirm that these Neumann conditions satisfy a solvability condition, which requires that
\begin{equation}\int_0^W -\left.\pd{\omega}{y}\right|_{z = h}  \rd y + \int_0^W \left. \pd{\omega}{y}\right|_{z=0}  \rd y = 0 \label{eq:solvability_p} \end{equation}
This naturally vanishes as, by lateral periodicity, we can turn the sum of these integrals into $\oint_C \pdl{\omega}{l} \rd l = 0$, where $C$ is the boundary of the rectangular domain $0 < y <W$, $0 < z < h$ and $l$ is an arc length coordinate.

In summary, the compressible Stokes flow problem can be reduced to \eqref{eq:Stokes_final1a} with \eqref{eq:bed_friction_final} and \eqref{eq:sliding_scaled}, \eqref{eq:Laplace_phi} with boundary conditions \eqref{eq:bc_phi}, \eqref{eq:biharmonic} with boundary conditions \eqref{eq:bc_Psi} and \eqref{eq:Laplace_p}, all of them with periodic boundary conditions in $u$, $\phi$, $\Psi$ and $p$.

\subsection{Numerical computation of a steady state solution} \label{sec:numerics}

On suppressing time dependence in the problem (which only appears in the form of time derivatives in \eqref{eq:heat_scaled1}, \eqref{eq:heat_scaled2}, \eqref{eq:basal_energy_scaled} \eqref{eq:kinematic_final} and \eqref{eq:masscons_int}), the downstream coordinate effectively takes on the role of time-like variable in \eqref{eq:heat_scaled1} and \eqref{eq:basal_energy_scaled} (with increasing $x$ being the forward time-like direction), while the steady-state version of \eqref{eq:masscons_int} also turns out to allow computation of $h$ by integrating from the ice divide, just as in the two-dimensional shallow ice counterpart of our model in \citet{Mantellietal2019}. The one thing that is harder here is that the velocity field is three- rather than two-dimensional, and needs to be computed in a somewhat convoluted way from the compressible Stokes flow problem analyzed in the previous section, rather than from an analytical formula as in the case of the shallow ice problem. The second, lesser complication arises from actually computing a varying bed temperature $T$ at $z = b$, rather than setting this to the melting point and using a subtemperate sliding formulation. Obviously, the increased dimensionality of the problem increases the computational demand of each step in the computation as well.

Our next step will be to semi-discretize in $x$. Before we do so, we introduce the stretched vertical coordinate
$$ \zeta = \frac{z}{h} $$
in order to map the ice domain onto a cuboid in $xy\zeta$; no stretching of the $x$-coordinate is needed since we do not consider the dynamic situation of an evolving ice margin, and there are no sharp boundaries between cold, subtemperate and temperate. Also note that we will retain the original variable $z$ to describe the temperature field in the bed. Letting $X = x$, we obtain (keeping $y$ unchanged for simplicity, though we should properly speaking put $Y = y$)
$$ \pd{}{x} = \pd{}{X} - \frac{\zeta}{h} \pd{h}{x}\pd{}{\zeta}, \qquad \pd{}{z} = \frac{1}{h}\pd{}{\zeta} $$

In terms of these, the steady-state version of the heat equation \eqref{eq:heat_scaled1} becomes, in divergence form,
\begin{equation} \Pe \left( \pd{(hvT)}{y} + \pd{ (w_{eff} T)}{\zeta} + \pd{(uhT)}{X}\right) + \left[\pd{}{y}\left(h\pd{T}{y}\right) + \pd{}{\zeta}\left(\frac{1}{h}\pd{T}{\zeta}\right)\right]  = \alpha h \left[\left(\pd{u}{y}\right)^2 + \frac{1}{h^2}\left(\pd{u}{\zeta}\right)^2 \right] \label{eq:heat_stretched} \end{equation}
where     
\begin{align} w_{eff} = & w - \zeta u \pd{h}{X} = \frac{1}{h}\pd{\phi}{\zeta} - \zeta u \pd{h}{X} - \pd{\Psi}{y}, \\
v = & \pd{\phi}{y} + \frac{1}{h}\pd{\Psi}{\zeta} \end{align}
The local mass conservation problem likewise changes: \eqref{eq:Laplace_phi} becomes in divergence form
\begin{equation} \pd{}{y}\left(h\pd{\phi}{y}\right) + \pd{}{\zeta}\left(\frac{1}{h}\pd{\phi}{\zeta}-\zeta u\pd{h}{X}\right) = - \pd{(uh)}{X} \end{equation}
while \eqref{eq:surface_gauge} becomes the corresponding vertical flux boundary condition
\begin{equation} \frac{1}{h}\pd{\phi}{\zeta} - \zeta u \pd{h}{X} = - \pd{Q}{X} = - a \end{equation}
at $\zeta = 1$, where the last equality results from being in steady state (an equivalent reformulation in divergence form is also possible if we dispense with the assumption of a steady state, in which case the coordinate transformation above also involves the time variable).

All other equations are either posed with $x = X$ as the only independent variable, such as \eqref{eq:masscons_diff} in steady state, or are posed on the transverse plane and $x$ plays no role, so a coordinate stretching is trivial or not required.

Below, we semi-discretize the steady state model in $x$ using a simple upwinding scheme, and an operator splitting to determine which parts of the bed are temperate and cold for the purposes of solving the hydrology portion of the model \citep[see also][]{SchoofHewitt2016}. Let subscripts $i$ and $i+1/2$ denote cell centres and cell boundaries in the $x$-direction, located at $x_i$ and $x_{i+1/2}$. Define the following notation
\begin{align} \nabla_{\perp,i}^2 = & \pd{^2}{y^2} + \frac{1}{h_i^2} \pd{^2}{\zeta^2}, \\ \nabla_{\perp,i+1/2}^2 = & \pd{^2}{y^2} + \frac{1}{h_{i+1/2}^2} \pd{^2}{\zeta^2},
\end{align}
Then
\begin{align} 
h_{i+1/2} - h_{i} - \frac{\left(h_{i+1} - h_{i}\right)(x_{i+1/2} -x_i) }{x_{i+1}-x_{i}} = & 0\\
\frac{Q_{i+1/2}-Q_{i-1/2}}{x_{i+1/2}-x_{i-1/2}} - W^{-1} \int_0^W a_i \rd y = & 0\\
 Q_{i+1/2} - \frac{h_{i+1/2}}{W}\int_0^W \int_0^{1} u_{i+1/2} \rd y \rd \zeta = & 0 \\
 \nabla_{\perp,i+1/2}^2 u_{i+1/2} \frac{h_{i+1}+b_{i+1}-h_i-b_i}{x_{i+1}-x_i} = & 0 & \mbox{on }0 < \zeta < 1 \\
 \frac{1}{h_{i+1/2}}\pd{u_{i+1/2}}{\zeta} = & 0 & \mbox{on } \zeta =  1 \\
 \frac{1}{h_{i+1/2}}\pd{u_{i+1/2}}{\zeta} - \gamma(T_i)u_{i+1/2} = & 0 & \mbox{on } \zeta = 0, \end{align}
 \begin{align}
 \nabla_{\perp,i}^2 \phi_i - \frac{1}{h_i}\pd{}{\zeta} \left[\frac{\zeta}{2}\left(u_{i+1/2}\frac{h_{i+1}-h_i}{x_{i+1}-x_i}+u_{i-1/2}\frac{h_i-h_{i-1}}{x_i-x_{i-1}}\right)\right]   +\frac{h_{i+1/2}u_{i+1/2}-h_{i-1/2}u_{i-1/2}}{h_i(x_{i+1/2}-x_{i-1/2})}  = & 0 \nonumber \\ \mbox{on } 0 < \zeta <  & 1 \label{eq:phi_Laplace_semidiscrete} \end{align}
 \begin{align}
 \nabla_{\perp.i}^2 \Psi_i -  \omega_i = & 0  & \mbox{on } 0 < \zeta < 1\\
  \nabla_{\perp.i}^2 \omega_i =&  0   & \mbox{on } 0 < \zeta < 1\\
  \nabla_{\perp.i}^2 p_i' =&  0   & \mbox{on } 0 < \zeta < 1\\
 \frac{1}{h_i} \pd{\phi_i}{\zeta} - \frac{\zeta}{2}\left(u_{i+1/2}\frac{h_{i+1}-h_i}{x_{i+1}-x_i}+u_{i-1/2}\frac{h_i-h_{i-1}}{x_i-x_{i-1}}\right) + a_i = &  0 & \mbox{on } \zeta = 1 \\
  \frac{1}{h_i} \pd{\phi_i}{\zeta} = & 0 & \mbox{on } \zeta = 0 \\
\Psi_i = & 0 & \mbox{on } \zeta = 1 \\
 \omega_i - \frac{1}{2}\left(\pd{u_{i-1/2}}{y} + \pd{u_{i+1/2}}{y}\right)\frac{h_{i+1}-h_i}{x_{i+1}-x_i} = & 0 & \mbox{on } \zeta = 1. \\
\Psi_i = & \Psi_b & \mbox{on } \zeta = 0 \\
 \omega_i -  \gamma(T_i)\left(\frac{1}{h_i}\pd{\Psi_i}{\zeta} + \pd{\phi_i}{y} \right) = & 0 & \mbox{on } \zeta = 0 \\
 \frac{1}{h_i} \pd{p_i'}{\zeta} + \pd{\omega_i}{y} = & 0 & \mbox{on } \zeta = 0 \mbox{ and } \zeta = 1   \end{align}
 \begin{align}
 \int_0^W \omega_i(y,0) \rd y = &  0 & \mbox{on } \zeta = 0 \\
 p_i - p_i' + \frac{u_{i+1/2}-u_{i-1/2}}{x_{i+1/2}-x_{i-1/2}} = & 0 & \mbox{on } 0 < \zeta < 1 \\
 v_i - \frac{1}{h_i}\pd{\Psi_i}{\zeta} - \pd{\phi_i}{y} = & 0 & \mbox{on } 0 < \zeta < 1   \\
 w_{eff,i} + \pd{\Psi_i}{y} - \frac{1}{h_i}\pd{\phi_i}{\zeta}  + \frac{\zeta}{2}\left(u_{i+1/2}\frac{h_{i+1}-h_i}{x_{i+1}-x_i}+u_{i-1/2}\frac{h_i-h_{i-1}}{x_i-x_{i-1}}\right) = & 0  & \mbox{on } 0 < \zeta < 1 
 \end{align}
 \begin{align}
 \Pe  \frac{1}{h_i} \left( \pd{(v_i h T_i)}{y} + \pd{(w_{eff,i} T_i)}{\zeta} + \frac{ u_{i+1/2}h_iT_i - u_{i-1/2}h_{i-1}T_{i-1}}{x_{i+1/2}-x_{i-1/2}}\right) - \nabla_{\perp,i}^2 T_i = & \nonumber \\
 \alpha \left\{\left[\pd{u_{i+1/2}}{y}\right]^2 + \frac{1}{h_i^2}\left[\pd{u_{i+1/2}}{\zeta}\right]^2\right\}  & & \mbox{on } 0 < \zeta < 1 \end{align}
 \begin{align}
 \nabla_{\perp}^2 T_i = & 0 & \mbox{on } - D < z < 0 \label{eq:heat_bed_discrete} \\
 T_i = & T_{s,i} & \mbox{on } \zeta = 1 \\
- \pd{T_i}{z} = & G & \mbox{on } z = -D \end{align}
\begin{align}
 I_{i-1}\left(\frac{q_{x,i+1/2}-q_{x,i-1/2}}{x_{i+1/2}-x_{i-1/2}} + \pd{q_{y,i}}{y}\right)  + \left[-\left.\frac{1}{h_i}\pd{T_i}{\zeta}\right|_{+} +  \left.\pd{T}{z}\right|_{-}\right] = & \nonumber \\
\alpha f(T_i,\mathcal{N}(\Pi), u_{i+1/2})u_{i+1/2}  & & \mbox{at } \zeta = z = 0 \label{eq:energy_conservation_bed_discrete} \end{align}
\begin{align}
 \left[T_i\right]_-^+ & = 0 & \mbox{on } \zeta = z = 0\\
 T_i = & 0 & \mbox{if } I_{i-1} > 0 \mbox{ on } \zeta = 0 \label{eq:temperature_bed_discrete} \\
 \Pi_i = &  0 & \mbox{if } I_{i-1} = 0 \mbox{ on } \zeta = 0 \label{eq:pressure_bed}  \\
 \sigma_{nn,i} - p_i - \frac{2}{h_i} \pd{^2\Psi}{y\partial \zeta} + \frac{2}{h_i^2} \pd{^2\phi_i}{\zeta^2} = & 0 & \mbox{at } \zeta = 0 \\
q_{x,i+1/2} + \tilde{\kappa}(\Pi_i)  \frac{h_{i+1} - h_i + r^{-1}(b_{i+1} - b_i)}{x_{i+1} - x_i} = & 0 \\
 q_{y,i} + \left[\tilde{\kappa}(\Pi_i)\pd{\sigma_{nn,i}}{y} +  \beta\kappa_2(\Pi_i) \pd{\Pi_i}{y}\right] = & 0 \label{eq:lateral_drain_discrete} \end{align}
 \begin{align}
 I_i =&  \left\{\begin{array}{l l} 1 & \mbox{if } T_i > 0 \\
 1 & \mbox{if } T_i = 0 \mbox{ and } \Pi_i > 0 \\
 0 & \mbox{otherwise} \end{array}\right. & \mbox{on } \zeta = 0,
\label{eq:indicator}
\end{align}
where the subscripts $|_-$ and $|_+$ indicate limits taken as $\zeta \rightarrow 0$ from above and $z \rightarrow 0$ from below, respectively. Note that we have also reduced the semi-infinite domain $z < 0$ for the heat equation \eqref{eq:heat_bed_discrete} with the finite strip $-D < z < 0$, where $D$ is large.
The perturbed surface elevation $s_{1,i}$ is computed diagnostically through
\begin{equation}
 s_{1,i} =  p_i + \frac{1}{h_i} \pd{^2\Psi}{y\partial \zeta} - \frac{1}{h_i^2} \pd{^2\phi_i}{\zeta^2}
\end{equation}
at $\zeta  = 1$.

The resulting set of differential equations in $y$, $\zeta$ and (for \eqref{eq:heat_bed_discrete}) $z$ are then discretized using finite volumes. 
The system above should be seen as a pseudo-time-step. The given ``initial'' values for the step are $T_i$, $h_i$ , $N_i$, as well as the upstream value of $h_{i-1}$ (which may need to be defined through a ghost node at the first step, see the ice divide solution in the main paper). The inflow velocity $u_{i-1/2}$, flux $Q_{i-1/2}$ and bed state indicator function $I_{i-1}$ are defined by these ``initial values'', with $u_{i-1/2}$ and $Q_{i-1/2}$ having been computed as part of the previous step and $I_{i-1}$ defined through \eqref{eq:indicator}. The primary dependent variables to solve for in the $i$th step defined by the system of equations above are $h_{i+1}$, $Q_{i+1/2}$,  $u_{i+1/2}$, $T_i$, $\phi_i$, $\omega_i$,  $\Psi_i$, $p_i'$ and $\Pi_i$. Note that the $a_i$, $T_{s,i}$ and $b_i$ are simply prescribed accumulation rates, surface temperatures and bed elevations, $D$ is a large value intended to emulate the infinite depth of the domain of the heat equation below the bed, while the $Q_{i+1/2}$, $q_{x,i+1/2}$ and $q_{y,i}$ fluxes, the averaged thickness $h_{i+1/2}$ and transverse velocities $v_i$ and $w_{eff,i}$ are auxiliary variables. By way of clarification (since the above retains a continuum formulation in $y$ and $\zeta$), note that the indicator function $I_{i-1}$ at any \emph{lateral} cell boundaries must be considered as zero if either of the adjacent cell boundaries remains cold (has a cell centre value of $I_{i-1} = 0$) so that there can be no fluxes into or out of cold cells. This is consistent with the description of the hydrology model in the main paper, and is essentially the same as the standard procedure of defining cell boundary diffusivities through harmonic averages of the diffusivities at the adjoining cell centres.

Note also that the enthalpy transport terms in \eqref{eq:energy_conservation_bed_discrete} employ a lagged version $I_{i-1}$ of the bed state indicator function $I_i$; this is the operator splitting advertised above for the determination of the temperate region in which flow of water is possible. An upshot of this procedure is that solutions at individual `time' steps $i$ may violate the physical constraints $T_i \leq 0$ and $\Pi_i \geq 0$; these violations are then fixed on the $i+1$th step by the cells in which this happens being set to temperate or subtemperate, respectively.

The use of the lagged indicator function is particularly relevant for the downstream flux difference   $I_{i-1}(q_{x,i+1/2}-q_{x,i-1/2})/(x_{i+1/2}-x_{i-1/2})$ in \eqref{eq:energy_conservation_bed_discrete}, where an `unlagged' version might be $(I_{i}q_{x,i+1/2}-I_{i-1} q_{x,i-1/2})/(x_{i+1/2}-x_{i-1/2})$, differing in the coefficient of the flux $q_{x,i+1/2}$ that flows downstream from the $i$th slice. The lagged indicator function $I_{i-1}$ will equal unity for a location upstream of which is an established portion of temperate bed ($T_{i-1} = 0$, $\Pi_{i-1} > 0$), in which case there can be a flux $I_{i-1}q_{x,i-1/2}$ into the present slice $i$ at that location. It will likewise equal unity for a location upstream of which a portion of bed has just become temperate. In the operator splitting approach adopted here, this implies $T_{i-1} > 0$, $\Pi_{i-1} = 0$, resulting in vanishing flux into the present location since $\tilde{\kappa}(\Pi_{i-1}) = 0$, and the flux `out' of that location (which, in the version of \eqref{eq:energy_conservation_bed_discrete} for index $i-1$ would be $I_{i-2}q_{x,i-1/2}$) also automatically vanishes. This indicates that the use of the lagged indicator function is unproblematic when going from cold to temperate, but it may seem more problematic when going from a temperate ($\Pi_{i-1} > 0$) to a cold ($\Pi_i < 0$) portion of the bed, since the `downstream' flux 
$$I_{i-1}q_{x,i+1/2} = - \tilde{\kappa}(\Pi_i) \frac{h_{i+1}-h_i+r^{-1}(b_{i+1}-b_i)}{x_{i+1}-x_i}$$ 
will be negative in that case (assuming the geometric hydraulic gradient remains positive as discussed further in section \ref{sec:implementation} below). In other words, enthalpy appears to be spuriously received from a downstream location that is, in fact, cold-based. That is however simply the price to pay for the operator splitting: the spurious removal of heat from downstream only occurs where the bed is freezing in any event, and the addition of enthalpy from downstream can do nothing to prevent freezing. In the next domain slice $i+1$, that removal does not make itself felt since $\Pi_i < 0$ implies that $I_i = 0$, the same location in the same slice is identified as frozen, and the upstream flux $I_iq_{x,i+1/2}$ for the $(i+1)$th slice vanishes (that is, by construction, it does not equal the negative downstream flux $I_{i-1}q_{x,i+1/2}$ out of the $i$th slice). 

In terms of replicating the expected continuum behaviour in the margins by means of the operator splitting, consider first the case of widening. In the semi-discretized model, this occurs, in the sense of water flux being able to reach a previously frozen portion of the bed, because part of the bed for which $I_{i-1} = 0$ was computed to have a basal temperature $T_i \geq 0$. The local behaviour of temperature in the margins then requires that the coefficient $a_0$ in \eqref{eq:temp_margin} be slightly positive; with a small step size $x_i-x_{i-1}$ that overshoot in the solution is minimized and the margin essentially migrates in such a way as to keep $a_0$ near zero. For an inward-migrating margin, the discretization is more transparent, simply emulating the weak form of \eqref{eq:Rankine_Hugoniot_supp} with \eqref{eq:margin_freeze_supp} implemented in lagged form.

Lastly, as described in the main paper, the forward integration developed here is started from an ice divide solution. That divide solution itself depends  on a single parameter, the ice thickness at the divide. In principle, that divide thickness must be solved for as part of the problem, so as to satisfy a boundary condition at a downstream ice margin. Different margin conditions are conceivable, though land based margins (at which $h \rightarrow 0$) are problematic since the model becomes singular in that limit. As in \citet{Mantellietal2019}, we envisage a marine-terminating ice sheet here. The ice sheet terminates at a location $x_g$ where $h(x_g) = -r^{-1}b(x_g)$, and simultaneously $Q(x_g) = Q_g(h(x_g))$ with $Q_g$ a function prescribed by a boundary layer treatment of the near-grounding line region \citep[e.g.][]{Schoof2012,Schoofetal2017}.

Starting with a given $h(0)$, both $Q(x)$ and $h(x)$ are uniquely defined, so the two constraints (on $h(x_g)$ and $Q(x_g)$) are sufficient to fix both, $h(0)$ and $x_g$, effectively as a shooting problem since the model requires forward integration in $x$. Each forward integration is however costly, and a simpler construction is to fix $h(0)$ and consider a one-parameter family of beds $b(x) + b_0$. Since the model we have formulated only includes $b$ (as opposed to $\pdl{b}{x}$) in the boundary condition at the margin, the solution for $h$ is the same regardless of the value of $b_0$, and we can fix $b_0$ such that $-r^{-1}(b(x_g)+b_0) = h(x_g)$ at the location where $Q(x_g) = Q_g(h(x_g))$ instead of solving for $h(0)$ for fixed $b_0$.

\subsection{Some notes on implementation} \label{sec:implementation}

 Note that, for the computations reported here and in the main paper, we have omitted the first term in square brackets in \eqref{eq:lateral_drain_discrete} because the discretization scheme we chose in $(y,\zeta)$ was not of sufficiently high order to permit the computation of second derivatives $\sigma_{nn}$, as is required by \eqref{eq:energy_conservation_bed_discrete} if the normal stress gradient is retained in the computation of $q_y$. This difficulty became apparent after implementation, and will need to be addressed by future work. Note that inclusion of $\sigma_{nn}$ need not fundamentally alter the nature  of the steady state problem as an initial value problem, stepping forward in $i$, but there are parameter combinations in which the linearized theory in section \ref{sec:hydraulic_supplementary} suggests the problem could become pathological, with a resonance-like effect and a (nearly) singular Jacobian, depending on the precise value of $W$ chosen.

There are some other basic issues worth pointing out immediately. The discretization scheme above is chosen to avoid some relatively easy-to-identify numerical stability issues that can arise: one simple one is the dissipation term in \eqref{eq:energy_conservation_bed_discrete}.  For a fully subtemperate bed, it is relatively straightforward to show that a lagged or averaged choice of this dissipation term can make the semi-discretized scheme unstable, for instance if we replaced the dissipation rate $\alpha f(T_i,\mathcal{N}(\Pi_i),u_{i+1/2}) u_{i+1/2}$ by $\alpha f(T_{i-1},\mathcal{N}(\Pi_{i-1}),u_{i-1/2}) u_{i-1/2}+ \alpha f(T_i,\mathcal{N}(\Pi_i),u_{i+1/2}) u_{i+1/2}]/2$, which may superficially seem appropriate as it averages the dissipation rate due to sliding just upstream and just downstream of the $i$th slice in which temperature is computed (since the velocity is computed upstream and downstream of that slice).

Also, mote that the Laplace-type problems for $\phi$ and $p'$ are subject to Neumann boundary conditions, and have been shown to satisfy the relevant solvability conditions. This of course still implies that their solutions are unique only up to an additive constant; numerically, their value needs to be fixed at a single point on the boundary (preferably at the ice surface $\zeta = 1$) to alleviate this non-uniqueness. In addition, the solvability condition \eqref{eq:solvability} for $\phi$ holds identically for the continuous problem, but this is not the case for all conceivable spatial discretizations in $y$ and $\zeta$. The $\zeta$-dependent part of the source term in \eqref{eq:phi_Laplace_semidiscrete} is deliberately formulated in divergence form to allow the discrete solvability condition to be satisfied, but the numerical quadratures involved in discretization must be chosen so as to satisfy equivalent discrete solvability conditions to the one derived for the continuous case.

In general, \eqref{eq:energy_conservation_bed_discrete} should be seen as a time step in a diffusive problem for $\Pi$, since the lateral flux $q_{y,i}$ can be regarded as a diffusion term (with $\kappa_2 > 0 $) and the divided difference $(q_{x,i+1/2}-q_{x,i-1/2})/(x_{i+1/2}-x_{i-1/2})$ mathematically represents a time-like divided difference derivative of an enthalpy change. Key to this is that the background hydraulic gradient term 
\begin{equation} \label{eq:hydraulic_gradient} - \frac{h_{i+1}-h_i+r^{-1}(b_{i+1}-b_i)}{x_{i+1}-x_i} \end{equation}
be positive, equivalent to the constraint
$$ \pd{b}{x} < -(r^{-1}-1)\pd{(h+b)}{x} $$
identified in the main paper: bed slopes cannot be allowed to be so retrograde as to cause ponding of water, or reverse its flow direction.

In the same vein, when solving \eqref{eq:energy_conservation_bed_discrete} by a Newton iteration, the background hydraulic gradient must not become negative during the iteration, since that would turn the problem into backward diffusion. Even if the hydraulic gradient remains positive at the actual solution, such a situation is possible to engineer through a bad choice of the initial guess for $h_{i+1}$ for instance: an obvious case is that of a flat bed ($b_{i+1} = b_i$) if we make the na\"ive initial guess of $h_{i+1} = h_i$. This must be avoided for instance by extrapolating from $h_{i-1}$ and $h_i$ forward to $h_{i+1}$ to create an initial guess.

For computational purposes, it is important (given a likely initial guess of $\Pi_i = 0$ in a cell that is about to become temperate, or that $\Pi_i$ may be predicted to be negative in a cell that is about to turn subtemperate) that the hydrology model and sliding law can be continued smoothly to negative $\Pi$. In particular, we require that $f(T,\Pi,u)$ and $\tilde{\kappa}(\Pi)$ can be extended to negative $\Pi$ and remain monotone. In particular, $\tilde{\kappa}$ must  be differentiable with a strictly positive derivative, including at $\Pi = 0$, in order to permit a Newton solver to work (a zero derivative of $\tilde{\kappa}$ at $\Pi = 0$ may render the Jacobian of the discretized system singular in the situation where for instance a single bed cell has just turned temperate). 
The following are suitable forms for $\tilde{\kappa}$ that mimics \eqref{eq:sample_transformation_final} for a power-law $\kappa$ (vanishing at $\Pi = 0$ but retaining a finite derivative):
\begin{equation} \label{eq:transformation_actual_supp1} \tilde{\kappa} = (\Pi^2 + \Pi_0^2)^{c/2} \Pi, \qquad \tilde{\kappa}_2 = c\end{equation}
where $\Pi_0$ is some regularizing constant, ideally small, and $c > 0$ is constant.  We can correspondingly take $\mathcal{N}$ as 
\begin{equation}\label{eq:transformation_actual_supp2}  \mathcal{N} = (\Pi^2 + \Pi_0^2)^{c/2-1} \Pi, \end{equation}
ensuring that $\mathcal{N}$ is monotone while retaining a finite derivative.
For $f$, we also require that $f$ remain monotone in $\Pi$, while remaining positive, and the simplest approach is to write $f$ in regularized form
$$ f_r(T,\mathcal{N}(\Pi),u) = f(T,\mathcal{N}(H_0(\Pi)\Pi),u) $$
where $H_0(\Pi)$ is a smooth,  monotone function satisfying $H_0(\Pi) = 0$ for $\Pi < 0$, $H_0(\Pi) = 1$ for $\Pi > \epsilon$. A plausible choice is the antidervative of the bump function
$$ H_0'(\Pi) = \left\{ \begin{array}{l l} C \exp\left(-\frac{1}{1-(2\Pi/\epsilon - 1)^2}\right) & \mbox{for } 0 <  \Pi < \epsilon, \\
                      0 & \mbox{otherwise},
                     \end{array} \right. $$
$C$ being chosen to ensure that $\int_0^{\epsilon} H_0'(\Pi) \rd \Pi = 1$.

Similarly, the friction coefficient $\gamma(T)$ needs to be extended to positive $T_i$ since our operator splitting permits $T_i$ to become positive in regions where $I_{i-1} = 0$, which are then diagnosed \emph{a posteriori} as being temperate, with $I_i = 1$ forcing $T_{i+1}$ back to the melting point in the next step. Key here is to ensure that $\gamma(T)$ remains monotone, and does not drop further as temperature is raised spuriously above the melting point, meaning the function $\gamma(T)$ should equal $\gamma(0)$ for $T > 0$. A plausible form for this is to implement numerically a regularized version $\gamma_r(T)$, of the form
$$ \gamma_r(T) = H_r(T)\gamma(0) + (1-H_r(T))\gamma(T) $$
where $H_r(T)$ is a smooth, monotone function satisfying $H_r(T) = 0$ for $T < -\epsilon$, $H_r(T) = 1$ for $T > 0$; this cut-off function can be constructed in the same way as $H_0$ above. 
%A plausible choice is the antidervative of the bump function
%$$ H_r'(T) = \left\{ \begin{array}{l l} C \exp\left(-\frac{1}{1-(2T/\epsilon + 1)^2}\right) & \mbox{for } -\epsilon <  T < 0, \\
%                      0 & \mbox{otherwise},
%                     \end{array} \right. $$
%$C$ being chosen to ensure that $\int_{-\epsilon}^0 H_r'(T) \rd T = 1$. 
For unbounded $\gamma(T)$, it may also be necessary to impose a low temperature cutoff to prevent numerical overflow errors during computation. One way to impose such a cut-off is to put
$$\gamma_{l} = H_l(T)\gamma_r(T_l) + (1-H_l(T))\gamma_r(T), $$
now with $H_l$ a smooth,monotone function satisfying $H_l(T) = 0$ for $T < T_l$, $H_l(T) = 1$ for $T > T_l + \epsilon$.

\section{Numerical solutions}

\begin{figure}
 \centering
 \includegraphics[width=\textwidth]{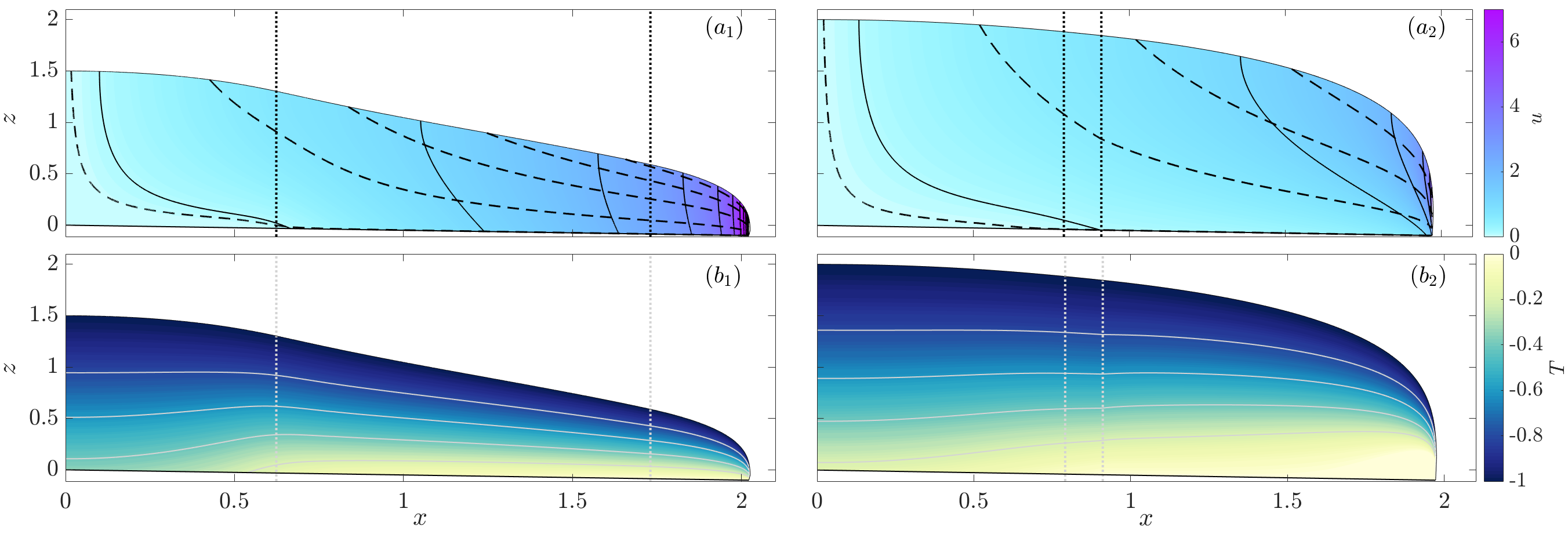}
\caption{Two-dimensional reference solutions, reproduced from the main paper.  Panels a$_1$, b$_1$ show velocity profiles (solid contours, contour interval 1, background shading) and streamlines (dashed lines) for two different, two-dimensional steady state solutions. Vertical dotted lines demarcate the region in which there is significant sliding at subtemperate basal temperatures. Panels a$_2$, b$_2$ show the corresponding temperature profiles (white contours, background shading, contour interval of 0.2). Both examples use $Pe = a = 1$, $G = 0.5$, $T_s = -1$, $\rd b / \rd x = 0.05$, $\delta = 0.03$. $\gamma_0 = 0.1$, $h(0) = 1.5$ (column a) and $\gamma_0 = 3$, $h(0) = 2$ (column b). Both examples use the temperate friction law $f(0,N,u) = \gamma_0u$ at $z = 0$} \label{fig:2D_results_supp}
\end{figure}

Here we provide additional detail on the numerical solutions reported in the main paper. As a reference point, we repeat figure \ref{fig:2D_results_supp} here, showing two-dimensional solutions with long (column a) and short (column b) regions of subtemperate sliding.

\begin{figure}
 \centering
 \includegraphics[width=0.75\textwidth]{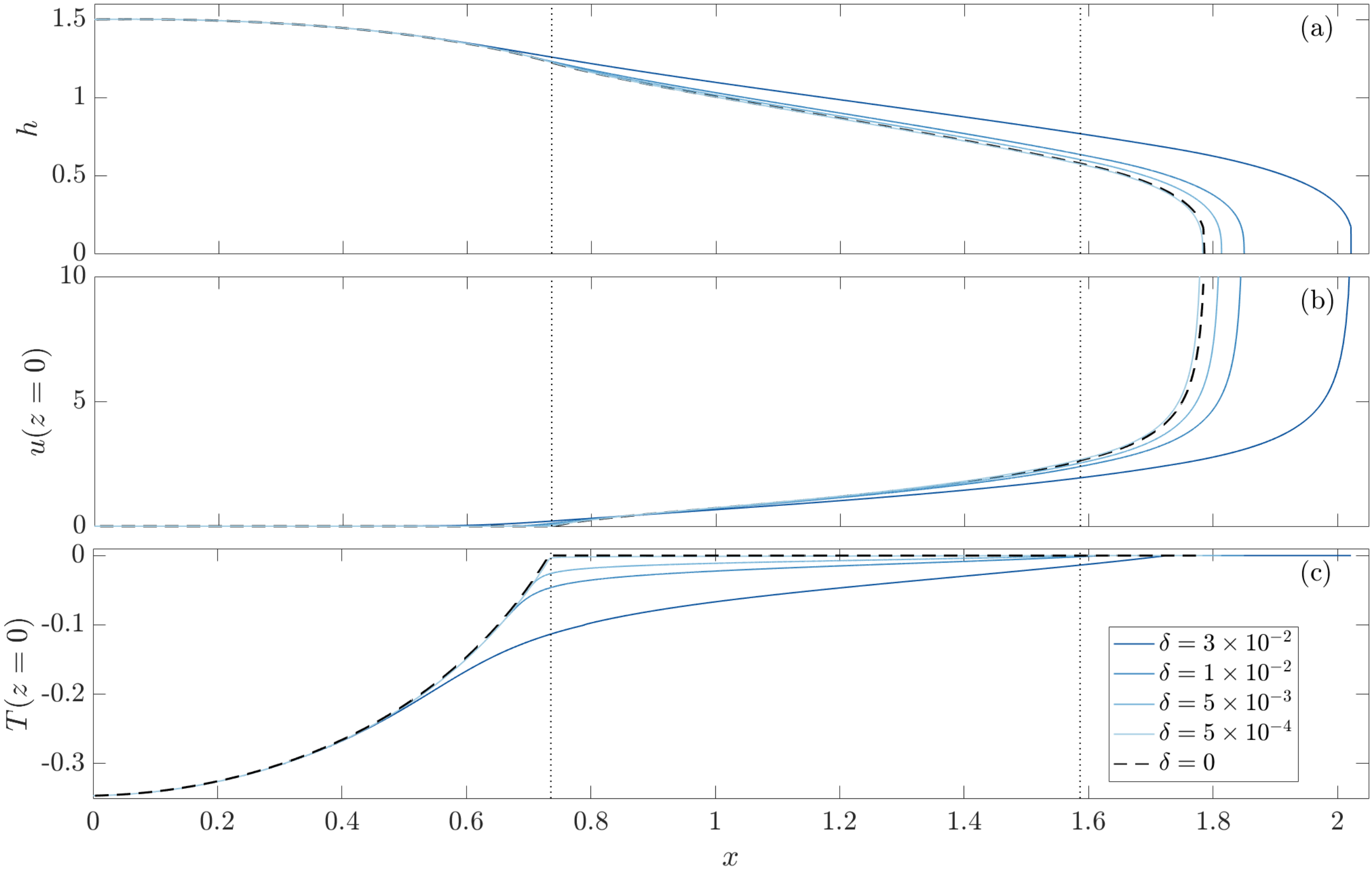}
\caption{Convergence of two-dimensional solutions as $\delta \rightarrow 0$.  Panel a: ice thickness $h(x)$ for different values of $\delta$ as indicated. Panel b: basal velocity $u$ at $z = 0$. Panel c: basal temperature $T$ at $z = 0$. All parameter values other than $\delta$ are the same as in figure \ref{fig:2D_results_supp}a.} \label{fig:converge}
\end{figure}

 Such solutions were previously considered in the asymptotic limit of $\delta \rightarrow 0$   by \citet{Mantellietal2019}. Figure \ref{fig:converge} demonstrates that the numerical solutions reported in the main paper indeed converge to the solutions computed for the leading order model in \citet{Mantellietal2019}. We see that the asymptotic limit $\delta \rightarrow 0$ underestimates the size of the ice sheet compared with finite $\delta$, and that quite small values of $\delta \lesssim 10^{-2}$ are required for reasonably good numerical agreement.

\begin{figure}
 \centering
 \includegraphics[width=\textwidth]{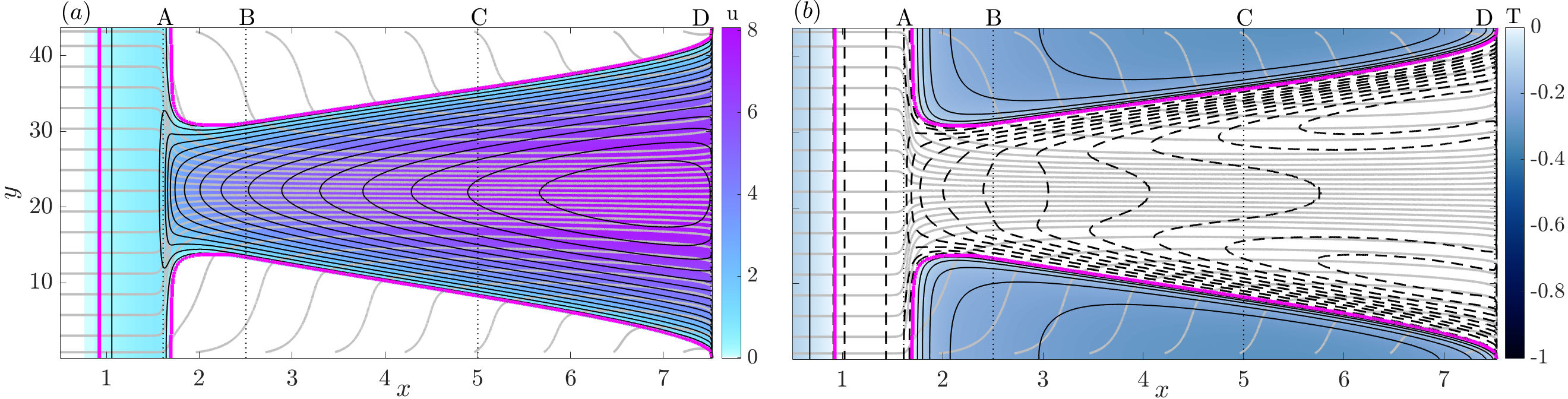}
\caption{Hydromechanical instability, reproduced from main paper: same parameters as figure \ref{fig:2D_results_supp}b but using the double power law $f(0,N,u) = \gamma_0uN/(N_s+N)$ and $N$ defined through \eqref{eq:transformation_actual_supp2}  with $\Pi_0 =  0.1$, $N_s = 0.1$, $k_0 = 4/3$, $\beta = 0.1$. 
Panel a shows contours of axial velocity $u$ at the bed (as well as background shading), with contour interval 0.5. The pink curve is still the subtemperate-temperate boundary, grey streamlines show velocity at the bed
 Panel b: contours of temperature $T$ (solid lines, also background shading) and $\Pi$ (dashed lines). Contour levels are 0.05 for $T$, 0.1 for $\Pi$, the subtemperate-temperate boundary is marked in pink. Solid grey lines are still streamlines of velocity at the bed.
} \label{fig:hydraulic_bedmap_supp}
\end{figure}
 
\begin{figure}
 \centering
 \includegraphics[width=\textwidth]{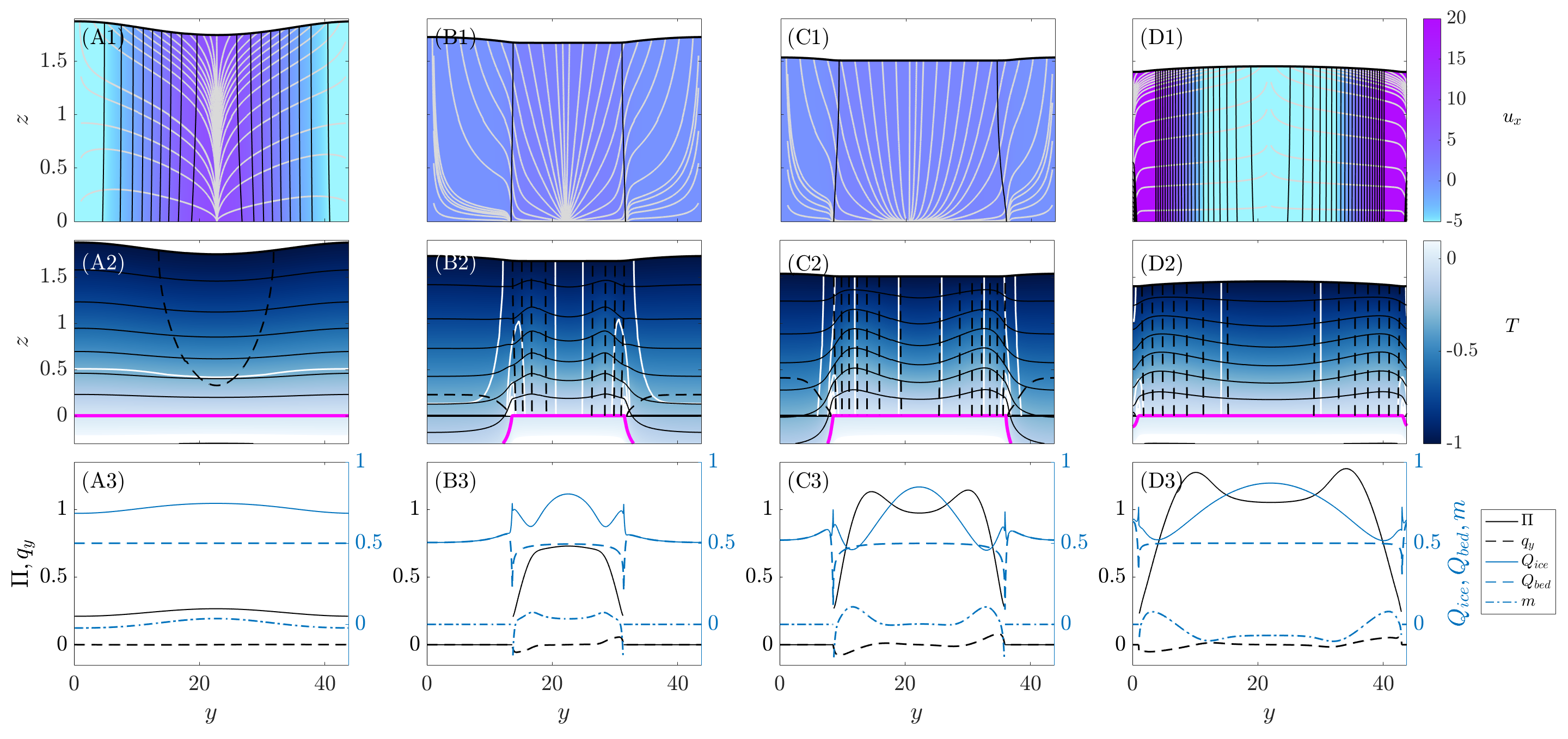}
\caption{Patterns of flow, dissipation and temperature in the ice, and fluxes at the bed. Each column corresponds to a cross-section indicated by the corresponding letter label in figure \ref{fig:hydraulic_supp}. Row 1 (top): contours of $\pdl{u}{x}$ (black, contour interval 1.5,  and background shading), streamlines of $\bm{v}_\perp$, Row 2 (middle): contours of temperature $T$ (black, contour interval 0.15, and background shading), contours of $u$ (dashed black, contour interval 1) and $\alpha |\nabla_\perp u|^2$ (white, contour interval 0.5), pink is the $T = 0$ contour. Row 3 (bottom): basal heat flux $Q_{ice} - \lim_{z\rightarrow 0^+}\pdl{T}{z}+$ in the ice (solid blue) and $Q_{bed} = -\lim_{z\rightarrow 0^+} \pdl{T}{z}$ in the bed (dashed blue), melt rate $m = Q_{bed} - Q_{ice}  + \alpha f(T,N,u)|u|$ (dot-dashed blue), effective pressure proxy $\Pi$ (solid black), lateral $q_y$ (dashed black).}
\label{fig:cross_section_supp}
\end{figure}

Next, we consider the hydromechanical instability of the main paper in greater detail. For completeness, we repeat figure \ref{fig:hydraulic_bedmap_supp} here, showing the formation of patterned flow after the onset of fully temperate sliding. Indicated in the figure are the locations of four cross-sections plotted in figure \ref{fig:cross_section_supp}, mirroring the cross-section plotted for patterned flow onset within the region of subtemperate sliding. 

The progression is similar to the subtemperate case: in the cross-section located furthest upstream (column A), a patch of bed at a higher value of $\Pi$ (and therefore a lower effective pressure) experiences faster sliding and draws in a transverse, secondary flow towards the region of faster sliding. A pattern of ice ridge and ice stream flow separated by a well-defined margin is established quickly (column B), with a single-peaked effective pressure proxy $\Pi$ at the bed. Noticeably, the pattern of shear heating at this point is actually concentrated singificantly inside the margins, as demonstrated by the shear heating concentration shown by white contours in panel B2, and the local minima in conductive heat flux $Q_{bed}$ into the ice in panel B3 (solid blue curve).

Further downstream, the margins are more strongly localized, but a combination of lateral advection of heat generated there, and shear heating due to lateral shear that is not concentrated near the bed (the closely bunched, near-vertical white contours inside the ice stream in panel C2 still causes local minima in $Q_{bed}$ that are some distance inside the ice stream. These lead to corresponding maxima in basal melt rate $m$ and hence the off-centre local maxima in basal effective pressure proxy $\Pi$. Note that the ice stream pattern here is significantly wider than those in the main paper, with $W = 45$, and the lateral diffusivity $\beta = 0.1$ is small and does not smooth out the basal effective pressure distribution as much as in most calculations in the main paper.

The last cross-section D is taken near the downstream end of the domain, see figure \ref{fig:hydraulic_bedmap_supp}, where the growth of ice stream width in the downstram direction has shrunk the ice stream to almost zero width. The original pattern of the ice ridge supplying ice to the ice stream through the inward-oriented secondary flow has reversed at this point: the ice stream is actually slowing marginally in the axial direction, and the secondary flow is one of lateral spreading. As a result the lateral surface slope has also reversed, with ice thickest at the centre of the ice stream.

This turns out to be a peculiarity of the solution here that is linked to its two-dimensional counterpart: the switch in basal friction law from the simple $f(0,N,u) = \gamma_0 u$ to $f(0,N,u) = \gamma_0 u N/(N_s+N)$ means that even in two dimensions, the ice surface is no longer convex near the right-hand edge of the domain, with ice thickness approaching zero: instead, the ice sheet is actually thickening slightly, with the surface slope flattening. Such solutions are not uncommon when integrating forward from a fixed ice divide thickness. In order for such a solution to satisfy a sensible boundary condition at its downstream (seaward), it would have to connect to an almost complete `buttressed' ice shelf, however \citep[e.g.][]{Schoofetal2017}.

\begin{figure}
 \centering
 \includegraphics[width=\textwidth]{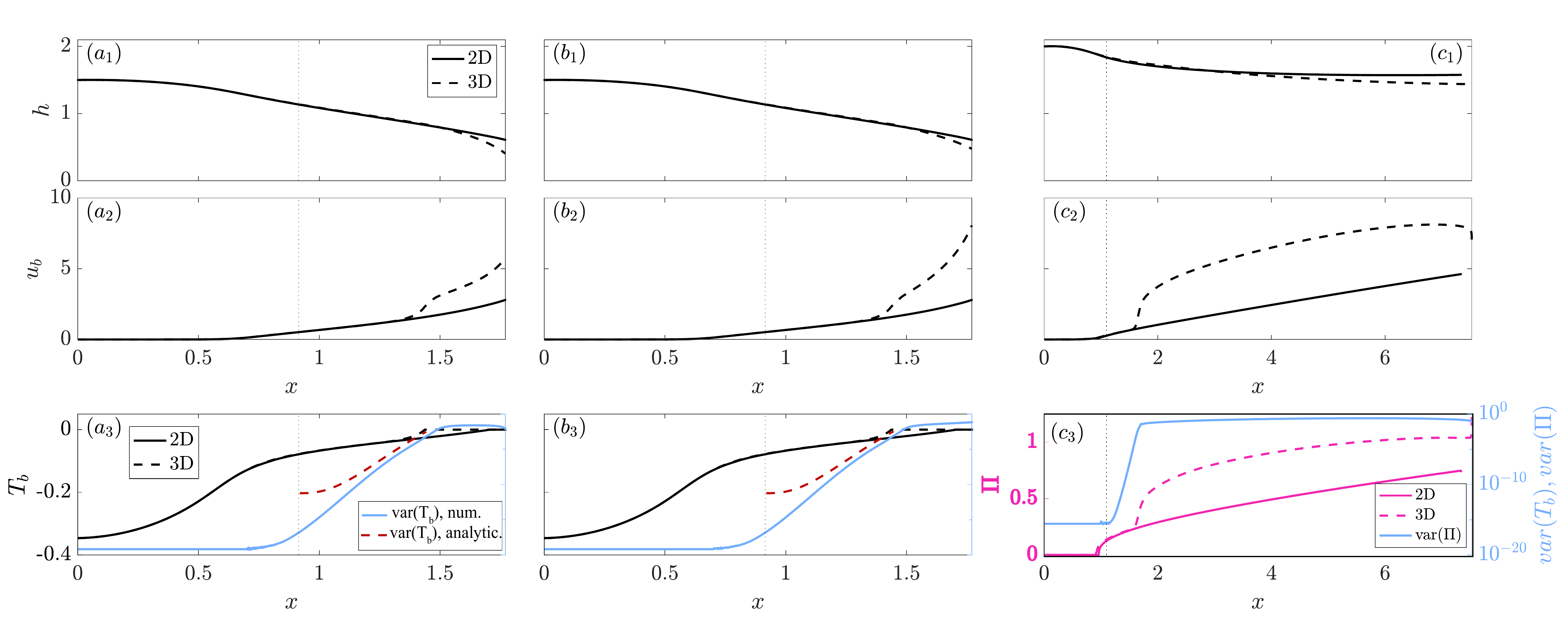}
 \caption{Comparison of patterned an unpatterned solutions. Each column corresponds to one set of parameter values, solid lines show the two-dimensional solution, dashed lines the three-dimensional solution along the central line of the ice stream. Column a uses the parameter values in figure \ref{fig:2D_results_supp}a, column b the same parameter values except that the temperate friction law is switched to $f(0,N,u) = \mu_0 \gamma_0 u N/(\gamma_0 u + \mu_0 N)$ with $\mu_0 = 0.005$, $\beta = 10$, $k_0 = 4/3$, and the relationship between $N$ and $\Pi$ defined by \eqref{eq:transformation_actual_supp2} with $Pi_0 = 0.1$. Column c uses the parameter values in figure \ref{fig:hydraulic_bedmap_supp}. }  \label{fig:centreprofile}
\end{figure}

The two-dimensional solution underlying figures \ref{fig:hydraulic_bedmap_supp}--\ref{fig:cross_section_supp} is shown in figure \ref{fig:centreprofile}, which provides additional detail on pattern growth in the numerical solutions. Columns a--c show comparisons between the patterned solutions generated from the two-dimensional solution in figure \ref{fig:2D_results_supp} and their unpatterned counterparts. Figure \ref{fig:centreprofile}a compares the pattern computed by solving the same problem as in figure \ref{fig:2D_results_supp}a in three dimensions and stochastically perturbing bed friction. Figure \ref{fig:centreprofile}b  replaces $f(0,N,u) = \gamma_0 u$ in figure \ref{fig:centreprofile}a with the regularized Coulomb friction law $f(0,N,u) = \mu_0 \gamma_0 u N/(\gamma_0 u + \mu_0 N)$ and computes two- and three-dimensional figures. Figure \ref{fig:centreprofile}c finally shows the solution in figures \ref{fig:hydraulic_supp}--\ref{fig:cross_section_supp} and its unpatterned, purely two-dimensional counterpart

The first row  of each column in figure \ref{fig:centreprofile} displays the `unpatterned' ice thickness $h$ against $x$ as a solid black line, and ice thickness for the patterned solution as a dashed line. As may be expected, the discrepancy between solid and dashed curves only becomes apparent where the pattern in the corresponding figures  in the main paper and in figure \ref{fig:hydraulic_bedmap_supp} does, and the ice thickness for the patterned solution with channelized ice stream flow is smaller than for the two-dimensional solutions.

The second shows the corresponding comparison for velocity at the ice sheet bed, shown for the patterned case along the centreline of the ice stream. Again, the solutions conform to expectation and ice velocity for the patterned solutions exceed those for the unpatterned ones. The third row shows basal temperature (panels a3, b3) and effective pressure proxy $\Pi$ (panel c3) for the unpatterned case and for the centreline of the patterned cases. In order to the initial growth of the pattern visible where this does not appear to the naked eye in figures such as \ref{fig:hydraulic_supp}, we also plot on a separate, logarithmic scale the variance in bed temperature (panels a3, b3) and the variance in $\Pi$ (panel c3) over the transverse coordinate y as a function of downstream position $x$. It should be apparent that growth of the pattern in the form of near-exponential growth in variance starts some distance upstream of the perceptible onset of streaming flow, as already commented on in the main paper. We will return to how and where that pattern growth originates in the next section.

\begin{figure}
 \centering
 \includegraphics[width=\textwidth]{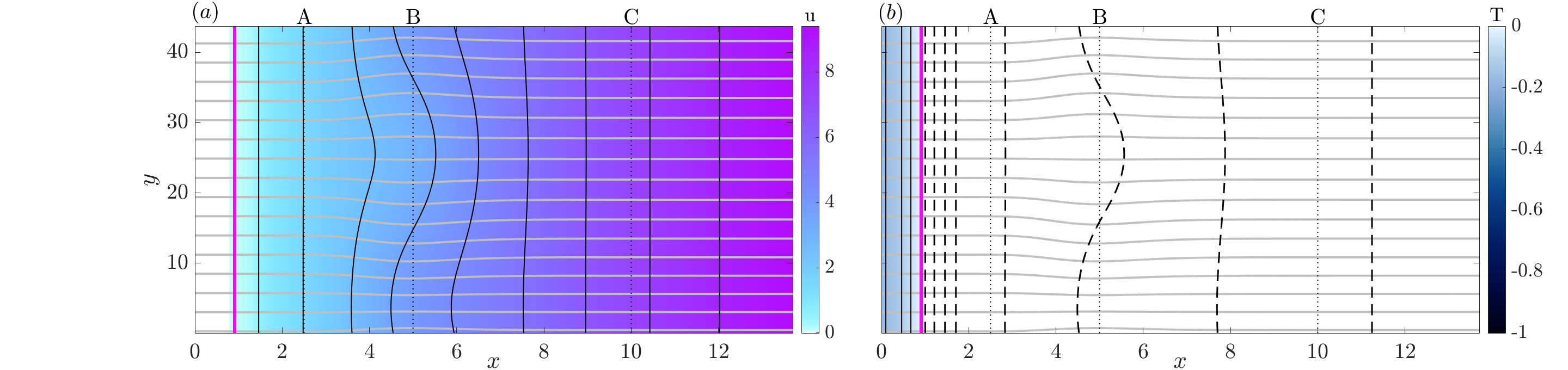}
\caption{Hydromechanical instability: same parameters as figure \ref{fig:2D_results_supp}b but using the double power law $f(0,N,u) = \mu_0 \gamma_0 u N/(\gamma_0 u + \mu_0 N)$ and $N$ defined through \eqref{eq:transformation_actual_supp2}  with $\mu_0 = 0.005$, $\Pi_0 =  0.1$, $k_0 = 4/3$, $\beta = 10$. The figure uses the same plotting scheme as figure \ref{fig:hydraulic_bedmap_supp}. Panel a: $u$ at the bed (solid black, contour interval 1, background shading), subtemperate-temperate boundary in pink, streamlines in grey streamlines show velocity at the bed.
 Panel b: contours of temperature $T$ (solid lines, contour interval 0.2, also background shading) and $\Pi$ (dashed lines, contour interval 0.1. Subtemperate-temperate boundary and basal velocity streamlines as in panel a.
} \label{fig:hydraulic_aborted}
\end{figure}

Here we conclude with the following observation: In the main paper and the examples shown here, we have focused on cases in which `pattern formation' has led to a clearly defined ice stream surrounded by equally clearly defined ice ridges. It is entirely possible to choose parameter combinations in our model that lead to something much less clear-cut. Figure \ref{fig:hydraulic_aborted} is an example, showing a solution based on figure \ref{fig:2D_results_supp}b, but replacing the temperate friction $f(0,N,u) = \gamma_0u$ with the regularized Coulomb friction law $f(0,N,u) = \mu_0 \gamma_0 u N/(\gamma_0 u + \mu_0 N)$ with $\mu_0 = 0.005$, and the hydraulic model defined through $\Pi_0 = 0.01$, $\beta = 10$, $k_0 = 4/3$. The plotting scheme is the same as in figure \ref{fig:hydraulic_bedmap_supp}.

We see that a pattern begins to form some distance downstream of the subtemperate-temperate boundary, but never evolves into a recongizable ice stream: velocity increases at the centre of the domain, and streamlines become focused there around $x \approx 4$. This pattern however disappears again around $x \approx 7$.

\section{Stability analysis} \label{sec:stability_supp}

\subsection{Spatial stability: a parallel-sided slab with subtemperate sliding} \label{sec:supplementary_stability_analysis}

The main paper describes the onset of a `spatial instability', exploiting the time-like nature of $x$ and the parametric limit of small $\delta$ (or high sensitivity of subtemperate sliding to temperature changes). Here we take a slightly different and more general approach, and  study the formation of patterning through a standard linear stability analysis, considering the simplest possible form of an underlying steady state solution, that of a parallel-sided, infinite slab of ice with uniform thickness $h$ and vanishing surface mass balance $a = 0$: in the sense of $x$ being time-like, this actually corresponds to a steady state \emph{in $x$}, while the two-dimensional solution in figure \ref{fig:2D_results_supp} are in fact non-trivial orbits of the dynamical system in $x$, for which the notion of stability is harder to define unambiguously. The main paper uses the parametric limit $\delta \ll 1$ to ensure that the underlying solution varies slowly compared with the length scale over which instability occurs, resulting in a linearized problem with approximately constant coefficients over the instability length scale.

With a constant bed slope $\pdl{b}{x} = -\theta < 0$ and surface temperature $T_s = -1$, we find a solution with no lateral flow $v = w = 0$, and velocity downstream velocity $u = \bar{u}(z)$ and temperature $T  = \bar{T}(z)$ dependent on $z$ only. First we focus on the case of subtemperate sliding, for which the solution takes the form
\begin{subequations} \label{eq:steady_uniform}
 \begin{align}
  \bar{u} = & \left\{\frac{1}{2}\left[h^2-(h-z)^2\right] + \gamma(T_b)^{-1} h \right\}\theta \\
  \bar{T}(z) = & -\frac{\alpha\theta^2}{12}(h-z)^4 + \left\{ \alpha\theta^2\left[\frac{1}{3}h^3 + \gamma(T_b)^{-1}h^2\right] + G\right\}(h-z) + T_s
 \end{align}
\end{subequations}
where the argument of $\gamma$ satisfies $T_b = \bar{T}(0)$:
$$ T_b =    \left\{ \alpha\theta\left[\frac{1}{4}h^3 + \gamma(T_b)^{-1}h^2\right] + G\right\}h + T_s. $$
This constraint determines $T_b$ and therefore $\bar{u}$ and $\bar{T}$; for general $\gamma$, a solution can only be found numerically.

We perturb that steady state as
$$ u = \bar{u}(z) + u'(z)\exp(\lambda x + i k y), \qquad v = v'(z)\exp(\lambda x + i k y), \qquad w = w'(z)\exp(\lambda x + i k y), $$ 
\begin{equation} p = p' \exp(\lambda x + i k y), \qquad T = \bar{T} + T'(z)\exp(\lambda x + i k y) \label{eq:spatial_perturbation} \end{equation}
and linearize the steady-state model of section \ref{sec:leading_order} in the primed variables. This leads to an eigenvalue problem for the spatial growth rate $\lambda$ whose solution determines whether structure emerges in the downstream direction. Below, we focus on non-zero transverse wavenumbers $k \neq 0$; the $k = 0$ case differs structurally because perturbations in $h(x,t)$ intrinsically have zero wavenumbers, and therefore only appear in the perturbed problem for $k = 0$. We discuss that case for completeness in section \ref{sec:zero_wavenumber_supplementary}, although it is not relevant to pattern formation.

Let
$$ \bar{\gamma} = \gamma(T_b), \qquad \bar{\gamma}_T = \od{\gamma}{T}|_(T=T_b) $$
Then the perturbed variables satisfy 
$$ \left(\od{^2}{z^2} - k^2\right) u' =  0, \qquad \left(\od{^2}{z^2} - k^2\right) v' - i k p' = 0, \qquad \left(\od{^2}{z^2} - k^2\right) w' - \pd{p'}{z} = 0, $$
\begin{equation} i k v' + \pd{w'}{z} = -\lambda u', \qquad  Pe\left(\lambda \bar{u} T' + w'\od{\bar{T}}{z}\right) - \left(\od{^2}{z^2} - k^2 \right)  T' =  2 \alpha \od{\bar{u}}{z} \od{u'}{z}, \label{eq:model_linear} \end{equation}
on $0 < z < h$, and
\begin{equation}  \left(\od{^2}{z^2} - k^2 \right)  T' = 0 \end{equation}
for $z < 0$, with boundary conditions
\begin{equation} \pd{u'}{z} = 0, \qquad \pd{v'}{z}+i k w' = 0, \qquad w' = 0, \qquad T' = 0 \end{equation}
at $z = h$, 
\begin{equation} \pd{u'}{z} = \bar{\gamma} u' + \bar{\gamma}_T \bar{u} T', \qquad \od{v'}{z} = \bar{\gamma} v', \qquad w' = 0, \qquad \left[-\od{T'}{z}\right]_-^+ = 2\alpha \bar{\gamma}\bar{u} u' + \alpha \bar{\gamma}_T \bar{u}^2 T', \qquad [T']_-^+ = 0 \end{equation}
at $z = 0$, and $\pdl{T'}{z} \rightarrow 0$ as $z \rightarrow -\infty$.

The solutions for $u'$ and $w'$ can be written in the form $u' = U T'(0)$, $w' = \lambda W T'(0)$, where
\begin{align} 
 U = & -\frac{\cosh[k(h-z)]}{k\sinh(kh) + \bar{\gamma}\cosh(kh)}\bar{\gamma}_T\bar{u}(0), \\
 W = & -  \frac{h\sinh(kz)-z \sinh(kh) \cosh[k(h-z)]}{2 k \sinh^2 (kh) + \bar{\gamma}\left[\sinh(kh)\cosh(kh) - kh\right]} \bar{\gamma}_T \bar{u}(0), \label{eq:W_scalefun}
\end{align}
while $ T'(z) = T'(0) \exp(|k|z) $ for $z < 0$.

The heat equation \eqref{eq:model_linear} therefore becomes the non-standard eigenvalue problem
\begin{equation} \left(\od{^2}{z^2} - k^2 \right)  T' =  \lambda Pe \, \bar{u} T' +  \left(  \lambda Pe \, W\od{\bar{T}}{z} - 2 \alpha \od{\bar{u}}{z} \od{U}{z}  \right) T'(0) \label{eq:eigval_stability1}  \end{equation}
for $0 < z < h$, subject to
\begin{equation} \left.\od{T'}{z}\right|^+ =  - \eta T' \qquad \mbox{at } z = 0, \qquad  T' = 0 \qquad \mbox{at } z = h \label{eq:bc_stability}
\end{equation}
where
\begin{equation} \eta = -|k| - \alpha \bar{\gamma}_T \bar{u}(0)^2 \frac{ \bar{\gamma} \cosh(kh)-k\sinh(kh)}{\bar{\gamma} \cosh(kh) + k \sinh(kh)} \label{eq:eta_def} \end{equation}
The main complication in \eqref{eq:eigval_stability1} is that the boundary value $T'(0)$ appears explicitly into the differential equation that $T'(z)$ satisfies.  As a brief digression, we describe next how this complicates a possible variational approach to the eigenvalue problem.

\subsection{Variational pitfalls} \label{sec:variational}

In this section, we touch on one of the main theoretical complications introduced by allowing the boundary value $T'(0)$ to appear in the eigenvalue problem \eqref{eq:eigval_stability1}: namely, that a simple variational approach does not help bound the eigenvalue $\lambda$, even in the limit of large wavenumbers $k$, where this would be possible if not for the vertical advection term. The latter limit is particularly relevant as it could provide a springboard for a formal proof of well-posedness of at least the linearized forward integration problem in $x$.

Multiplication of \eqref{eq:eigval_stability1} by the complex conjugate ${T'}^*$ of $T'$ and integration over $[0,h]$, using integration by parts, leads to 
\begin{align} \int_0^h -\left|\od{T'}{z}\right|^2 - k^2\left| T'\right|^2 \rd z + \eta \left|T'(0)\right|^2 = & \lambda Pe \left\{ \int_0^h \bar{u}\left|T'\right|^2 \rd z + \int_0^h W \od{\bar{T}}{z} {T'}^* \rd z \, T'(0) \right\} \nonumber \\ - 2\alpha \int_0^h \od{\bar{u}}{z} \od{U}{z} {T'}^* \rd z \, T'(0).  \label{eq:variational_supplementary}  \end{align}
 Define the usual $L^2$ norm
 $$ \norm{f} = \sqrt{\int_0^h |f|^2 \rd z}. $$
Importantly, we cannot bound $|T(0)|$ in terms of the $L^2$ norm $\norm{T}$, but only in terms of an $H^1$ norm: the following trace inequality holds 
\begin{equation} \label{eq:trace} |T'(0)|^2 \leq \norm{\od{T'}{z}}^2 \end{equation}
since $T'(h) = 0$. We also have the following $L_\infty$ bounds
$$ \inf \bar{u} = \bar{u}(0) > 0, \qquad \sup \left|\od{\bar{u}}{z}\right| = \theta, \qquad \sup \left| \od{\bar{T}}{z}\right| \leq \alpha \theta \left[ \frac{1}{3}h^3 + \bar{\gamma}^{-1}h^2\right] + G $$
where $\bar{\gamma} = \gamma(T_b)$. Rearranging \eqref{eq:variational_supplementary},
\begin{equation} \label{eq:variational_supplementary_bis} \lambda Pe \left\{ \int_0^h \bar{u} |T'|^2 \rd z + \int_0^h W \od{\bar{T}}{z} T' \rd  z T'(0) \right\} = - \norm{\od{T'}{z}}^2 - k^2 \norm{T'}^2 + \eta |T'(0)|^2 + 2 \alpha \int_0^h \od{\bar{u}}{z} \od{U}{z} T' \rd z \, T'(0). \end{equation}
We can bound the right-hand side, since by the Cauchy-Schwarz inequality and \eqref{eq:trace}
\begin{equation} \label{eq:shear_bound} \left| 2 \alpha \int_0^h \od{\bar{u}}{z} \od{U}{z} T' \rd z \, T'(0) \right| \leq  2 \alpha \sup\left|\od{\bar{u}}{z}\right| \norm{\od{U}{z}} \norm{T'} \norm{\od{T'}{z}} \end{equation}
and, for instance in the limit of large $k$, we can make $\norm{\rd U / \rd z} \rightarrow 0$. Poincar\'e's inequality then makes sure that the last term on the right-hand side of \eqref{eq:shear_bound} is therefore dominated by $\norm{\rd T'/\rd z}^2$, and the right-hand side of \eqref{eq:variational_supplementary_bis} is guaranteed to be negative for large $k$. Ordinarily one would now proceed by proving that the coefficient of $\lambda$ on the left-hand side of \eqref{eq:variational_supplementary_bis} has positive real part; that however turns out to be not feasible by simple means, since we can put
\begin{align} \int_0^h \bar{u} |T'|^2 \rd z \geq & \bar{u}(0) \norm{T'}^2 \\
\left| \int_0^h W \od{\bar{T}}{z} {T'}^* \rd  z T'(0) \right| \leq & \sup \left| \od{\bar{T}}{z}\right| \norm{W} \norm{T'} |T'(0)| %\nonumber \\
%\leq & \frac{2h |\bar{\gamma}_T|\bar{u}(0)}{k} \sup \left| \od{\bar{T}}{z} \right| \norm{T'} |T'(0)| \\
\end{align}
but while again $\norm{W}$ goes to zero as $k \rightarrow \infty$, we have to recognize that $|T'(0)|$ is only bounded by $\norm{\rd T' / \rd z}$, and there is nothing \emph{a priori} to prevent the latter from being much larger than $\norm{T'}$. (Note that if we had $\norm{W} = 0$, we could immediately prove that the eigenvalue has negative real part for large $k$.) In short, the most obvious, simple bounds on hte variaous integrals in \eqref{eq:variational_supplementary} do not help to establish that $\lambda$ has negative real part for large wavenumbers. This is perhaps not surprising: in the analysis of the spatial stability problem for small $\delta$ in the main paper, it is not so much that scaling of $\norm{W}$ with $k$ that leads to stabilization as the fact that $W < 0$ and  $\rd W / \rd z < 0$; the sign of $W$ and its derivatives is however lost in a simple variational approach to the problem, suggesting that something more sophisticated is needed.

\subsection{Numerical solution of stability problem}

Equation \eqref{eq:eigval_stability1} defies closed-form solution except, since the steady state solution is non-constant and hence the coefficients $\bar{u}$ and $\rd \bar{T}/\rd z$ depend on $z$. Analytical progress is possible some parametric limits: we explore high temperature sensitivity $-\bar{\gamma}_T \sim \delta^{-1} \gg 1$ in the main paper, and rapid sliding $\bar{\gamma} \ll 1$ below. In general, however, we have to solve the eigenvalue problem numerically. To do so, we reformulate \eqref{eq:eigval_stability1} first as an integral equation. The Green's function for the operator $\rd^2 / \rd z^2 - k^2$ on $0 < z < h$, subject to the homogeneous boundary conditions \eqref{eq:bc_stability}, is
\begin{equation}  G(z,z') = \left\{\begin{array} {l l} \theta_2(z)\theta_1(z') & z' < z, \\
                                 \theta_1(z)\theta_2(z') & z' \geq z, \end{array} \right.\qquad
\theta_1(z) =  \frac{k\cosh(kz)-\eta\sinh(kz)}{k[k\cosh(kh)-\eta \sinh(kh)]}, \qquad 
\theta_2(z) =  -\sinh[k(h-z)]  \label{eq:Greensfunction} \end{equation}
except when $k \cosh(kh) - \eta \sinh(kh) = 0$, which can happen at most at a single value of $k$ (at which the operator has a zero eigenvalue), and \eqref{eq:eigval_stability1} can be written as
\begin{align} T'(z) = &  \lambda \int_0^h Pe \, G(z,z') \bar{u}(z')  T'(z') \rd z' + \lambda\int_0^h Pe\, G(z,z')W(z') \left.\od{\bar{T}}{z}\right|_{z'} \rd z' T'(0) \nonumber \\
&  - \int_0^h 2\alpha G(z,z')\left.\left( \od{\bar{u}}{z}\od{U}{z}\right)\right|_{z'} \rd z' T'(0) \label{eq:eigval_stability2}
\end{align}
We discretize $T'(z)$ using its values $T_m$ at Gauss-Lobatto integration nodes $z_m$ in the interval $[0,h]$, where $m = 0,\ldots, n$, with $z_0 = 0$ and $z_n = h$. Then
\begin{align} T_m = &  \lambda \sum_{l=1}^{n-1} G_{m,l}\bar{u}_l T_l + \lambda \sum_{l=1}^n Pe\, G_{m,l}\left.\left(W\od{\bar{T}}{z}\right)\right|_{z=z_l} T_0 - \sum_{l=0}^n 2 \alpha G_{m,l} \left.\left(\od{\bar{u}}{z}\od{U}{z}\right)\right|_{z=z_l} T_0
\end{align}
for $m = 0,1,\ldots n-1$, where $\bar{u}_l = \bar{u}(z_l)$,
\begin{equation} G_{kl} = G(z_k,z'_l)w_l,
\end{equation}
and $w_l$ is the integration weight corresponding to the $l$th node. We have made use of the fact that the Green's function $G_{nl}$ vanishes on account of the Dirichlet condition at $z = h$, which also requires that $T_n = 0$. The discretized system takes the form $\bm{A}\bm{T} = \lambda \bm{B} \bm{T}$, where $\bm{T} = (T_0,T_1,\ldots,T_{n-1})^\mathrm{T}$ and $\bm{A}$ and $\bm{B}$ are $n$-by-$n$ dense matrices. We solve resulting generalized eigenvalue problem  using the MATLAB routine eigval.

\subsection{Results}

The first thing to emphasize is that temperature-dependent friction ($\gamma_T \neq 0$) is essential in causing the instability, and this is mediated in the stability problem primarily through the coefficient $\gamma$. Consider the counterexample of $\bar{\gamma}_T = 0$, in which case $W = U = 0$ and $\eta = -|k| < 0$, and the coupling with $T'(0)$ in \eqref{eq:eigval_stability1} disappears.  We obtain a standard Sturm-Liouville problem with negative eigenvalues: \eqref{eq:variational_supplementary} with $\bar{\gamma}_T = 0$ becomes.
\begin{equation} \label{eq:eigval_variational}  \int_0^h -\left|\od{T'}{z}\right|^2 - k^2\left| T'\right|^2 \rd z + \eta \left|T'(0)\right|^2 = \lambda Pe \int_0^h \bar{u}\left|T'\right|^2 \rd z. \end{equation}
With $\bar{u} > 0$ and $\eta < 0$ (the latter because $\bar{\gamma}_T = 0$), it follows that $\lambda < 0$.  The sign of $\eta$ is key to this argument; if $\eta$ were positive, which can happen for  $\bar{\gamma}_T \neq 0$, \eqref{eq:eigval_variational} leaves open the possibility of finding positive $\lambda$ even if we continue to ignore vertical advection $W$ and englacial dissipation $\rd U / \rd z$. In fact, with $W$ and $\rd U / \rd z$ omitted, the linearized problem remains of Sturm-Liouville type and resembles a reaction-diffusion term, albeit with the reaction term appearing on the boundary rather than inside the domain. As the main paper shows for large $|\bar{\gamma}_T|$, vertical advection in fact plays a stabilizing role, while englacial dissipation is a higher order effect when $\delta \ll 1$. 

\begin{figure}
 \centering
 \includegraphics[width=\textwidth]{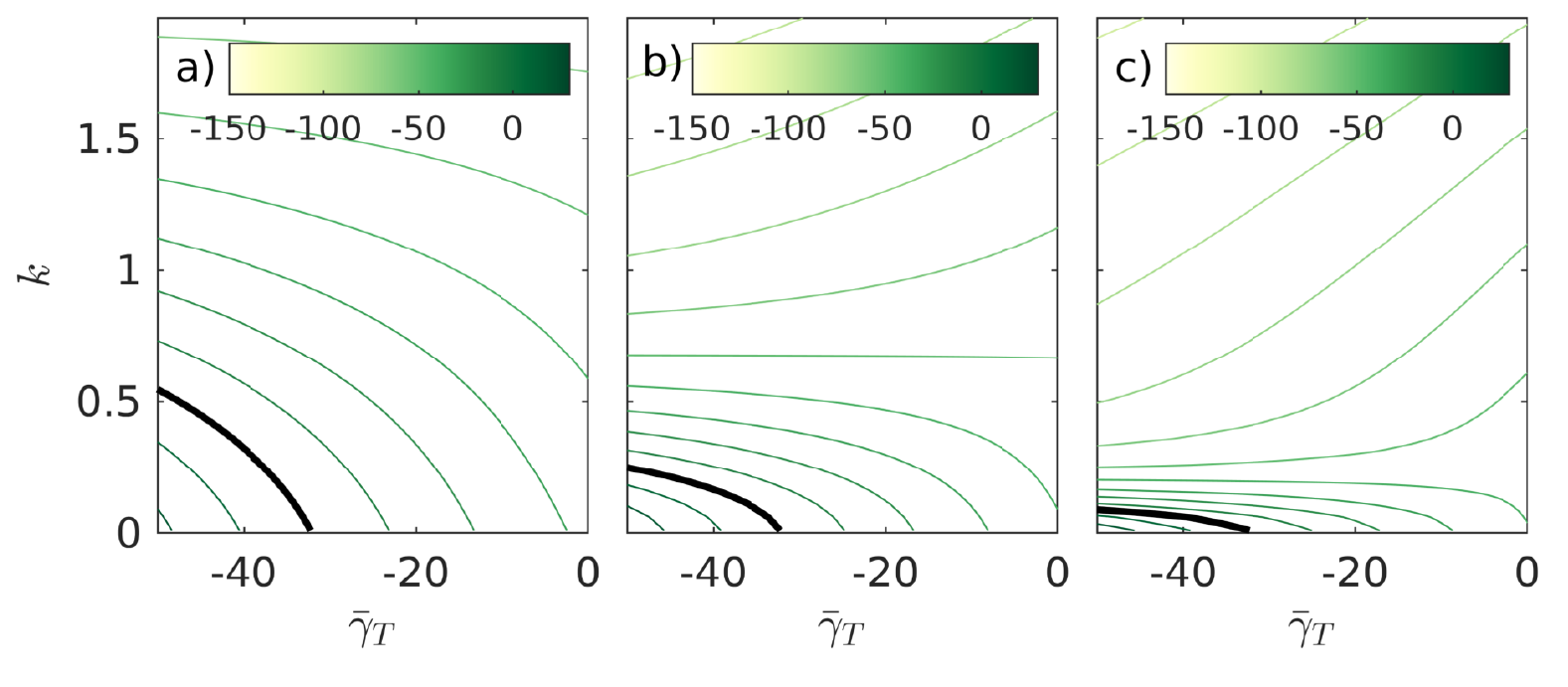}
 \caption{Contour plots of the largest real part of any eigenvalue against wavenumber $k$ and temperature sensitivity $\bar{\gamma}_T$. $Pe = \alpha = 1$, $G = 0.1$, $h = 0.5$ and $\theta$, $\bar{\gamma}$ are constrainted through $h\theta/\bar{\gamma} = 1/4$ in each panel, with $\bar{\gamma}h = 1$ (a), $\bar{\gamma}h = 0.1$ (b) and $\bar{\gamma}h = 0.01$ (c). The solid black line indicates the zero contour in each case.} \label{fig:gammaprime}
\end{figure}

The numerical solution of the full stability problem  \eqref{eq:eigval_stability1}  confirms the critical role played by temperature sensitivity $\bar{\gamma}_T$ in controlling instability. Figure \ref{fig:gammaprime} shows contours plots of the maximum over the real part of all the eigenvalues computed for a given wavenumber $k$ and set of parameters, with the spectrum computed as $\bar{\gamma}_T$ and $k$ are varied. The different panels are computed for fixed $G = 0.1$, $Pe = 1$, $\alpha = 1$ and $h = 0.5$, but show different combinations of $\bar{\gamma}$ and slope $\theta$, adjusted so that the sliding velocity $\bar{u}(0) = h\theta/\bar{\gamma} =  1/4$ in each panel. This amounts to changing the ratio of sliding to shearing (the ratio of flux due to sliding to flux due to shearing being equal to $3/\bar{\gamma}h)$ between the panels while keeping the dimensionless sliding velocity at unity. Panel (a) has $\bar{\gamma}h = 1$, panel (b) $\bar{\gamma}h = 0.1$, panel (c) $\bar{\gamma}h = 0.01$.

Note that while the range of unstable wavenumbers $k$ differs between the examples, in each case instability first occurs when $-\bar{\gamma}_T$ exceeds a critical value that appears not to change (and in fact remains constant at approximately 32). We see that non-zero $\bar{\gamma}_T$ is key to the instability.; we will return in section \ref{sec:rapid_sliding_stability} to the question of why instability appears first at an appparently constant critical value of $-\bar{\gamma}_T$.

\begin{figure}
 \centering
 \includegraphics[width=\textwidth]{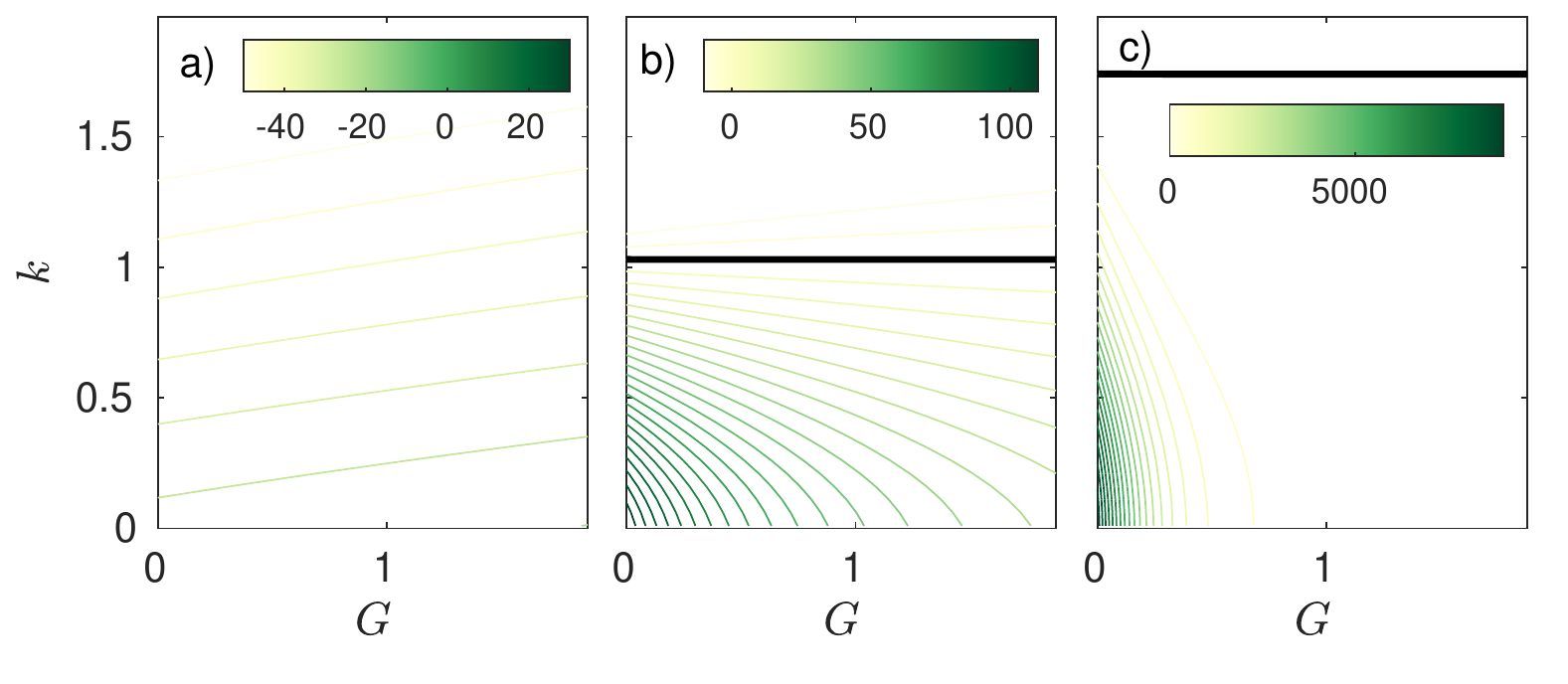}
 \caption{Contour plots of the largest real part of any eigenvalue against wavenumber $k$ and geothermal flux $G$. $Pe = \alpha = \theta = 1$, $h = 0.5$ and $\bar{\gamma} = 2$, with $\bar{\gamma}_T = -10$ (a), $\bar{\gamma}_T = -100$ (b) and $\bar{\gamma}_T = -1000$ (c). The solid black line indicates the zero contour in each case.} \label{fig:G}
\end{figure}

The main paper identifies a different parameter combination, $G h /(\alpha \bar{u}(0)^2)$ as controlling the onset of instability, in the limit of large $\delta^{-1} = |\bar{\gamma}_T|$. Here we investigate this by varying $G$ while keeping $\bar{\gamma}$, $h$, $\theta$ as well as $Pe$, $\alpha$ and $\bar{\gamma}_T$ constant. $G$ is not a `natural' parameter to consider since geothermal heat flux is unlikely to vary much along an ice sheet, at least compared with for instance sliding velocity $\bar{u}(0)$, but using it as a control parameter makes a visual comparison easier. In figure \ref{fig:G}, contours of the largest real part of an eigenvalue are plotted against $k$ and $G$ for $Pe = \alpha = \theta = 1$, $h = 0.5$ and $\bar{\gamma}_T = -10$ (panel a), $-100$ (b) and $-1000$ (c).

Superficially, $G$ does not seem to control the onset of stability: when we do see a stability boundary in $(k,G)$-space, the boundary is invariably parallel to the $G$-axis: shorter wavelengths are stable, long wavelengths are stable or unstable independently of $G$ but dependent on $\bar{\gamma}_T$, in line with figure \ref{fig:gammaprime}. However, for larger $|\bar{\gamma}_T|$, eigenvalues remain moderate (of $O(1)$ for larger $G$), and increase dramatically for $G$ somewhere below unity. As we shall see in the next section, this is in fact the meaning of the stability criterion derived in the main paper: for large temperature sensitivities, moderate growth rates occur for larger $G h /(\alpha \bar{u}(0)^2)$, but these growth rates may merely cause an perturbation to reach a moderate multiple of its original, upstream magnitude along the length of the ice sheet, and this is likely too little to lead to a fully former ice stream emerging. For $G h /(\alpha \bar{u}(0)^2)$ below its critical value of 3, rapid amplification occurs in the downstream direction, leading to a recognizable ice stream pattern.

One of the challenges in using the numerical linear stability analysis here is that the underlying steady state \eqref{eq:steady_uniform} is not generally realized in practice: in the parlance of dynamical systems, we are not looking at a steady state \emph{in the time-like variable} $x$ becoming unstable, but at a non-trivial trajectory of a potentially \emph{non-autonomous} version of the problem (if e.g. $a$ or $b$ depends on $x$) becoming `unstable'. This is easy to understand in the limit of small $\delta$, where the growth rates are sufficiently large that the underlying solution can be treated locally as uniform in $x$ and a standard constant-coefficients linear problem arises. For finite $\delta$, it is generally not possible to conduct the stability analysis for the actual underlying solution, as the linearization would then lead to coefficients that do depend on $x$; that is the reason for using the somewhat unrealistic parallel-sided slab solution \eqref{eq:steady_uniform}.

What we would really like to determine in practice is where instability first appears as we travel downstream along a laterally uniform ice sheet, without having to confine ourselves to a slab solution. We can at best emulate this here. What we do is to fix ice thickness $h$ and vary basal sliding coefficient $\bar{\gamma}$, as would typically occur along the region of subtemperate sliding. To account for the ice flow solution itself also changing as the basal sliding coefficient does, we simultaneously vary $\theta$ to enforce a constant ice flux $(h^2/\bar{\gamma} + h^3/3)\theta$, mimicking the behavour of a relatively short subtemperate sliding zone in which ice thickness and ice flux do not vary by muchm but the surface slope flattens as the bed warms, so as to ensure an constant flux. This surface flattening is evident in the two-dimensional solutions in the main paper.

\begin{figure}
 \centering
 \includegraphics[width=\textwidth]{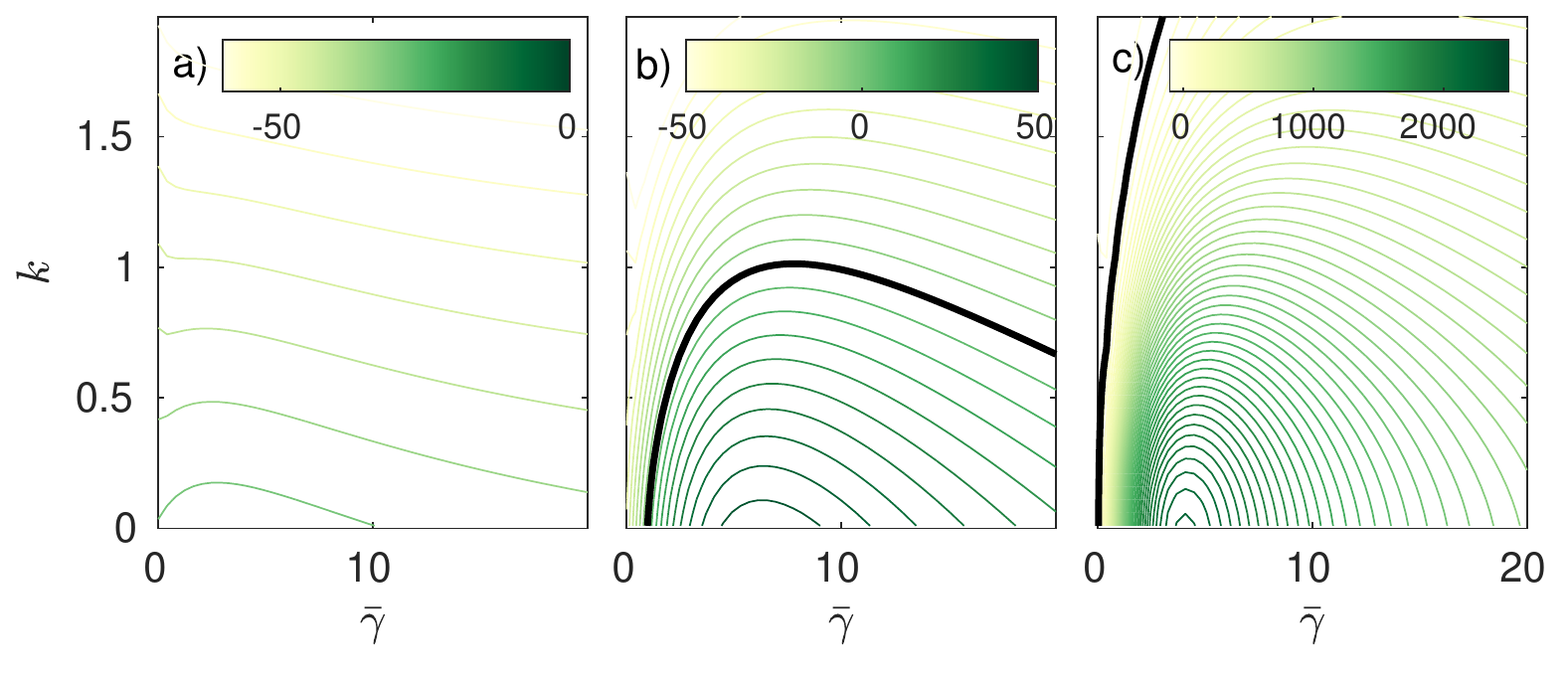}
 \caption{Contour plots of the largest real part of any eigenvalue against wavenumber $k$ and friction coefficient $\bar{\gamma}$. $Pe = \alpha = \theta = 1$, $h = 0.5$ while $\theta$ is given through the constraint $(h^2/\bar{\gamma} + h^3/3)\theta) = 1/6$, while $\bar{\gamma}_T = -5\bar{\gamma}$ (a), $\bar{\gamma}_T = -50\bar{\gamma}_T$ (b) and $\bar{\gamma}_T = -500\bar{\gamma}$ (c). The solid black line indicates the zero contour in each case.} \label{fig:gamma_flux}
\end{figure}

Alongside the variation in $\bar{\gamma}$, we also vary $\bar{\gamma}_T$ to keep a constant ratio $\bar{\gamma}_T/\bar{\gamma}$; with an exponential sliding law $\gamma(T) = \exp(-\delta^{-1} T)$, that ratio is simply $-\delta^{-1}$. Figure \ref{fig:gamma_flux} shows contours of the largest real part of an eigenvalue against $k$ and $\bar{\gamma}$ for different values of $\delta = 0.2$ (panel a), $0.02$ (panel b) and $0.002$ (panel c). In agreement with figure \ref{fig:gammaprime}, there is no unstable combination of $(\bar{\gamma},k)$ at all for the largest value of $\delta$. For smaller $\delta$, instability appears at large $\bar{\gamma}$ (corresponding to relatively low basal temperatures in the region of subtemperate sliding), with growth rates rapidly increasing as $\bar{\gamma}$ decreases, equivalent to a warming bed; further reductions in $\delta$ lead to more unstable combinations of $(\bar{\gamma},T)$ and faster growth rates.

Figure of numerical results from numerical eigenvalue computation: short wavelength stabilization, $k = 0$ grows fastest, larger velocities more likely to be unstable, often stable for all $k$ at lower velocities. Also, oscillatory eigenfunctions for stable solutions with small $\delta$ as advertised in the main text.

\subsection{The limit of high temperature sensitivity revisited} \label{sec:small_delta_revisit}

In the main paper, a stability problem for the case of high temperature sensitivity in the sliding law ($\delta \ll 1$) was derived directly from a rescaled version of the nonlinear, thermomechanical ice flow model. Here we reconcile that stability problem with the finite-depth   eigenvalue problem \eqref{eq:eigval_stability1}--\eqref{eq:bc_stability} for a parallel-sided slab, and explore the caveats to the results in the main paper that the previous section has already foreshadowed.. 

The definition of the parameters $\gamma_0$ and $\delta$ implies that the sliding law coefficients in the stability analysis scale as  $\bar{\gamma}_T \sim \gamma_0 \delta^{-1}$ and $\bar{\gamma} \sim \gamma_0$ for small $\delta$. Assuming again that sliding is highly temperature-sensitive while actual sliding velocities are $O(1)$, we have $\delta \ll 1$ while $\gamma_0 = O(1)$. Then $\eta$ becomes large, and a rescaling of $z$ and $\lambda$ becomes appropriate in \eqref{eq:eigval_stability1} and \eqref{eq:bc_stability}. We can use this to obtain the simplified eigenvalue problem for small $\delta$ in the main paper.

Specifically, we put
\begin{equation} \label{eq:small_delta_rescale}  \Lambda = \delta^2 \lambda, \qquad Z = \delta^{-1}z, \qquad \Theta'(Z) = \delta^{-1}T'(z) \end{equation}
The rescaling of $\lambda$ implies that we are looking at pattern growth over short length scales $x \sim \delta^{-2}$ as in the main paper.
The rescaling in $z$ implies we are again looking at the same a boundary layer near the bed, and once more need to use an appropriate version of the advection and dissipation terms. In particular, we need to Taylor expand $W$ around the bed $z = 0$. As in the main paper, $W$ is large when $-\bar{\gamma}_T \ll 1$, and if we write
\begin{equation} Q_0 = \left. -\od{\bar{T}}{z}\right|_{z = 0} = \alpha \bar{\gamma} \bar{u}(0)^2 + G, \qquad W_{z0} =  \delta \left.\od{ W}{ z}\right|_{z = 0}, \qquad \eta_0 = \delta\eta,  \label{eq:small_delta_coefficients} \end{equation}
we recover the same definitions of $Q_0$, $Q_{z0}$ and $\eta_0$ as in the main paper, with $Q_0$, $W_{z0}$ and $\eta_0$ being $O(1)$ quantities. Specifically, we find from \eqref{eq:W_scalefun} and \eqref{eq:eta_def}
\begin{subequations} \label{eq:small_delta_coefficients2}
\begin{align} W_{z0} = &  \delta \bar{\gamma}_T  \bar{u}(0)\frac{\sinh(kh)\cosh(k h)-kh}{2k \sinh^2(kh) + \bar{\gamma}[ \sinh(kh) \cosh (kh) - kh],} \label{eq:small_delta_advection_coefficient} \\
 \eta_0 = &  -\alpha \delta \bar{\gamma}_T  \bar{u}(0)^2 \frac{ \bar{\gamma} \cosh(kh)-k\sinh(kh)}{\bar{\gamma} \cosh(kh) + k \sinh(kh)} \label{eq:eta_0_def_supplementart}
\end{align}
\end{subequations}
at leading order, where $\delta \bar{\gamma}_T = O(1)$. As before  $W_{z0}$ is always negative, while $\eta_0$ is positive for small $k$ and becomes negative for large $k$. 

With those in place, writing $W \sim \delta^{-1}W_{z0}z = W_{z0}Z$ near the bed, \eqref{eq:eigval_stability1} becomes inside the thermal boundary layer
\begin{equation} \od{^2\Theta'}{z^2} =  \Lambda Pe \left( \bar{u}(0) \Theta' - W_{z0} Z Q_0  \Theta'(0) \right) \label{eq:eigval_stability1_small_delta} \end{equation}
while boundary conditions at the bed \eqref{eq:bc_stability} and matching with the advection-dominated upper part of the slab of ice require that
\begin{align}  \od{\Theta'}{Z} + \eta_0 \Theta' =  0 \qquad \mbox{at } Z = 0, \qquad \od{\Theta'}{ Z} \sim & W_{z0}Q_0/\bar{u}(0) \qquad \mbox{as } Z \rightarrow \infty. \label{eq:small_delta_bc_app}
\end{align}
On cosmetically replacing $\bar{u}(0)$ with $\bar{U}_b$ and $\delta \bar{\gamma}_T$ with $\bar{\Gamma}_\Theta$, we obtain the same eigenvalue problem for spatial instability as in the main paper.

As shown there, the leading order problem has solution
\begin{equation} \Theta(Z) = \Theta(0)\left[ \exp\left(-\sqrt{\Lambda Pe\, \bar{u}(0)}Z\right) + \frac{W_{z0}Q_0 Z}{\bar{u}(0)}\right], \label{eq:supplementary_bl_solution}  \end{equation}
where the eigenvalue $\Lambda$ satisfies $ \sqrt{\Lambda Pe\, \bar{u}(0)} = \eta_0 + W_{z0}Q_0/\bar{u}(0)$, and $\sqrt{\Lambda Pe  \, \bar{u}(0)}$ must have positive real part (and therefore must be real and positive)
so that instability occurs if
\begin{equation} \eta_0 + \frac{Q_0 W_{z0}}{\bar{u}(0)} > 0, \label{eq:supplementary_stability_criterion} \end{equation}
with the first term (representing the net feedback between warming and dissipation at the bed) being positive for small wavenumbers $k$ and negative for large $k$, while the seecond term is always negative and represents the stabilizing effect of cold ice being drawn down towards the bed as the axial flow accelerates. 

Note that we can reduce the set of model parameters involved in determining stability to the two combinations, $\bar{\gamma}h$ and $G h /(\alpha \bar{u}(0)^2)$.  To see this, note from \eqref{eq:small_delta_coefficients2} that we can define  $\breve{\eta}_0(\bar{\gamma}h,kh) = \eta_0/(-\alpha \delta \bar{\gamma}_T\bar{u}(0)^2)$ and $\breve{W}_{z0}(\bar{\gamma}h,kh) = W_{z0}/(-\bar{\Gamma}_{\Theta}\bar{u}(0)h)$ as functions of $\bar{\gamma}h$ and $kh$ only. As $\delta \bar{\gamma}_T < 0$, $\breve{\eta}_0$ and $\breve{W}_{z0}$ have the same signs as $\eta_0$ and $W_{z0}$. Recall that the unperturbed basal heat flux takes the form $Q_0 = \alpha \bar{\Gamma} \bar{u}(0)^2 + G$. Then the solvability condition \eqref{eq:supplementary_stability_criterion} becomes
$$\breve{\eta}_0 + \left(\bar{\gamma}h + \frac{Gh}{\alpha\bar{u}(0)^2}\right)\breve{W}_{z0} > 0 $$
and the dispersion relation 
can be written as
\begin{equation}\frac{\sqrt{Pe \bar{u}(0) \Lambda}}{-\alpha \delta \bar{\gamma}_T \bar{u}(0)^2} = \breve{\eta}_0(\bar{\gamma}h, kh) + \left(\bar{\Gamma} h + \frac{Gh}{\alpha \bar{u}(0)^2}\right) \breve{W}_{z0}(\bar{\gamma}h.kh),  \label{eq:smalldelta_growth_rate_solution_v2} \end{equation}
or equivalently, when \eqref{eq:supplementary_stability_criterion} is satisfied, the eigenvalue  $\Lambda$ satisfies
\begin{equation} \Lambda = \frac{1}{Pe \bar{u}(0)} \left(\eta_0 + \frac{W_{z0}Q_0}{\bar{u}(0)} \right)^2 = \frac{(-\alpha \delta \bar{\gamma}_T \bar{u}(0)^2)^2}{Pe \bar{u}(0)} =\left[  \breve{\eta}_0(\bar{\gamma}h, kh) + \left(\bar{\Gamma} h + \frac{Gh}{\alpha \bar{u}(0)^2}\right) \breve{W}_{z0}(\bar{\gamma}h.kh)\right]^2, \label{eq:eigval_small_delta_supplementary} \end{equation}
and it suffices to treat $\bar{\Gamma} h$ and $G h /(\alpha\bar{u}(0)^2)$ as the only parameters that control stability for a given scale wavenumber $kh$. As $\delta \bar{\gamma}_T < 0$, the requirement that the exponent $\sqrt{Pe \bar{u}(0) \Lambda}$ have a positive real part is again the same as requiring the right-hand side to be positive.

Using the specific form of $\breve{\eta}_0$ and $\breve{W}_{z0}$, it is also possible to show that \eqref{eq:supplementary_stability_criterion} is satisfied and a rapidly-growing instability with $\Lambda \sim \delta^{-2}$ first appears at $k = 0$, which is always the fastest growing wavenumber, and does so when
\begin{equation} \frac{Gh}{\alpha \bar{u}(0)^2} < 3 \label{eq:small_delta_criterion} \end{equation}

When \eqref{eq:supplementary_stability_criterion} is not satisfied, this does not imply the absence of an asymptotic solution, but merely that the solution cannot be of boundary layer type. A regular expansion proceeds by putting $T'(z) = T'_0(z) + \delta T'_1(z)$, and straightforward manipulations in \eqref{eq:eigval_stability1}--\eqref{eq:bc_stability} lead to a reduced problem of the form
\begin{equation} \left(\od{^2}{z^2} - k^2 \right)  T'_0 =  \lambda Pe \, \bar{u} T'_0 +  \left( - \lambda Pe \, \eta^{-1} W \od{\bar{T}}{z} + 2 \alpha  \eta^{-1}\od{\bar{u}}{z} \od{U}{z}  \right) \left. \od{T'_0}{z}\right|_{z=0} \label{eq:eigval_stability1_reduced}  \end{equation}
for $0 < z < h$, subject to
\begin{equation}  T'_0(0) = 0  \qquad \mbox{at } z = 0, \, h\label{eq:bc_stability_reduced}
\end{equation}
where $\eta^{-1}\rd U/\rd z$ and $\eta^{-1} W$ are generally $O(1)$ since the factors of $\bar{\gamma}_T \sim \delta$ in their definitions cancel. Further complications arise in the vicinity of the wavenumber $k$ at which $\bar{\gamma} h \cosh(kh) = k \sinh(kh)$, in which case  $\eta$ becomes small and  these two terms become large again. Without pursuing this further by analytical means, suffice it to say that there is no reason why the situations in which \eqref{eq:supplementary_stability_criterion}  are not satisfied are fundamentally defective; this simply means that there are then no boundary-layer type of solution \eqref{eq:supplementary_bl_solution} with large growth rates $\lambda = \delta^{-2}\Lambda$.

\begin{figure}
 \centering
 \includegraphics[width=\textwidth]{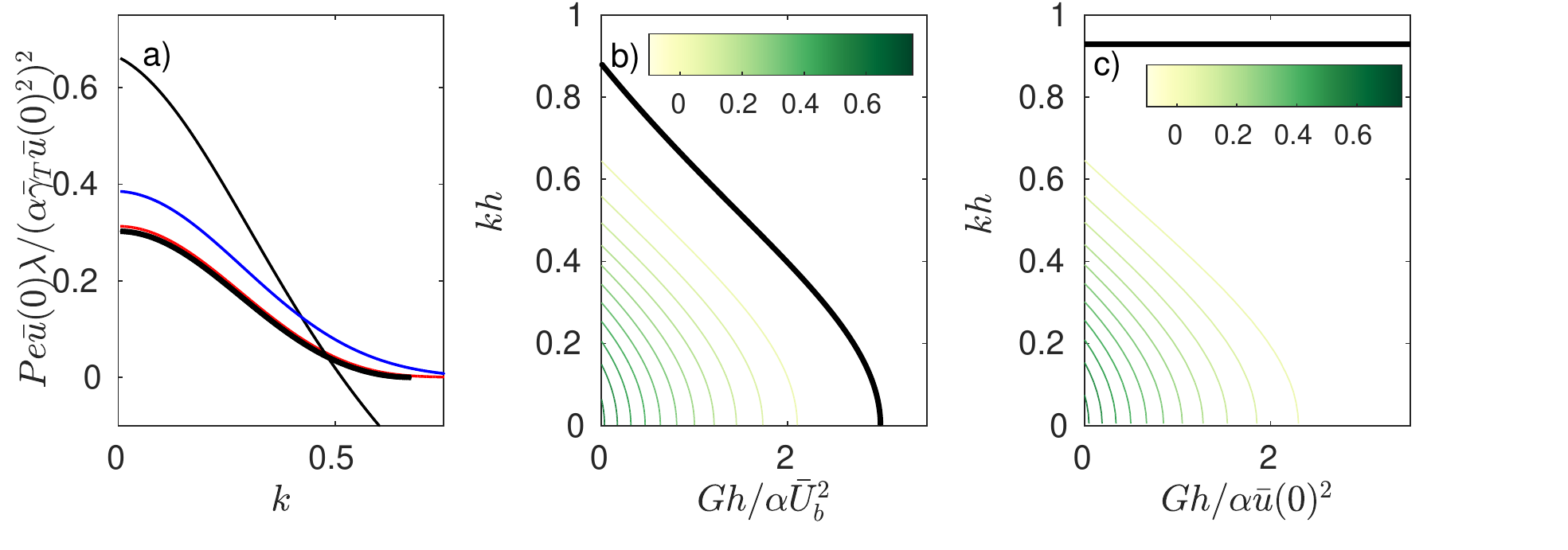}
 \caption{Panel a) shows $Pe \bar{u}(0) \lambda / (-\alpha \bar{\gamma}_T \bar{u}(0)^2)^2) = Pe \bar{U}_b \Lambda/(-\alpha \bar{\Gamma}_\Theta \bar{U}_b^2)^2$ against $kh$ for the asymptotic result \eqref{eq:eigval_small_delta_supplementary}  and computed using \eqref{eq:eigval_stability2} with $\bar{\gamma}h = 1$ and $Gh/(\alpha \bar{u}(0)^2) = 0.8$ as a thick black line, and for $Pe = \alpha = \theta = 1$, $G = 0.1$, $h = 0.5$, $\bar{\gamma} = 2$ and $\bar{\gamma}_T = -100$ (thin black), $-1000$ (blue), $-1000$ (red), so that $\bar{\gamma}h = 1$, $Gh/(\alpha \bar{u}(0)^2$  in all cases, and the thick black line should be the limit of the other lines as $-\bar{\gamma}_T \rightarrow \infty$. The eigenvalue with the largest real value is shown in each case. Panel b): contours of $Pe \bar{u}(0) \Lambda/(-\alpha \delta \bar{\gamma}_T \bar{u}(0)^2)^2$ against $G h/(\alpha \bar{u}(0)^2) = Gh/(\alpha \bar{U}_b^2)$ and $kh$ at $\bar{\gamma}h = 1$, computed from  \eqref{eq:eigval_small_delta_supplementary}. Panel c) shows contours of  $Pe \bar{u}(0) \lambda/(-\alpha  \bar{\gamma}_T \bar{u}(0)^2)^2$ against $G h/(\alpha \bar{u}(0)^2)$ and $kh$ for  $Pe = \alpha = \theta = 1$, $h = 0.5$, $\bar{\gamma} = 2$ and $\bar{\gamma}_T = -10000$. The solid black line is the zero contour in both cases. } \label{fig:small_delta}
\end{figure}

We can in fact pursue the stability problem at small $\delta$ by numerical means. Figure \ref{fig:small_delta} shows a comparison between eigenvalues computed using the asymptotic solution \eqref{eq:eigval_small_delta_supplementary} and the largest real parts of eigenvalues computed using \eqref{eq:eigval_stability2}. As show in the main paper, a rescaling of \eqref{eq:eigval_small_delta_supplementary} shows that  should be a function of $\bar{\gamma}h$, $Gh/(\alpha \bar{u}(0))^2)$ and $kh$ only in the limit of small $\delta$, or of $\bar{\gamma}_T \rightarrow -\infty$. Panel a) plots  $Pe \bar{u}(0) \lambda / (-\alpha \bar{\gamma}_T \bar{u}(0)^2)^2)$ against $kh$ for the asymptotic result \eqref{eq:eigval_small_delta_supplementary} with $\bar{\gamma}h = 1$ and $Gh/(\alpha \bar{u}(0)^2) = 0.8$ as a thick black line, and for $Pe = \alpha = \theta = 1$, $G = 0.1$, $h = 0.5$, $\bar{\gamma} = 2$ and $\bar{\gamma}_T = -100$ ($\delta = 50$, thin black line), $-1000$ ($\delta = 500$, blue line) and $-10000$ ($\delta = 5000$, red line). Very large values of $\delta$ are obviously required to achieve close numerical agreement (as is evident for similar parameteric limits explored in \citet{Haseloffetal2018}), while noticeable discrepancies arise for moderately large (and probably practically realistic) values of $\delta$ (like $\delta = 50$).

Panel b) shows contour plots of $Pe \bar{u}(0) \lambda / (-\alpha \bar{\gamma}_T \bar{u}(0)^2)^2)$  against $kh$ and $Gh/(\alpha \bar{u}(0)^2)$ computed from the asymptotic solution  \eqref{eq:eigval_small_delta_supplementary}, while panel c) shows the same contours computed for
$Pe = \alpha = \theta = 1$, $G = 0.1$, $h = 0.5$, $\bar{\gamma} = 2$ and $\bar{\gamma}_T = -1000$  using \eqref{eq:eigval_stability2}. (corresponding to the red line in panel a). The contour levels are almost identical, \emph{except} for the zero contour shown as a thick black line. Where the asymptotic solution predicts no viable solution, the full solution based on \eqref{eq:eigval_stability2} still computes eigenvalues that can have positive real part (in fact, are positive and real), but these are numerically of $O(1)$ rather than $O(|\bar{\gamma}_T|^2)$, corresponding to modest fractional growth of an upstream perturbation along the entire length of the ice sheet --- and potentially no perceptible 'ice streaming'. Panel c) in fact qualitatively replicates panel c) of figure \ref{fig:G}, where we previously noted that $G$, and hence $Gh/(\alpha \bar{u}(0)^2)$ does not seem to play an actual role in controlling stability. This is true, in the sense that large values of  $Gh/(\alpha \bar{u}(0)^2)$ seem to still admit positive eigenvalues, but eigenvalues are not large when $Gh/(\alpha \bar{u}(0)^2) > 3$, so that the scaled eigenvalue $Pe \bar{u}(0) \lambda / (-\alpha \bar{\gamma}_T \bar{u}(0)^2)^2)$ is positive but close to zero.

\begin{figure}
 \centering
 \includegraphics[width=\textwidth]{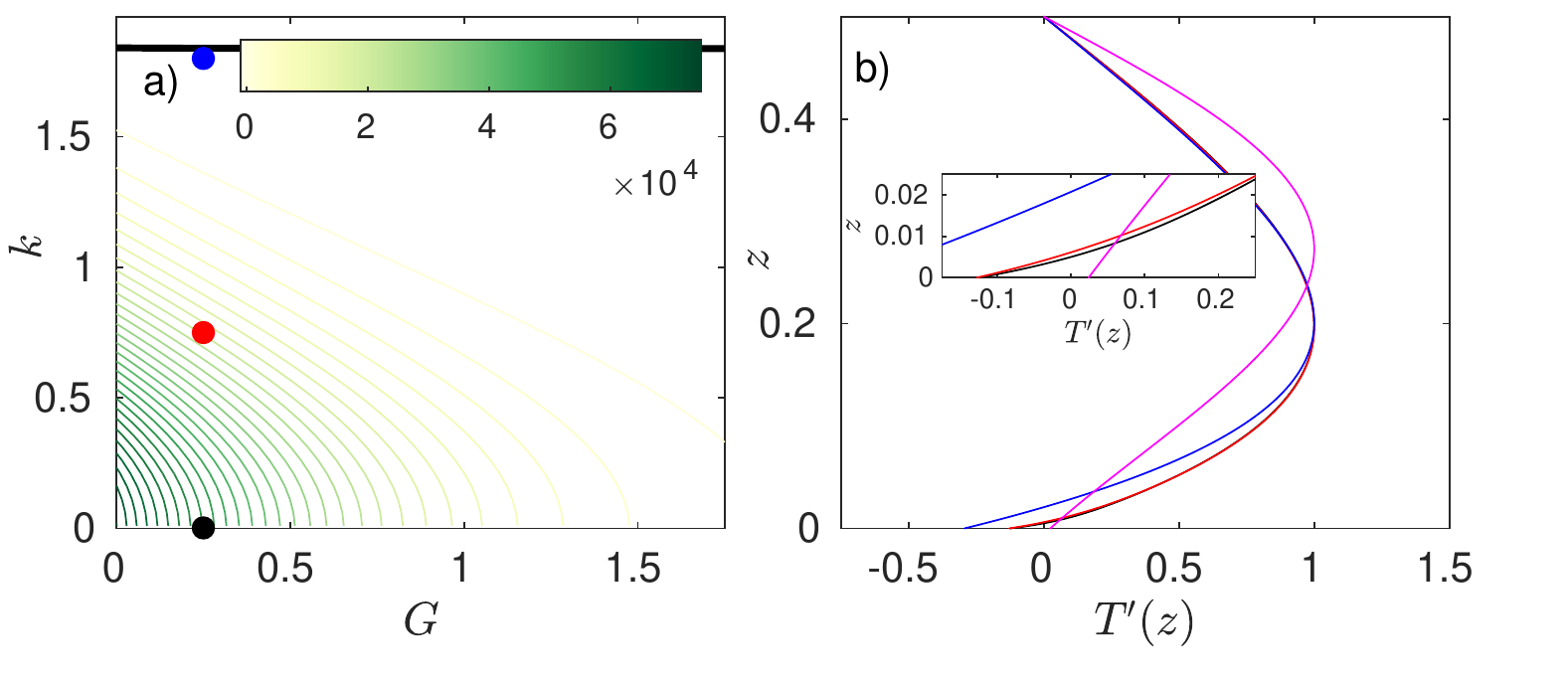}
 \caption{Panel a) replicates figure \ref{fig:small_delta}c) but showing contour lines of $\lambda$ against $G$ and $k$ without any scaling applied. The coloured dots indicate parameter combinations for which we also plot the corresponding eigenfunctions $T'(z)$  using the same colour code in panel b), see text for numerical parameter values.} \label{fig:eigfun}
\end{figure}

As discussed above, we can tie this phenomenon to the boundary layer structure of the eigenmodes. Figure \ref{fig:eigfun}a) shows the dispersion relation of figure \ref{fig:small_delta}c) (but plotting contours of $\lambda$ against $G$ and $k$), with coloured dots indicating combinations $(Gh/(\alpha\bar{u}(0)^2),kh)$ for which we have also computed the corresponding eigenfunctions; these all have $Gh/(\alpha\bar{u}(0)^2) = 0.5$ and $kh = 0$ (black) $0.375$ (red) $0.9$ (blue) and 5 (magenta, outside the plot box). Panel b) plots the corresponding eigenfunctions in the same colours, with $z$ on the vertical. The inset box shows detail near the bed, where we see that the rapidly growing eigenmodes (red and black) have pronounced boundary layer structure while appearing virtually identical above the bed despite having eigenvalues that differ by an order of magnitude; the advection-dominated outer solution is common to both. The very slowly growing eigenmode (blue) lacks the distinct boundary layer structure, but differs noticeably from the rapidly growing eigenmodes only in the lower half of the domain; for this eigenmode (which has a small value of the dissipation-feedback coefficient $\eta$), basal temperature differs noticeably from $0$. The decaying eigenmode (magenta) also lacks a boundary layer structure and in fact has $T'(0) \approx 0$ to a close approximation as discussed above; this mode also differs substantially from the growing eigenmodes throughout the domain.

\begin{figure}
 \centering
 \includegraphics[width=\textwidth]{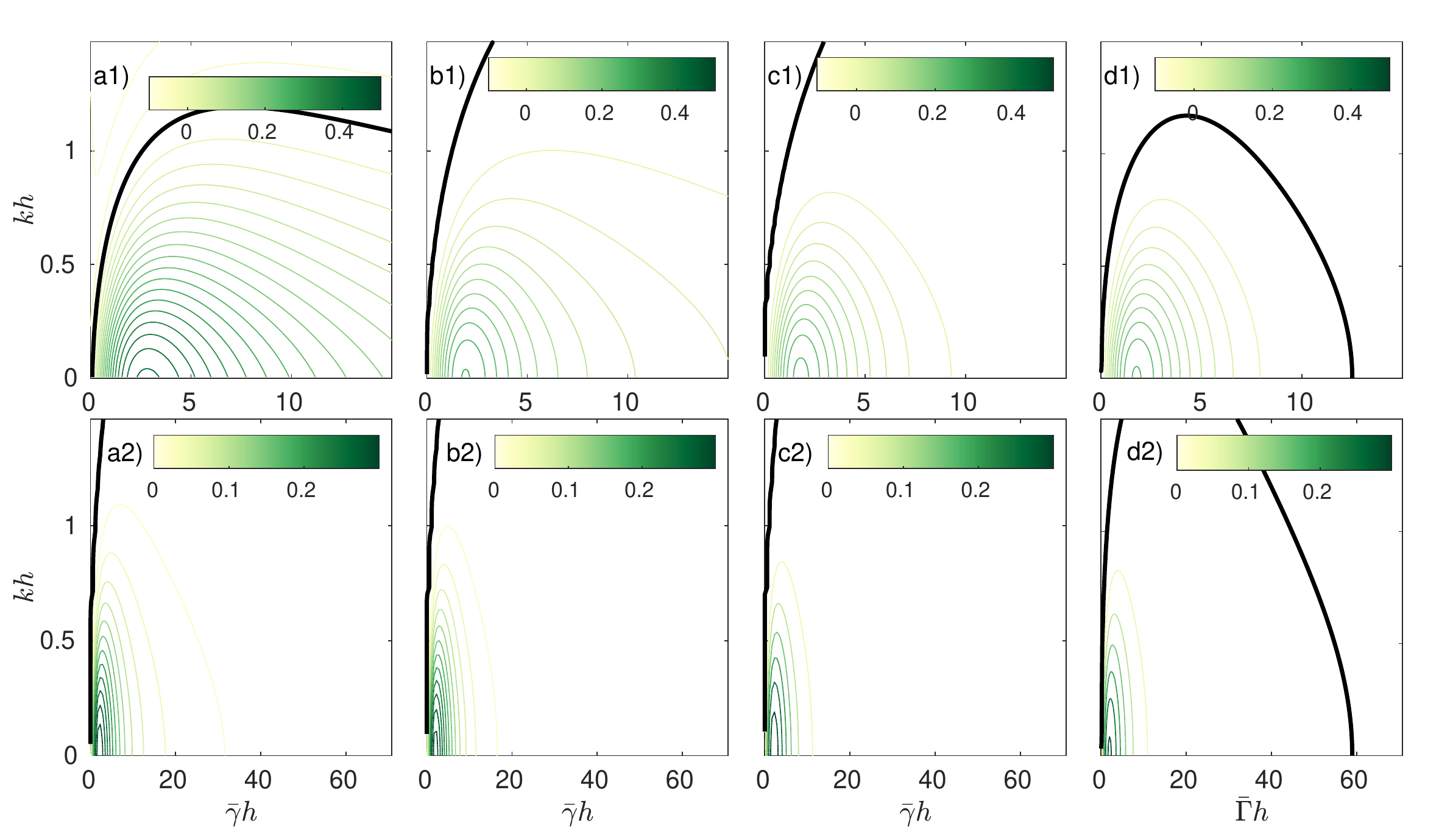}
 \caption{Each panel shows contours of $Pe \bar{u}(0) \lambda / (-\alpha \bar{\gamma}_T \bar{u}(0)^2)^2) = Pe \bar{U}_b \Lambda/(-\alpha \bar{\Gamma}_\Theta \bar{U}_b^2)^2$ against $kh$ and $\bar{\gamma h}$ with slope $\theta$, or equivalently the coefficient $Gh/(\alpha \bar{u}_(0)^2)$, adjusted as $\bar{\gamma}h$ is varied so as to ensure a constant flux $Q_0 = (h^2/(\bar{\gamma}h)+h^3/3)\theta)$; equivalently, $G h/(\alpha \bar{u}_(0)^2) = Gh^3(1+\bar{\gamma}h/3)^2/(\alpha Q_0^2)$. As always, the eigenvalue with the largest real part is shown. Panels a-c) in each row show results computied using \eqref{eq:eigval_stability2} with $\gamma_T = -50\bar{\gamma}$ (panels a1,a2) $-500\bar{\gamma}$ (panels b1,b2) and $-5000\bar{\gamma}$ (panels c1,c2), while panels d1,d2 show the asymptotic result \eqref{eq:eigval_small_delta_supplementary}. Row uses $Q_0 = 0.333$, row 2 uses $Q_0 = 1.333$; both rows use $Pe = \alpha = 1$, $G = 0.1$, $h = 0.5$. The heavy black line is the zero contour in each case } \label{fig:small_delta_flux}
\end{figure}

We can revisit the results of figure \ref{fig:gamma_flux} in the context of large temperature sensitivity, in order to explain why numerical solutions of the full, nonlinear model appear to grow noticeably at slower sliding speeds than are predicted by the asymptotic result \eqref{eq:small_delta_criterion}. Figure \ref{fig:small_delta_flux}. We again compute eigenvalues with largest real parts for fixed $Pe = \alpha = 1$, $h = 0.5$, $G = 0.1$, varying $\bar{\gamma}$ and $\theta$ simultaneously to achieve a fixed flux $Q_0 = (h^2/\bar{\gamma} + h^3/3)\theta$; a little algebra reveals that this is equivalent to ensuring that $Gh/(\alpha \bar{u}(0)^2) = Gh^3(1+\bar{\gamma}h/3)^2/(\alpha Q_0^2)$ in the asymptotic formula \eqref{eq:eigval_small_delta_supplementary}.

Each column a--c) of figure \ref{fig:small_delta_flux} corresponds to a different fixed ratios $\delta^{-1} = -\bar{\gamma}_T/\bar{\gamma}$, with temperature sensitivity increasing from left to right, while column d) shows the asymptotic result. The top row corresponds to $Q_0 = 0.333$, the bottom row to $Q_0 = 1.333$. The contours show scaled eigenvalues of $Pe \bar{u}(0) \lambda / (-\alpha \bar{\gamma}_T \bar{u}(0)^2)^2)$. In each case, there is an actual cut-off value of $\bar{\Gamma}h = \bar{\gamma}h$ above which there is no viable solution (and hence no instability) in the asymptotic model (column d), while such a cut-off is absent in the finite depth model \eqref{eq:eigval_stability2} except in panel a1) (although even there, the cut-off does not appear in the plot. However, the non-zero contours agree of the asymptotic result agree well with those for the largest finite $\bar{\gamma}_T$. As $\bar{\gamma}_T$ is reduced, the scaled eigenvalue of $Pe \bar{u}(0) \lambda / (-\alpha \bar{\gamma}_T \bar{u}(0)^2)^2)$ systematically increases, and significant values of the eigenvalue occur at progressively larger values of the friction coefficient $\bar{\gamma} h$; in practice, this means that the asymptotic formula underestimates growth at low bed temperatures, for which the friction coefficient is still large. This is evident in the plots of the variance of bed temperature against downstream position for the solutions of the nonlinear model in the main paper.

\subsection{Spatial stability for rapid sliding} \label{sec:rapid_sliding_stability}

The main paper and the previous section detail an asymptotic analysis in the limit of highly temperature-dependent sliding $\delta \ll 1$ as sketched immediately above, for which a simple analytical stability criterion can be derived. Recall that this criterion states that a minimum sliding velocity threshold must be reached in order to overcome the negative feedback associated with the drawing down of cold ice caused by accelerated sliding.

Here we investigate the alternative limit $\bar{\gamma} \ll 1$ while allowing $\delta \sim O(1)$. In this limit, sliding is rapid, but the onset of sliding occurs very gradually: we have a block of ice that moves almost entirely by sliding even at very low temperatures. That is the limit in which \citet{Hindmarsh2009} operates.  Starting with the assumption that ice flux remains of $O(1)$,  we must have $\bar{\gamma}^{-1}h^2\theta  = O(1)$ or $h \sim \bar{\gamma}^{1/2}$: a slippery bed leads to thinner ice. With $\bar{\gamma}$ small, note that we also expect $\bar{\gamma}_T \sim \delta^{-1}\bar{\gamma} \ll 1$ if $\delta = O(1)$. If we still want to maintain a bed temperature $\bar{T}(0)$ that differs from the surface temperature by an $O(1)$ amount despite a thin ice column with $h \ll 1$, we have to assume that the Brinkmann number $\alpha$ is in fact large, with $\alpha \sim \bar{\gamma}^{-1/2}$.

The form of $\eta$ in \eqref{eq:eta_def} also suggests that we consider small wavenumbers $k^2h \sim \bar{\gamma}$, since this is the appropriate scale at which $\eta$ changes sign: essentially, with rapid sliding, lateral shear stresses permit significant lateral velocity variations only at long lateral wavelengths. Using that as motivation, we rescale
$$ K = \bar{\gamma}^{-1/4} k, \qquad H = \bar{\gamma}^{-1/2} h, \qquad Z = \bar{\gamma}^{-1/2} z \qquad \bar{U} = \bar{\gamma}^{1/2}\bar{u},    $$
$$ \tilde{W} = \bar{\gamma}^{-3/2} W,  \qquad \tilde{U} = \bar{\gamma}^{1/2} U, \qquad \tilde{\Lambda} = \bar{\gamma}^{1/2} \lambda, \qquad  \tilde{\alpha} = \bar{\gamma}^{1/2}\alpha, \qquad \tilde{\eta} = \bar{\gamma}^{1/2} \eta $$
we find at leading order that
\begin{equation} \od{^2 T'}{Z^2} = \tilde{\Lambda} \Pe  \bar{U} T' \end{equation}
subject to
\begin{equation} \od{T'}{Z} = -\tilde{\eta}T \quad \mbox{ at } Z = 0, \qquad T' = 0 \quad \mbox{ at } z = 1. \label{eq:Sturm_Liouville_fast_sliding} \end{equation}
where
$$ \tilde{\eta} = -\tilde{\alpha} \frac{\bar{\gamma}_T}{\bar{\gamma}}\bar{U}^2\frac{1-K^2H}{1+K^2H}, \qquad \bar{U} = H\theta $$
are both of $O(1)$. This has solution
$$ T'(z) = A \sinh\left(\sqrt{Pe \, \bar{U} \tilde{\Lambda}}(Z-H)\right), $$
and the boundary condition at $Z = 0$ requires
\begin{equation} \sqrt{Pe \,  \bar{U} \tilde{\Lambda}}H \cosh\left(\sqrt{Pe \, \bar{U} \tilde{\Lambda}}H\right) = \tilde{\eta}H \sinh \left(\sqrt{Pe \, \bar{U} \tilde{\Lambda}} H\right) \label{eq:fast_slide_asymptotic} \end{equation}
Note that \eqref{eq:Sturm_Liouville_fast_sliding} is of Sturm-Liouville form and has real eigenvalues, so the only question to answer here is whether these are positive, in which case the root $\sqrt{Pe\, \bar{U} \tilde{\Lambda}}$ is real and non-zero. Straightforwardly, for an equation of the form $y \cosh(y) = c \sinh(y)$, we find real non-zero roots when $c > 1$. Equivalently, we find positive $\tilde{\Lambda}$ when
$$ \tilde{\eta}H > 1. $$
As in the case of $\bar{\gamma} \sim O(1)$, $\delta \ll 1$ treated in the main paper; instability will first occur at small $K$, where $\tilde{\eta}$ is smallest. The instability for $\bar{\gamma} \ll 1$, $\delta \sim O(1)$ first occurs when
\begin{equation} \label{eq:fast_sliding_criterion} -\frac{\tilde{\alpha} \bar{\gamma}_T H \bar{U}^2}{\bar{\gamma}} > 1, \end{equation}
or, in the original notation, when
\begin{equation} -\alpha \bar{\gamma}_T h \bar{u}^2  > 1. \end{equation}
The stability problem here differs fundamentally from that for  $\bar{\gamma} \sim O(1)$, $\delta \ll 1$  of the main paper: \eqref{eq:Sturm_Liouville_fast_sliding} does not include a downward advection term at leading order, whereas downward advection is the key control on stability in the limit $\bar{\gamma} \sim O(1)$, $\delta \ll 1$ considered in the main paper. Instead, stabilization here occurs due to diffusion of heat across the thickness of the ice, instead of being confined to a boundary layer. As in the main paper, we find a velocity threshold for instability, now of the form $\bar{U} > \sqrt{\bar{\gamma}/(\tilde{\alpha} \bar{\gamma}_T H)}$. If we take the limit of large $\bar{\gamma}_T$ \emph{after} imposing $\bar{\gamma} \ll 1$, equivalent to putting $\bar{\gamma} \ll \delta \ll 1$, then we see that the instability criterion \eqref{eq:fast_sliding_criterion} is automatically satisifed provided the sliding velocity $\bar{U}$ is not small: essentially, for fast sliding and basal friction that is moderately sensitive to temperature, we expect the instability to occur unconditionally. At issue then is only how fast the instability grows in the downstram direction, and that depends critically on the advection rate $Pe \tilde{U}$: the faster that advection rate, the smaller $\tilde{\Lambda}$. In a system of finite extent, this leaves the possibility that growth may occur, but too slowly to allow an ice stream pattern to emerge. fully.

By contrast, for the limit $\delta \ll 1$ considered in the main paper, we found that instability occurs when $G < 3 \alpha \bar{u}(0)^2/h$  in the original notation used in section \ref{sec:supplementary_stability_analysis}. This result, which holds provided $\delta \ll \bar{\gamma}$, is not affected by making $\bar{\gamma}$ small. As a result, we have a case of non-commuting limits, and have to view fast sliding and a highly temperature-sensitive friction law as distinct cases. This is particularly relevant when we contrast the onset of spatial instability with that for temporal instability: recall that for $\delta \ll 1$ the system is more prone to temporal than spatial instability, and it is unlikely that the steady spatial patterns computed from the nonlinear model in the main can persist over time. That situation may differ from the case of small $\bar{\gamma}$ and $\delta \sim O(1)$, which was investigated by \citet{Hindmarsh2009}, whose numerical solutions approach a spatially patterned steady state for large times.

\begin{figure}
 \centering
 \includegraphics[width=\textwidth]{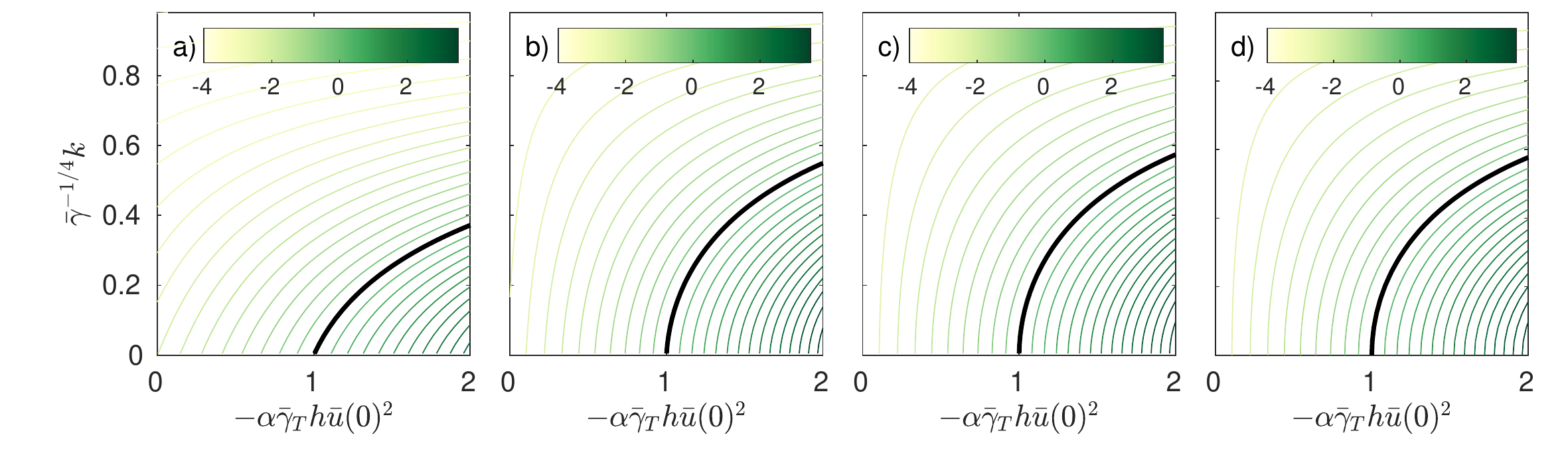}
 \caption{Contours of $Pe h^2 \bar{u}(0) \lambda = Pe \tilde{U} \tilde{\Lambda}$ against $\tilde{\eta}H = -\alpha \bar{\gamma}_T \bar{u}(0)^2$, and $\sqrt{H}K = \bar{\gamma}^{-1/2} h^{1/2}k$, showing the eigenvalue with largest real part in each case. The first three panels use different values of $\bar{\gamma} = 2$ (a), $1/8$ (b) $1/128$ (c), with $Pe = \theta = 1$, $G = 0.1$ and $\alpha  =\bar{\gamma}^{-1/2}$, $h = 0.5\bar{\gamma}^{1/2}$. Panel d) shows the result computed from  \eqref{eq:fast_slide_asymptotic}.} \label{fig:fast_sliding}
\end{figure}

We can confirm the asymptotic result above against solutions of \eqref{eq:eigval_stability2}. Figure \ref{fig:fast_sliding} plots contours of $Pe h^2 \bar{u}(0) \lambda$ against $\bar{\gamma}^{-1/2}h^{1/2}k$ and $-\alpha \bar{\gamma}_T h \bar{u}(0)^2$. Panels a--c) show results computed using \eqref{eq:eigval_stability2} with $Pe = \theta = 1$, $G = 0.1$, $\alpha  =\bar{\gamma}^{-1/2}$, $h = 0.5\bar{\gamma}^{1/2}$, with $\bar{\gamma} = 2$ (panel a), $1/8$ (b) and 128 (c)  Panel d) shows the largest eigenvalue computed computed from the asymptotic formula \eqref{eq:fast_slide_asymptotic}  Note that panels a)--c) are essentially rescaled versions of figure \ref{fig:gammaprime}.

Positive eigenvalues occur when $-\alpha \bar{\gamma}_T h \bar{u}(0)^2 > 1$ as predicted by the asymptotic result \eqref{eq:fast_sliding_criterion}. Note that this remains true even when $\bar{\gamma} = 2$ is not exactly small. This also explains why instability occurs for the same value of $\bar{\gamma}_T$ in each plot in figure \ref{fig:gammaprime}. The range of wavenumbers for which instability occurs and the growth rates that result does depend on $\bar{\gamma}$, however, and close agreement with the asymptotic results is only achieved for the smallest value of $\bar{\gamma}$ used, in panel c).

For fast sliding, we can also rewrite the instability criterion  $-\alpha \bar{\gamma}_T h \bar{u}(0)^2 > 1$ in alternative ways that make it easier to understand where along a flow path instability might first appear. Suppose as above that the friction coefficient $\bar{\gamma}$ and its temperature sensitivity $\bar{\gamma}_T$ vary along a flow path, but the ratio $\delta^{-1} = -\bar{\gamma}_T/\bar{\gamma}$ is fixed. Recognizing also that, with fast sliding, ice flux is $Q = h\bar{u}(0)$ and $\bar{u}(0) = h\theta/\bar{\gamma}$, we find instability when
$$  \alpha \delta^{-1}Q h \theta > 1; $$
instability is favoured by greater temperature sensitivity, larger ice fluxes, as well as thicker and steeper ice (the last three not being mutually independent, of course). For fixed $\delta$, we expect instability to appear first (if at all) at some distance from the ice divide, when ice flux $Q$ becomes sufficiently large.

\subsection{Spatial instability for temperate sliding} \label{sec:hydraulic_supplementary}

We can also consider a base state analogous to \eqref{eq:steady_uniform}, but with a temperate bed, so $\bar{T}(0) = 0$. Using the same notation as above, we end up with
\begin{subequations} \label{eq:temperate_uniform_supplementary}
 \begin{align}
  \bar{u} = & \frac{1}{2}\left[h^2-(h-z)^2\right]\theta + \bar{u}(0)  \\
  \bar{T}(z) = & -\frac{\alpha\theta^2}{12}\left[(h-z)^3-h^3\right](h-z) - \frac{z}{h}
 \end{align}
\end{subequations}
where $\bar{u}(0)$ and the steady state proxy variable $\bar{\Pi}$ for basal effective pressure are defined implicitly through
$$ f(0,\mathcal{N}(\bar{\Pi}),|\bar{u}(0)|)\sgn(\bar{u}(0)) = h\theta, \qquad G + \alpha h \theta \bar{u}(0) = \frac{1}{h} - \frac{\alpha \theta^2}{4}h^3. $$

We again perturb as
$$ u = \bar{u}(z) + u'(z)\exp(\lambda x + i k y), \qquad v = v'(z)\exp(\lambda x + i k y), \qquad w = w'(z)\exp(\lambda x + i k y), $$ 
\begin{equation} \label{eq:temperate_perturbation}  p = p' \exp(\lambda x + i k y), \qquad T = \bar{T} + T'(z)\exp(\lambda x + i k y), \qquad \Pi = \bar{\Pi} + \Pi' \exp(\lambda x + i k y),  \end{equation}
and linearize. Defining
$$ \bar{f} = f(0,\mathcal{N}(\bar{\Pi}),|\bar{u}(0)|), \qquad f_\Pi = \left.\pd{f}{N}\pd{\mathcal{N}}{\Pi} \right|_{(T,\Pi,u) = (0,\bar{\Pi},\bar{u}(0)}, \qquad f_u = \left.\pd{f}{u} \right|_{(T,\Pi,u) = (0,\bar{\Pi},\bar{u}(0)}, $$
$$ \bar{\kappa} = \tilde{\kappa}(\bar{\Pi}), \qquad \kappa_\Pi = \left.\pd{\tilde{\kappa}}{\Pi}\right|_{\Pi = \bar{\Pi}}. $$
the linearized model becomes, for $k \neq 0$,
$$ \left(\od{^2}{z^2} - k^2\right) u' =  0, \qquad \left(\od{^2}{z^2} - k^2\right) v' - i k p' = 0, \qquad \left(\od{^2}{z^2} - k^2\right) w' - \pd{p'}{z} = 0, $$
\begin{equation} i k v' + \pd{w'}{z} = -\lambda u', \qquad  Pe\left(\lambda \bar{u} T' + w'\od{\bar{T}}{z}\right) - \left(\od{^2}{z^2} - k^2 \right)  T' =  2 \alpha \od{\bar{u}}{z} \od{u'}{z}, \label{eq:temperate_linear_supplementary} \end{equation}
on $0 < z < h$,
\begin{equation}  \left(\od{^2}{z^2} - k^2 \right)  T' = 0 \end{equation}
for $z < 0$, with boundary conditions
\begin{equation} \pd{u'}{z} = 0, \qquad \pd{v'}{z}+i k w' = 0, \qquad w' = 0, \qquad T' = 0 \end{equation}
at $z = h$, 
$$  \pd{u'}{z} = f_u u' + f_\Pi \Pi', \qquad \od{v'}{z} = \bar{f}\bar{u}(0)^{-1}  v', \qquad w' = 0, \qquad T' = 0, \qquad \sigma_{nn}' = p - 2\od{w}{z} $$
\begin{equation}  \lambda r^{-1} \theta \kappa_\Pi \Pi' +\beta  k^2 \kappa_2 \Pi' + k^2 \bar{\kappa} \sigma_{nn}' =   \left[\od{T'}{z}\right]_-^+ +  \alpha \left[\bar{f} + f_u\bar{u}(0)\right] u' +\alpha f_\Pi \bar{u}(0) \Pi',  \end{equation}
at $z = 0$, and $\pdl{T'}{z} \rightarrow 0$ as $z \rightarrow -\infty$.

Following the same scheme as for the spatial stability analysis for subtemperate sliding, we can solve once more for velocity perturbations as well as the normal stress perturbation at the bed. These are now proprtional to the perturbation $\Pi'$,
\begin{equation} u' = \breve{U} \Pi', \qquad w' = \lambda \breve{W} \Pi', \qquad \sigma_{nn} = \lambda \Sigma_{nn} \Pi' \end{equation}
where
\begin{align} 
 \breve{U} = & -\frac{\cosh[k(h-z)]}{k\sinh(kh) + f_u\cosh(kh)} f_\Pi, \label{eq:U_scalefun_supplementary} \\
\breve{W} = & -  \frac{h\sinh(kz)-z \sinh(kh) \cosh[k(h-z)]}{2 k \sinh^2 (kh) + \bar{f}\bar{u}(0)^{-1}\left[\sinh(kh)\cosh(kh) - kh\right]} \chi f_\Pi , \label{eq:W_scalefun_supplementary} \\
 \Sigma_{nn} = & \frac{2 k h}{2 k \sinh^2 (kh) + \bar{f}\bar{u}(0)^{-1}\left[\sinh(kh)\cosh(kh) - kh\right]} \chi f_\Pi
\end{align}
and
\begin{equation} \chi = \frac{k \sinh(k h) + \bar{f}\bar{u}(0)^{-1} \cosh(kh)}{k \sinh(kh) + f_u \cosh(kh).} \end{equation}
In addition, we have $ T'(z) = 0 $ for $z < 0$.
By analogy with the stability problem for a subtemperate bed in the main paper, we can reduce \eqref{eq:temperate_linear_supplementary} to
\begin{equation} \left(\od{^2}{z^2} - k^2 \right)  T' =  \lambda Pe \, \bar{u} T' +  \left(  \lambda Pe \, \breve{W}\od{\bar{T}}{z} - 2 \alpha \od{\bar{u}}{z} \od{\breve{U}}{z}  \right) \Pi' \label{eq:eigval_temperate}  \end{equation}
subject to $T' = 0$ at $z = 0,\,h$, and
\begin{equation} \lambda \left(r^{-1} \theta \kappa_\Pi  + k^2 \bar{\kappa}\Sigma_{nn}\right) \Pi' +  k^2 \beta \kappa_2  \Pi' =   \breve{\eta} \Pi' + \left.\od{T'}{z}\right|_{z = 0} \label{eq:reaction_diffusion}
\end{equation}
where
\begin{equation} \breve{\eta} = -  \alpha f_\Pi \bar{u}(0) \frac{ \bar{f}\bar{u}(0)^{-1}\cosh(kh)-k\sinh(kh)}{f_u \cosh(kh) + k \sinh(kh)} \label{eq:eta_def_supplementary} \end{equation}

Before we proceed further, we have to note a feature of \eqref{eq:reaction_diffusion} that can render the linearized problem ill-posed: the coefficient of $\lambda$ can vanish, which implies an infinite eigenvalue: this is effectively an odd form of resonance. Equation \eqref{eq:reaction_diffusion} is similar to a reaction-diffusion problem, but the effective (wavelength-dependent) `heat capacity' $\theta \kappa_\Pi + \bar{\kappa} k^2\Sigma_{nn}$ can vanish, or even become negative at certain small wavenumbers $k$. We have  $\kappa_\Pi > 0$ and $f_\Pi < 0$ since  $\mathcal{N}(\Pi)$ decreases as $\Pi$ increases, and permeability as well as friction decrease with increasing effective pressure. It is therefore straightforward to see that $\bar{\kappa}k^2\Sigma_{nn} < 0$ is an increasing function of $k$ that behaves as 
$$ \bar{\kappa} k^2 \Sigma_{nn} \sim \frac{\bar{\kappa} \bar{f}\bar{u}(0)^{-1}f_\Pi }{f_uh(1 + \bar{f}\bar{u}(0)^{-1} h/3)} $$
Consequently, we expect that $\theta \kappa_\Pi + \bar{\kappa} k^2\Sigma_{nn}$ vanishes for finite $k$ if
\begin{equation} r^{-1}\theta \kappa_\Pi +\frac{\bar{\kappa} \bar{f}\bar{u}(0)^{-1}f_\Pi }{f_uh(1 + \bar{f}\bar{u}(0)^{-1} h/3)} < 0, \end{equation}
becoming negative at small $k$. We give a more refined account of this problem shortly, but first describe a numerical method for solving the eigenvalue problem.

Ignoring this complication for a moment, we re-write the stability problem as an integral equation, using the Green's function
\begin{equation} \breve{G}(z,z') =  \left\{ \begin{array}{l l} \theta_2(z)\theta_1(z') & z' < z, \\
\theta_1(z)\theta_2(z') & z' \geq z,
                       \end{array}\right. \qquad \theta_1(z) = \frac{\sinh(kz)}{k\sinh(kh)}, \qquad \theta_2(z) = -\sinh[k(h-z)], \end{equation}
The eigenvalue problem \eqref{eq:eigval_temperate} can be written in the same fashion as its counterpart for subtemperate sliding,
\begin{align} T'(z) = &  \lambda \int_0^h Pe \, \breve{G}(z,z') \bar{u}(z')  T'(z') \rd z' + \lambda\int_0^h Pe\, \breve{G}(z,z')\breve{W}(z') \left.\od{\bar{T}}{z}\right|_{z'} \rd z' \Pi' \nonumber \\
&  - \int_0^h 2\alpha \breve{G}(z,z')\left.\left( \od{\bar{u}}{z}\od{\breve{U}}{z}\right)\right|_{z'} \rd z' \Pi' \label{eq:eigval_stability2_temperate}
\end{align}
and \eqref{eq:reaction_diffusion} becomes
\begin{align} \lambda r^{-1} \left(\theta \kappa_\Pi + k^2 \Sigma_{nn} \bar{\kappa}\right)\Pi' +  k^2 \beta \kappa_2  \Pi' = &  \breve{\eta} \Pi' +  \lambda \int_0^h Pe \, \left.\pd{\breve{G}}{z}\right|_{(0,z')} \bar{u}(z')  T'(z') \rd z' \nonumber \\
 &+ \lambda\int_0^h Pe\,\left. \pd{\breve{G}}{z}\right|_{(0,z')}W(z') \left.\od{\bar{T}}{z}\right|_{z'} \rd z' \Pi'  \\ 
 & - \int_0^h 2\alpha \left. \pd{\breve{G}}{z}\right|_{(0,z')}\left.\left( \od{\bar{u}}{z}\od{U}{z}\right)\right|_{z'} \rd z' \Pi'
\end{align}
We discretize this analogously to the problem for subtemperate sliding, using $n$ Gaussian integration nodes in the domain $(0,h)$ and an additional degree of freedom to represent $\Pi'$.

\begin{figure}
 \centering
 \includegraphics[width=\textwidth]{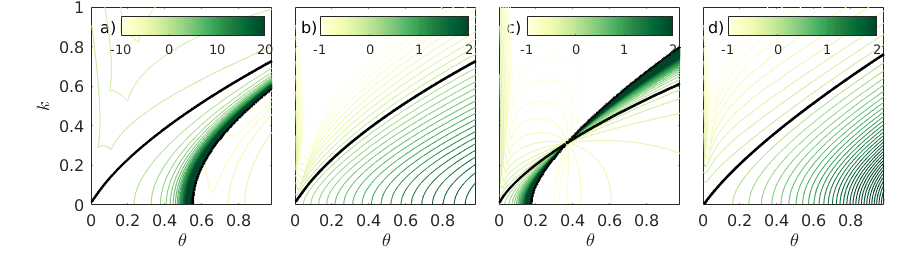}
 \caption{Contours of $\lambda$ against $\theta$ and $k$, showing the eigenvalue with largest real part in each case. In each case, $h$ and $\bar{\Pi}$ are varied to maintain a fixed flux $Q_0$, based on a sliding law $f(u,N) = uN$ with $N = \Pi^{1/(k_0-1)}$ for constant $k_0$, and  $\tilde{\kappa}(\Pi) = \Pi^{k_0/(k_0-1)}$ and $\kappa_2 = 1/(k_0-1)$. $Pe = \alpha = \beta = 1$, $G = 0.1$, in all cacluations. Panel a) has $Q_0 = 0.9833$, $k_0 = 1.3333$, b) uses the same parameter values as a) but omits the normal stress term $\sigma_{nn}$ in the calculation, c) uses $Q_0 = 3.8208$  and $k_0 = 1.3333$, d) uses $Q_0 = 0.9833$, $k_0 = 3$. Black lines are either zero contours or (where green contours are closely bunched) singularities in the spectrum.} \label{fig:hydraulic_supp}
\end{figure}

For the temperate sliding problem, it is even harder to emulate the onset of instability along a flowline in a meaningful way, as the solution being perturbed really is not the parallel-side slab \eqref{eq:temperate_uniform_supplementary}. In particular, once the bed is temperate, we expect to find a positive basal energy balance, with the water flux $q_x$ increasing as we move downstream. Simply for the sake of illustration, we again vary the slope $\theta$ while allowing $h$ and $\bar{\Pi}$ to vary simultaneously in such a way as to maintain a constant flux $Q_0$ and also maintain energy balance at the bed; we assume that $\bar{\Pi}$ is determined by a friction law of the form $f(N,u) = uN$ with $N = \mathcal{N}(\Pi) = \Pi^{1/(k_0-1)}$, while $\tilde{\kappa}(\Pi) = \Pi^{k_0/(k_0-1)}$ and $\kappa_2 = 1/(k_0-1)$.

Figure \ref{fig:hydraulic_supp} shows contours of the eigenvalues with largest real parts against $\theta$ and $k$. Panel a) uses $Q_0 = 0.9833$, $Pe = \alpha = \beta = 1$, $G = 0.1$, $k_0 1.3333$. Panel b) uses the same parameter values, but omits the normal stress term $\sigma_{nn}$ from the calculation. Panel c) uses $Q_0 = 3.8208$ but retains all other parameter values, while panel d) reverts to $Q_0 = 0.9833$ but uses $k_0 = 3$.

A singularity in the spectrum is clearly visible in panels a) and c) as a solid black line near one side of which contours are closely bunched. In addition to the singularity, there is a regular stability boundary at some finite $k$, and the locus of the singularity can cross that stability boundary as in panel c). The calculation in panel b), omitting the effect of normal stress $\sigma_{nn}$ on drainage, shows $k = 0$ as the fastest growing wavenumber as in the corresponding instability in the case of subtemperate sliding, and a stability boundary at a finite $k$ that closely matches the corresponding stability boundary in panel a) (which has otherwise identical parameter values). Note that the singularity in the spectrum first appears at some finite, critical slope $\theta$, and is either absent or off-plot in panel d.

The singularity in $\lambda$ at finite $k$ that we find for certain parameter combinations deserves closer examination. As $\lambda \rightarrow \infty$, the appropriate balance in the heat equation \eqref{eq:eigval_temperate} is that
$$ Pe \, \bar{u} T' + Pe \, \breve{W} \od{\bar{T}}{z} \Pi' \sim 0 $$
and hence
$$ T' \sim - \bar{u}^{-1} \breve{W} \od{\bar{T}}{z} \Pi'. $$
At the bed, this implies
$$ \left. \od{T'}{z}\right|_{z = 0} \sim \frac{\breve{W}_{z0} Q_0}{\bar{u}(0)} \Pi' $$
where
$$ \breve{W}_{z0} = \left.\od{\breve{W}}{z}\right|_{z=0}, \qquad Q_0 = - \left.\od{\bar{T}}{z} \right|_{z = 0}. $$
Equation \eqref{eq:reaction_diffusion} becomes
$$ \lambda \left(r^{-1} \theta \kappa_\Pi + k^2 \bar{\kappa} \Sigma_{nn} \right) \sim \breve{\eta} - k^2 \beta \kappa_2  + \frac{\breve{W}_{z0} Q_0}{\bar{u}(0)} .  $$
If the `heat capacity' term $r^{-1} \theta \kappa_\Pi + k^2 \bar{\kappa} \Sigma_{nn}$ is positive, we obtain a modified version of the stability criterion for small $\delta$ for the case of subtemperate sliding,
\begin{equation} \label{eq:temperate_stability_criterion} \breve{\eta} + \frac{\breve{W}_{z0} Q_0}{\bar{u}(0)} - k^2 \beta \kappa_2 > 0. \end{equation}
Recall that $\breve{\eta}(k)$ represents the net feedback between increased basal effective pressure and dissipation, which now has to overcome not only the stabilizing effect of downward advection of cold ice represented by $\breve{W}_{z0}(k) Q_0/\bar{u}(0)$, but also of lateral drainage of water $-k^2 \beta \kappa_2$. In particular, stabilization need not occur because $\breve{\eta}$ becomes sufficiently small for larger $k$ due to the effect of lateral shear stresses as discussed in section 4 of the main paper: it can occur simply because the effect of diffusivity in the basal drainage system as envisaged by \citet{FowlerJohnson1996}, who omit the effect of lateral shearing entirely from their model; the analysis here is closer to the model in \citet{KyrkeSmithetal2014,KyrkeSmithetal2015} in that the latter retain a more complete representation of stresses in the ice, although they (as in the main paper here) omit the effect of dynamic normal stress variations $\sigma_{nn}$ on pattern formation.

The latter can clearly be important, and in fact raises questions over the validity of the model for certain parameter combinations. The instability becomes rapid when $r^{-1} \theta \kappa_\Pi + k^2 \bar{\kappa} \Sigma_{nn}$ is small because the second term (representing the effect of normal stress on lateral drainage) is negative and makes the sum small. More importantly, the instability criterion \eqref{eq:temperate_stability_criterion} can be reversed entirely if $r^{-1} \theta \kappa_\Pi + k^2 \bar{\kappa} \Sigma_{nn}$ becomes negative. The possibility of $\theta \kappa_\Pi + \bar{\kappa} k^2\Sigma_{nn}$ changing sign for some wavelength $k$ is a real physical effect, associated with the secondary flow in the transverse plane. If velocity $u$ increases with $x$ in some part of the domain, a transverse flow has to compensate, and that transverse flow must be driven by a transverse pressure gradient. The latter also expresses itself at the bed as a transverse gradient of normal stress $\sigma_{nn}$, and the latter will tend to drive flow of water into the region of accelerated axial velocity, acting seemingly as a positive feedback. The strength of the transverse pressure gradient however is itself proportional to the acceleration of the axial flow: instead of leading positive but bounded growth rates, the latter fact can lead to infinite growth rates at the wavelength $k$ for which $r^{-1}\theta \kappa_\Pi + \bar{\kappa} k^2\Sigma_{nn} = 0$, and counterintuitively lead to negative growth rates where $r^{-1}\theta \kappa_\Pi + \bar{\kappa} k^2\Sigma_{nn}$ is itself negative. Conversely, it can lead to instability at small $k$ even though the system is stable in the absence of the normal stress term in the definition of $q_y$, and that instability is then purely caused by lateral drainage forced by the same surface topography that drives lateral ice flow towards regions of faster sliding. These possibilities are clearly visible in the numerical results above.

Note that this behaviour is suppressed in the numerical results of the nonlinear problem in the main paper because we have simply omitted the normal stress term $\sigma_{nn}$ from the computation of the lateral flux $q_y$. Future work will need to address this omission.

\subsection{Instability at zero wavenumber} \label{sec:zero_wavenumber_supplementary}

In the analysis above, we have explicitly excluded consideration of the zero wavenumber case $k = 0$, since the latter requires us to allow for perturbations in ice thickness $h$ (with $h$ having no dependence on $y$, any perturbations in $h$ cannot admit a mode of the form $h' \exp( \lambda x + i k y)$ unless $k = 0$.

In this section, we briefly consider the case of $k = 0$, restricting ourselves to the simpler case of a large P\'eclet number $\Pe \ll 1$. The point is that the $k = 0$ is \emph{always} unstable, and that instability does not even rely on the basal friction coefficient being dependent on $T$ or any other variable; the 'instability' is simply a reflection of the fact that, with a given flux $Q$, a perturbation in $h$ requires a perturbation in slope $\pdl{h}{x}$ that cases the surface perturbation in $h$ to increase in size in the downstream direction: this corresponds to the familiar observation that ice sheets, like other thin film flows, typically steepen as they approach the edge of the fluid, which requires the shape of the flow to depart from a parallel-sided slab as the edge is approached.  This ``instability'' is, then, quite natural, and underscores the fact that the construction of a spatial instability is different from the usual temporal instability the reader may be tempted to rely on for intuition. The point of the spatial stability analysis is that, while the parallel-sided slab investigated here is invariably unstable to geometrical departures from its slab-like shape in the downstream direction, but it will not break lateral symmetry without some positive feedback that requires temperature-dependent or water-pressure-dependent sliding.

In the large P\'eclet number limit, we can reduce the temperature equation to a pure advection problem, and basal heat flux $q_b = -\pdl{T}{z}|_{z=0}$ satisfies
$$ \left.\pd{u}{z} \right|_{z = 0} \pd{T_b}{x}  - u_b \pd{q_b}{x} + \pd{u}{x} q_b = 0, $$
where $T_b = T|_{z=0}$; this was referred to as the \emph{$Q$-equation} in  \citep{Mantellietal2019}, amended here to allow for temperature gradients $\pdl{T_b}{x}$ along the bed. In addition, omitting lateral perturbations, we have
\begin{align*} \left.\pd{u}{z} \right|_{z = 0} =\tau_b = & \gamma(T_b) u_b = h\left(\theta-\pd{h}{x}\right), \\
 Q = & u_b h + \frac{1}{3} h^3 \left(\theta - \pd{h}{x}\right), \\
 q_b = & \alpha \tau_b u_b - G,
\end{align*}
where we employ the same notation as in section \ref{sec:supplementary_stability_analysis}. We again denote the parallel-sided slab solution by by overbars, and perturbations by primes, as $T_b = \bar{T}_b + T_b' \exp(\lambda x)$, $u_b = \bar{u}_b + u_b' \exp(\lambda x)$ etc. This leads to the linearized problem
\begin{align} 0 = &\bar{\tau}_b T_b' + \bar{u}_b q_b'+ \bar{q}_b u_b ' \\
 \tau_b' = & \tau_{bT} T_b' + \tau_{bu} u_b' \nonumber \\
 = & \theta h'  - \lambda  \bar{h} h' \\
 0 = & u_b' \bar{h} + \bar{u}_b h' + Q_{sh} h' - \lambda q_{s\theta} h' \\
 q_b ' = & \alpha\bar{u}_b \tau_b' + \alpha \bar{\tau}_b u_b'
\end{align}
where $\tau_{bT} = \gamma_T(\bar{T}_b) \bar{u}_b$, $\tau_{bu} = \gamma(\bar{T}_b)$, $Q_{sh} = \bar{h}^2 \bar{\theta}$ and $Q_{s\theta} = h^3/3$ are the derivatives of basal friction and ice flux due to shearing with respect to basal temperature, basal velocity, ice thickness and surface slope, respectively, all evaluated at the uniform base state.

The solution for growth rate $\lambda$ is
\begin{equation} \lambda = \frac{(\bar{\tau}_b - \alpha \bar{u}_b^2 \tau_{bT})(\bar{u}_b + q_{sh}) + \bar{h}\theta}{h^2 + q_{s\theta}(\bar{\tau}_b - \alpha \bar{u}_b^2 \tau_{bT})}; \end{equation} 
note that since $\tau_{bT} < 0$ and all other quantities are positive, $\lambda$ is invariably positive: that in fact remains the case even if we make basal sliding independent of temperature with $\tau_{bT} = 0$. As advertised, the $k = 0$ mode (without lateral structure) is unconditionally unstable, at least in the limit of a large P\'eclet number.

\subsection{Lack of wavelength selection}

The stability results above and in the main paper underline one basic fact: in the confines of the model constructed in section \ref{sec:model_construction}, we invariably find that when there is instability, spatial or temporal, the fastest growing transverse wavenumbers (denoted $k$ or $k_y$) are zero. Strictly speaking, since the $k = 0$ mode differs intrinsically from non-zero wavenumbers, the fastest growth rates occur in the limit $k \rightarrow 0$ or $k_y \rightarrow 0$. In other words, the model allows for instability at asymptotically large wavelengths, and does not select for a finite wavelength in pattern formation. This is also the reason why our numerical solutions invariably show a single wavelength per domain width.

The fact that the model above permits features with unbounded wavelengths to grow unstably does however not imply that this is true in reality, merely that the model cannot capture the mechanisms that suppress the instability at long lateral length scales. One of the reason for the lack of a long-wavelength cut-off is that the model allows for excessively large secondary flow velocities $v$ at large transverse wavelengths.  The model has transverse distances scaled to ice thickness, and along-flow distances to the length of the ice sheet. Based on a leading-order balance for such short transverse length scales, the secondary flow is then required to balance any local increase in axial velocity $u$ in the along-flow direction $x$, so as to keep the surface elevation $h+b$ independent of $y$. The transverse flow velocity scales as transverse wavelength for the same axial rate of speed-up $\pdl{u}{x}$, and requires lateral surface slopes represented by the diagnostic correction term $h_1$ defined in equation \eqref{eq:surface_correction_supp} (denoted by $s_1$ in the main paper).

For very long transverse wavelegnths, the lateral slopes required by the increasing rapid transverse flow become comparable to slopes in the along-flow direction, violating the assumptions behind our leading order model (see section \ref{sec:model_construction} in this supplementary material). For instance, once the slopes are comparable, the pressure gradient $\pdl{(h+b)}{x}$ driving flow in the axial direction can no longer be treated as being independent of transverse position $y$.

Next, we can ask how the excessively large lateral velocities allowed for by our transverse flow model at large wavelengths facilitate instability. This is easiest to do in the context of the limit $\delta \ll 1$ considered in the main paper and in section
\ref{sec:small_delta_revisit}, for which the stability analysis is particularly simple. Recall that the growth rate $\Lambda$ for spatial pattern growth is given by
$$ \Lambda = \frac{1}{Pe \bar{u}(0)} \left( \eta_0 + \frac{W_{z0} Q_0}{\bar{u}(0)}\right)^2, $$
with the requirement that $\eta_0 + W_{z0}Q_0/\bar{u}(0) > 0$. $\eta_0$ is a decreasing function of $k$, positive for $k = 0$ and becoming negative at large $k$. The behaviour of $W_{z0}$ is key to the instability: it represents the stabilizing effect of cold ice being pulled down towards the bed by an accelerating flow.

Consider this in the limit of large wavelengths. Recall that
\begin{subequations}
\begin{align} W_{z0} = & \delta \left.\pd{w'}{z}\right|_{z=0} \lambda^{-1}T'(0)^{-1} =   \left.\pd{w'}{z}\right|_{z=0} \Lambda^{-1} \Theta'(0)^{-1}, \label{eq:supplementary_advection_sensitivity}  \\
 \eta_0 = & \delta \alpha [2\bar{\gamma}\bar{u}(0)u'(0) + \bar{\gamma}_T T'(0) \bar{u}(0)^2 ]  T'(0)^{-1}
\end{align}
are the sensitivities of vertical advection near the bed and basal dissipation to basal temperature perturbations for a given wavenumber $k$, where $\Lambda = \delta^2 \lambda$ and $\Theta' = \delta^{-1} T'$ are the rescaled spatial growth rate and temperature perturbation in the basal boundary layer. Note also that for small $k$, lateral shear stresses are insignificant in force balance at the bed, and the net perturbation in basal shear stress vanishes at leading order, so
$$ \bar{\gamma} u'(0) + \bar{\gamma}_T T'(0) \bar{u}(0) = 0, $$
or, in rescaled terms
\begin{equation} \label{eq:supplementary_basal_sliding_sensitivity}  u'(0) \sim -\delta \bar{\gamma}_T \bar{u}(0)/\bar{\gamma} \Theta'(0). \end{equation}
By the same token,
\begin{equation} \eta_0 \sim -\delta \bar{\gamma}_T \alpha \bar{u}(0)^2 \label{eq:supplementary_dissipation} \end{equation}
for small $k$. 
\end{subequations}

Now suppose that there were no lateral flow involved in the the patterning instability for large transverse wavelengths.  In that case, vertical strain rate would have to be $\pdl{w}{z} = -\pdl{u}{x}$, and in the linearized model, this would become $\rd w'/\rd z \sim -\Lambda u'$. Evaluating this at $z = 0$ and using the relationship \eqref{eq:supplementary_basal_sliding_sensitivity} in \eqref{eq:supplementary_advection_sensitivity} this equivalent to  
$$W_{z0} \sim \delta \bar{\gamma}_T \bar{u}(0)/\bar{\gamma}.$$
With $Q_0 = \alpha \bar{\gamma} \bar{u}(0)^2 + G$, we would then have
$$ \eta_0 + \frac{W_{z0}Q_0}{\bar{u}(0)} =   \frac{\delta \bar{\gamma}_T G}{\bar{\gamma}}  < 0, $$
and long wavelengths would always be stable. In reality however, the formula \eqref{eq:small_delta_advection_coefficient} reduces to
$$W_{z0} \sim \delta \bar{\gamma}_T \bar{u}(0) /(\bar{\gamma} + 3/h)$$
for small $k$. The downdraw rate of ice $W_{z0}$ is therefore smaller in magnitude than the value $\delta \bar{\gamma}_T \bar{u}(0)/\bar{\gamma}$ just derived on the basis of assuming no lateral flow. The reduction in magnitude is accounted for by the negative divergence of lateral flow in regions where the axial velocity $u$ is increasing. That lateral flow convergence reduces the need to pull ice down towards the bed and therefore reduces the tendency to cool the bed to the point where the instability is suppressed. As explained, however, the large lateral flow velocities involved are not accurate because the model is valid only for wavelengths comparable to ice thickness (and smaller).

The key assumption involved in the construction of the model is that the surface perturbation $s^1$ does not feed back into mass balance component of the model,
while the secondary flow simply acts to keep the surface flat at the leading order.
The transverse wavelength at which that assumption breaks down is the length scale over which the instability grows in the axial direction. We give a formal derivation in the next subsection \ref{sec:outer}, showing that the leading order version of the appropriately rescaled the model in section \ref{sec:leading_order} then has comparable transverse and axial gradients in the ice surface, although variations in ice thickness still remain small.  As a result, rather than being forced to flow sideways in order to maintain a leading order flat upper surface, the local surface correction $h_1$ can equally cause axial ice flow to accelerate locally. 

%The relevant instability length scale is $\lambda^{-1} \sim \delta^2$ relative to the ice sheet length, or $\varepsilon^{-1}\delta^2$ relative to the ice sheet thickness scale, where $\varepsilon$ is the ice sheet aspect ratio. In using the ice flow model of section \ref{sec:model_construction} to analyze the instability, we are implicitly assuming that the instability length scale is much longer than the ice thickness scale, since the description of ice flow in the $x$-direction is of thin-film form.

The rescaling in subsection \ref{sec:outer} retains the constraint $\delta^2 \gg  \varepsilon$, and the leading order model obtained only makes formal sense: as section \ref{sec:small_delta_revisit} shows, the short-wavelength cut-off to the spatial instability in the leading order model of section \ref{sec:leading_order} occurs at lateral wavelengths comparable to ice thickness, there is no hope that a leading order model at the `outer' lateral scale $\varepsilon^{-1}\delta^2$ as in section \ref{sec:outer} can capture the behaviour of these short wavelengths. 

We can however ask whether any simpler alterations of the model in section \ref{sec:leading_order} are possible that will capture the same physics as the model in section \ref{sec:outer} if we consider lateral length scales $\sim \varepsilon^{-1} \delta^2$. As the model in section \ref{sec:outer} shows, we still do not expect to see $O(1)$ ice thickness variations in the lateral direction over that length scale, which is intermediate between ice thickness and ice sheet length scale; all we want to see is the surface elevation correction, $h_1$ in the model in section \ref{sec:leading_order} and $H_1$ in section \ref{sec:outer}, feeding back into the calculation of ice velocity in the $x$-direction. The obvious modification the model in section \ref{sec:leading_order} that allows gradients of the surface elevation correction $h_1$ to contribute to the axial flow is
\begin{equation} \label{eq:axial_modified} \nabla_\perp^2 u = \pd{(h+b+\varepsilon^2 h_1)}{x}. \end{equation}
With this modification and with $\varepsilon \ll \delta^2 \ll 1$, the long-wavelength limit of the model in the main paper and in section \ref{sec:leading_order} in fact becomes the same as the model in section \ref{sec:outer}, and in principle we now have a hope of seeing multiple ice streams emerge in a single periodic domain, provided that domain is wide enough. the price to pay is that we now have to resolve a relatively long ($y \sim \varepsilon^{-1}\delta^2 \gg 1$) and a shorter ($y \sim 1$) length scale simultaneously: as in \citet{MantelliSchoof2019}, we find poor separation of length scales.

Moreover, it also turns out that the seemingly small alteration to the axial flow model in \eqref{eq:axial_modified} changes the nature of the system of equations: $x$ is no longer a time-like direction. We can then no longer `step forward' from slice $i$ to slice $i+1$, but need to compute the solution in all slices simultaneously. In a loose sense, we have turned a problem that was `hyperbolic' in the $x$-direction into one that is `diffusive' in the $x$-direction, albeit with a small diffusivity; this is supported further by the fact that the surface geometry variable $H_1$ in section \ref{sec:outer} satisfies an elliptic problem, \eqref{eq:depth_int_outer}. The analysis of the main paper, treating $x$ as time-like, no longer applies:  consequently, we cannot apply the spatial stability analysis developed there to the modified model, nor can we use the forward integration routine of section \ref{sec:numerics} without admitting that the computation of the solution in the $i$th slice requires knowledge not only of the $(i-1)$th slice but of the $(i+1)$th slice. This obliterates one of the most desirable features of the model, the ability to compute steady states efficiently.

There is another parameter regime to consider. If we take $\delta^2 \sim  \varepsilon$, the onset of sliding occurs over a single ice thickness, and both, onset and transverse stabilization then require a model in which all three coordinates are scaled with ice thickness, and the stability analysis of the previous section breaks down entirely, precisely because it is based on thin-film scalings in the $x$-direction. In that case, we have to use a full, three-dimensional Stokes flow model at least locally to capture the transition from subtemperate sliding to a fully developed ice-stream-ice-ridge pattern. Computationally this is likely to be feasible at present only if a relatively limited part of the ice sheet needs to be modelled this way: effectively, we would have to be able to treat the onset region as an internal layer in the ice sheet as already suggested in section 4 of the main paper (that is, a boundary layer that is not attached to one of the boundaries), and model only that internal layer using a Stokes flow solver.

\subsection{A larger lateral scale} \label{sec:outer}

Persisting here with with the assumption that $\delta^2 \gg \varepsilon$, we formally derive a model that captures dynamics at a larger lateral length scale, where surface $\pdl{h}{x}$ driving the axial flow is not the same across the width of the domain. To do so, we rescale both horizontal coordinates in section \ref{sec:leading_order} with the instability length scale, and time $\tau$ correspondingly
\begin{equation} \label{eq:rescale} Y = \delta^{-2} \varepsilon y^*, \qquad X = \delta^{-2} x^*, \qquad \tau = \delta^{-2} t^*.  \end{equation}
Note that $X$ is the same as its counterpart in section 4 of the main paper; all that differs from the internal layer model developed there (the transition from unpatterned to patterned being an internal layer in $x$ for small $\delta$) is that we also introduce a rescaled later coordinate $Y$, and time variable $\tau$. In addition, we require the same thermal boundary layer coordinate as in section 4 of the main paper and in section \ref{sec:small_delta_revisit}, since we are effectively replicating the same internal layer formulation as in section 4 of the main paper, but with a wider lateral extent:
\begin{equation} Z = \delta^{-2}z^*, \end{equation}
and define $\Theta(X,Y,Z,\hat{t}) = T^*(x^*,y^*,z^*,t^*) $ as temperature in the boundary layer. We also write 
\begin{equation} h(x,y,t) = \bar{H} + \delta^2H_1(X,Y,\tau)\end{equation}
with $\bar{H}$ constant to account for the fact that the surface slope scale remains the same, but a short horizontal length scale $\sim \delta^2$ relative to ice sheet length implies only small variations in surface elevation. We also put 
\begin{equation} U = u^*,  \qquad V = \varepsilon v^*, \qquad  W = \delta^{2} w^*,\qquad  P =  \delta^{-2} \varepsilon^2 p_1^*,\end{equation}
which replaces the definitions of $U_o$, $V_o$, $W_o$  and $P_o$ in section 4 of the main paper, bearing in mind that `$p$' in the main paper is the pressure correction $p_1^*$ relative to the cryostatic $p_0^* = h_0^*-z^*$ (reintroducing the asterisk superscripts throughout for clarity) in section \ref{sec:leading_order}.

These rescalings formally lead to a restricted `shallow-ice' type gravity current problem at leading order:
\begin{subequations} \label{eq:large_scale}
\begin{align} \pd{^2 U}{z^2} = & -\theta +   \pd{H_1}{X} & \mbox{for } 0 < z < H, \\
\pd{^2 V}{z^2} = &  \pd{H_1}{Y} & \mbox{for } 0 < z < H, \\
 \pd{U}{X} + \pd{V}{Y} + \pd{W}{z} = & 0 & \mbox{for } 0 < z < H, \label{eq:incompressible_outer}  \\
\pd{U}{z} = & 0 & \mbox{at } z = H, \\
\pd{V}{z} = & 0 & \mbox{at } z = h, \\
\pd{U}{z} = & \Gamma(\Theta)U & \mbox{at } z = 0, \\
\pd{V}{z} = & \Gamma(\Theta)V & \mbox{at } z = 0, \\
W = & 0 & \mbox{at } z = 0, \,H \label{eq:kinematic_outer}  \\
Pe \,\left(\pd{\Theta}{\tau} + U|_{z = 0}\pd{\Theta}{X} + B|_{z=0}\pd{\Theta}{Y} + \left.\pd{W}{z}\right|_{z=0} Z \pd{\Theta}{Z}\right) - \pd{^2 \Theta}{Z^2} = & 0 & \mbox{for } Z > 0, \label{eq:heat_outer}  \\
Pe \, \pd{\Theta}{\tau} - \pd{^2\Theta}{Z^2} = & 0 & \mbox{for } Z < 0, \\
\left[-\pd{\Theta}{Z}\right]_-^+ = & \alpha \Gamma(\Theta) U^2  & \mbox{at } Z = z = 0, \\
\pd{\Theta}{\tau} + U|_{z = 0}\pd{\Theta}{X} + B|_{z=0}\pd{\Theta}{Y} + \left.\pd{W}{z}\right|_{z=0} Z \pd{\Theta}{Z} \sim & 0 & \mbox{for } Z \rightarrow \infty, \\
\pd{\Theta}{Z} \rightarrow & G & \mbox{as } Z \rightarrow -\infty, 
\end{align}
with
$$ P = H_1. $$
Here $\theta = -\rd b / \rd x$  is the local bed slope as previously used, and $\Gamma(\Theta) = \gamma(T)$ is the rescaled friction factor function also used in section 4 of the main paper. We have for simplicity also only formulated this model for subtemperate sliding. A generalization to incorpoprate fully temperate sliding is staightforward.
\end{subequations}

Note that our primary interest is in the heat equation \eqref{eq:heat_outer}. The ice flow problem being of shallow-ice type, it can be depth-integrated to yield
\begin{equation} U_{z = 0} = \Gamma^{-1} H \left(\theta-\pd{H_1}{X}\right), \qquad V|_{z = 0} = -\Gamma^{-1} h \pd{H}{Y},\label{eq:slide_outer} \end{equation}
while the leading-order kinematic condition \eqref{eq:kinematic_outer} combined with \eqref{eq:incompressible_outer} requires the depth-integrated flux to have zero divergence at leading order, leading to an elliptic problem for $H_1$
\begin{equation} \pd{}{X} \left[\left(\Gamma^{-1}H^2 + \frac{H^3}{3}\right) \left(\theta - \pd{H_1}{X}\right)\right] + \pd{}{Y} \left[\left(\Gamma^{-1}H^2 + \frac{H^3}{3}\right) \left( - \pd{H_1}{Y}\right)\right] = 0. \label{eq:depth_int_outer} \end{equation}

In an ideal world, we would like to repeat our spatial stability analysis of section \ref{sec:supplementary_stability_analysis} at this longer transverse length scale. Unfortunately, we are out of luck: while the original model of section \ref{sec:model_construction} allows $x$ to be treated as being time-like, this is no longer the case at larger lateral length scales, precisely because \eqref{eq:depth_int_outer} is elliptic in $X$ and $Y$ as independent variables; $C$ is no longer time-like. The exercise of deriving the reduced model \eqref{eq:large_scale} was primarily to determine that the relevant outer length scale is indeed $\varepsilon \delta^{-2}$ in the lateral direction, relative to the original transverse coordinate $y^*$.

%\bibliography{references}